\documentclass[epsfig,a4paper,twocolumn,noshowpacs,floatfix,groupedaddress,
amsmath,prl,preprintnumbers,nofootinbib,citeautoscript]{revtex4}
\usepackage[T1]{fontenc}
\usepackage[latin1]{inputenc}
\usepackage{amssymb,amsmath,bm,dcolumn,graphicx}
\usepackage{color,booktabs,wrapfig,microtype,afterpage} \graphicspath{{Figures/}}
\usepackage[charter,greekuppercase=italicized]{mathdesign}
\usepackage{sidecap}
\usepackage{url}

\makeatletter\renewcommand{\fnum@figure}[1]{\textbf{\sffamily\figurename~\thefigure~|\,}}\makeatother
\makeatletter\renewcommand{\fnum@table}[1]{\textbf{\sffamily\tablename~\thetable~|\,}}\makeatother

\definecolor{NatureBlue}{rgb}{0.012,0.3,0.63}
\definecolor{NiceOrange}{rgb}{0.85,0.42,0.21}
\usepackage[colorlinks,plainpages=false,linkcolor=NatureBlue,urlcolor=NatureBlue,
citecolor=NatureBlue,pdfpagemode=UseNone,pdfstartview=FitBH]{hyperref}

\makeatletter
\newcommand{\bibstyle@supplement}{\bibpunct[, ]{[S}{]}{;}{n}{,}{,}%
    \gdef\bibnumfmt##1{[S##1]}}
\makeatother

\newcommand{\Red}{\textcolor{black}}
\newcommand{\Blue}{\textcolor{NatureBlue}}

\graphicspath{{figures/}}

\begin{document}

\title{\flushleft\fontsize{15pt}{15pt}\selectfont\sffamily \textcolor{NiceOrange}
{Computational models for active matter}}

\author{\sffamily M.\ Reza Shaebani$^{1}$, Adam Wysocki$^{1}$, Roland G.\
Winkler$^2$, Gerhard Gompper$^{2}$ \& Heiko Rieger$^{1}$\smallskip}

\affiliation{\flushleft
\mbox{\sffamily $^1$\hspace{0.5pt}Department of Theoretical Physics \& Center
for Biophysics, Saarland University, 66123 Saarbr\"ucken, Germany}
\mbox{\sffamily $^2$\hspace{0.5pt}Theoretical Soft Matter and Biophysics, Institute of Complex Systems \& Institute
for Advanced Simulation,}
\mbox{\sffamily $ $\hspace{0.5pt} Forschungszentrum J\"ulich, 52425 J\"ulich, Germany}\\
\vspace*{3pt}}

\begin{abstract}\citestyle{nature}
\parfillskip=0pt\relax\fontsize{9pt}{11pt}\selectfont\noindent\textbf{
A variety of computational models have been developed to describe active 
matter at different length and time scales. The diversity of the methods 
and the challenges in modeling active matter---ranging from molecular motors 
and cytoskeletal filaments over artificial and biological swimmers on 
microscopic to groups of animals on macroscopic scales---mainly 
originate from their out-of-equilibrium character, multiscale 
nature, nonlinearity, and multibody interactions. In the present review, 
various modeling approaches and numerical techniques are addressed, compared, 
and differentiated to illuminate the innovations and current challenges in
understanding active matter.  The complexity increases from minimal microscopic 
models of dry active matter toward microscopic models of active matter in 
fluids. Complementary, coarse-grained descriptions and continuum models are 
elucidated. Microscopic details are often relevant and strongly affect collective 
behaviors, which implies that the selection of a proper level of modeling is 
a delicate choice, with simple models emphasizing universal properties and 
detailed models capturing specific features. Finally, current approaches to 
further advance the existing models and techniques to cope with real-world 
applications, such as complex media and biological environments, are discussed.
}\end{abstract}

\pagestyle{plain}
\makeatletter
\renewcommand{\@oddfoot}{\hfill\bf\scriptsize\textsf{\thepage}}
\renewcommand{\@evenfoot}{\bf\scriptsize\textsf{\thepage}\hfill}
\makeatother

\maketitle

\renewcommand\bibsection{\section*{\sffamily\bfseries\footnotesize References\vspace{-10pt}\hfill~}}

\begin{figure*}
\centering
\includegraphics[width=0.9\textwidth]{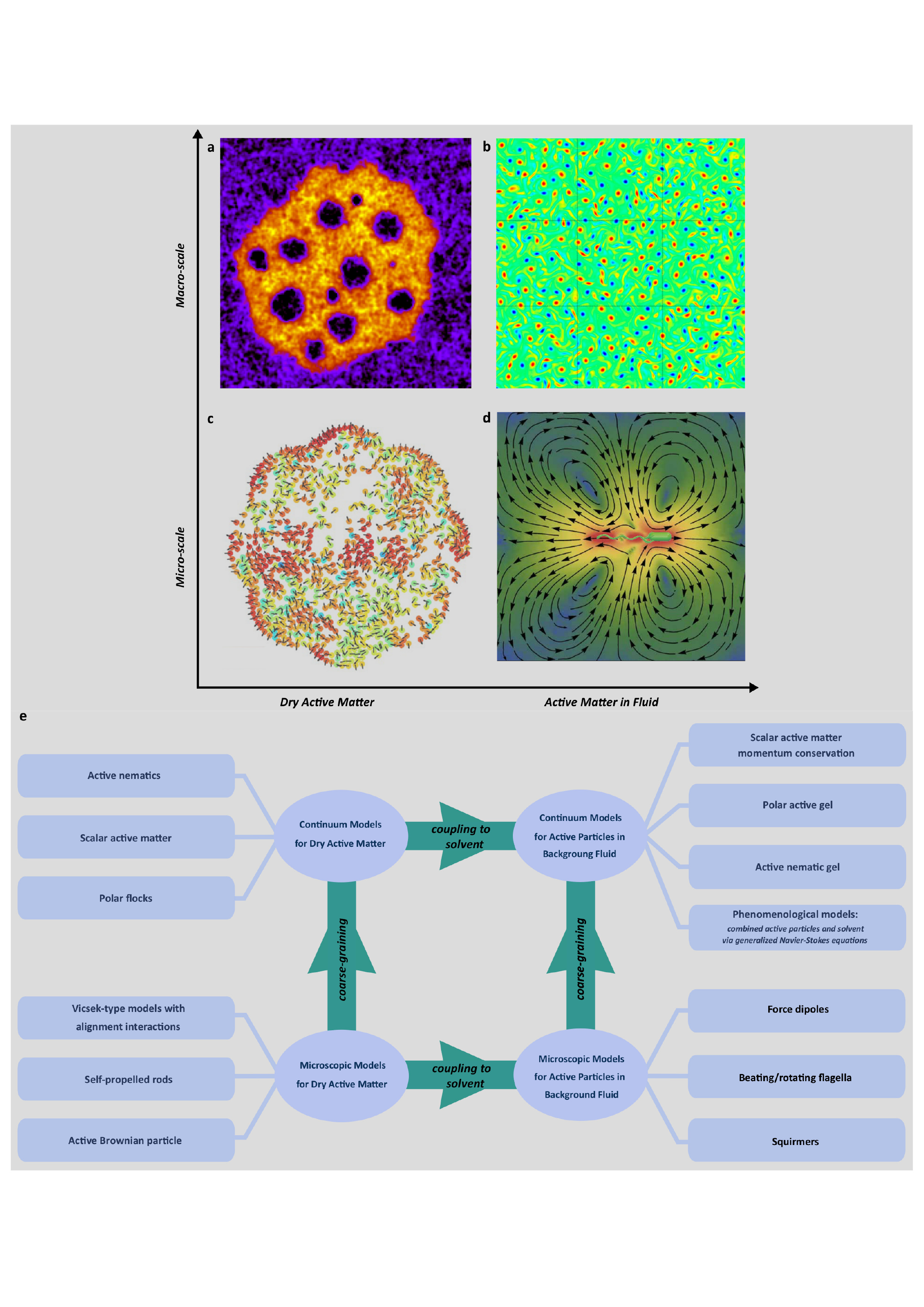}
\caption{{\bf Modeling active matter. a-d}| Typical numerical snapshots
of active systems in (dry-wet, micro-macro) phase space. \Red{{\bf a}| 
Bubbly phase separation in the density profile of a continuum dry active 
matter model. {\it Boiling liquid} and {\it vapor} phases are indicated 
with yellow and blue colors, respectively. Adapted from [\cite{Tjhung18}]. 
{\bf b}| Vorticity field in the turbulent regime of a continuum active
fluid model. Adapted from [\cite{Bratanov15}]. {\bf c}| Ordering of 
vibrated polar disks in confinement. The color code denotes the relative 
alignment of dry active particle to the neighboring particles, ranging 
from blue (anti-parallel) to red (parallel). Adapted from [\cite{Weber13}]. 
{\bf d}| Flow field generated by a single swimming bacterium. Adapted 
from [\cite{Hu15}].} {\bf e}| Schematic drawing of the main computational 
models and methods for active matter. The arrows indicate the direction 
of generalization or increasing complexity in approaches, from microscopic 
to continuum models, and from dry active matter to active motion with 
coupling to a momentum-conserving solvent.}
\label{Fig1}
\end{figure*}

Active matter consists of particles, agents, or constituents that consume energy
and convert it into directed motion,  generate forces and shape deformations, 
or even proliferate and annihilate. Active matter systems are therefore
manifestly out of equilibrium, and the local energy consumption discriminates
them from other out-of-equilibrium systems, in which energy is injected, e.g., 
via the boundaries like in shear flow. Some of the basic features of active matter 
are depicted in Box~1a. New models, methods, and computational techniques have been
developed in the last two decades to understand and unravel the unique physical
principles governing active matter [\cite{Marchetti13,Ramaswamy10,Toner05,Elgeti15b,
Bechinger16}]. Of particular interest are novel many-particle effects or collective 
phenomena, like motility-induced phase separation, spontaneous rotational symmetry 
breaking in two dimensions, pattern formation, and self-organization. Living systems 
are paradigmatic examples of active matter, where active units reproduce, adapt, and
dynamically respond to environmental changes. The huge diversity of active agents
and their wide range of behaviors are major challenges in developing a comprehensive
theoretical description of living matter. A large variety of numerical methods with 
different levels of resolution, ranging from micro- to macroscale, have been developed 
and employed to model living and artificial active matter [\cite{Ramaswamy10,Toner05,
Marchetti13,Elgeti15b,Bechinger16}], as summarized in Fig.~\ref{Fig1}. The goal of 
this review is to summarize, compare, and differentiate between currently available 
models and to elucidate existing challenges in computational modeling of active matter.

We first consider dry active matter, i.e., systems where hydrodynamic interactions
are absent and momentum is not conserved. Next, approaches to model hydrodynamics
of active suspensions are discussed. Here, the dynamics of the solvent is incorporated 
in the model, ensuring local momentum conservation. Moreover, we present an overview 
of the continuum models employed for active fluids, and elucidate the practical 
relevance of the numerical approaches. The degree of coarse-graining determines 
the details of real systems that can be captured, which becomes most evident in 
the modeling of cells, tissues, and animal groups. Finally, we address the open 
challenges that modeling active matter is currently facing, and present an outlook 
on model developments toward real-world applications.

\Blue{\section{Dry active matter}}
\label{Sec:Dry}

Dry active systems are characterized by the absence of momentum conservation.
This can be due to contact with a momentum-absorbing medium, as in gliding of
bacteria or vibration of granular beads on frictional surfaces. The omission
of momentum conservation may also originate from the minor relevance of 
hydrodynamic interactions in systems where other effects, such as fluctuations, 
volume exclusion, and short-range or metric-free interactions, dominate. Relevant
examples include animal flocks and swimming of dense bacterial collections.
Specifically, we discuss two paradigmatic minimal models of dry active matter
and their variants and extensions: active Brownian particles and Vicsek-type
models with alignment interactions. In addition, we briefly address 
continuum-modeling approaches. Neglecting birth, division, and death processes, 
we here only consider systems in which the number of particles is conserved.

\Blue{\subsection{\it Active Brownian particles}}

A rather generic model of an active agent is the {\it active Brownian particle}
(ABP), a self-propelled spherical particle whose dynamics is described by the
overdamped Langevin equations
\begin{align}
\centering
&\frac{\text{d}\bm{r}(t)}{\text{d}t}=\frac{D}{k_BT}\big[-\bm\nabla
U(\bm{r})+\bm{F}_{\text{active}}(\bm{\theta})+\bm\xi(t)\big], \nonumber \\
&\frac{\text{d}\bm{\theta}(t)}{\text{d}t}=-\frac{D_r}{k_BT}\bm\xi_r(t),
\label{Eq:ABP}
\end{align}
for the position $\bm{r}(t)$ and the orientation vector $\bm{\theta}(t)$. $D$
and $D_r$ are the translational and rotational diffusion coefficient, $\bm\xi$
and $\bm\xi_r$ uncorrelated random force and torque (white noise), $\bm{F}_{
\text{active}}$ a self-propulsion force (usually along the current direction
of motion, i.e., $\bm{F}_{\text{active}}{=}F\!\!_\circ\frac{\bm \theta}{|\bm \theta|}\,$),
and $U(\bm r)$ is either the interaction energy with other particles
or an external potential. The overdamped limit is suitable for motion in
low-Reynolds-number fluids as, e.g., in biological environments. In order to
model experimental systems, the equations of motion \ref{Eq:ABP} often have to
be generalized and adapted to cope with specific applications. Some of the
extensions of the ABP model are discussed in Box~1b [\cite{Romanczuk12,
Nguyen14,Lowen16,Peruani16,tenHagen15,Kaiser15,Eisenstecken16,Eisenstecken17}].
Other approaches to simulate self-propelled particle systems include active 
lattice gas model and (kinetic) Monte Carlo approach [\cite{Kourbane-Houssene18,
Klamser18,Sadjadi15,Shaebani14,Levis14,Najafi18,Hafner16}].
\\
\begin{center}
\includegraphics[width=0.48\textwidth]{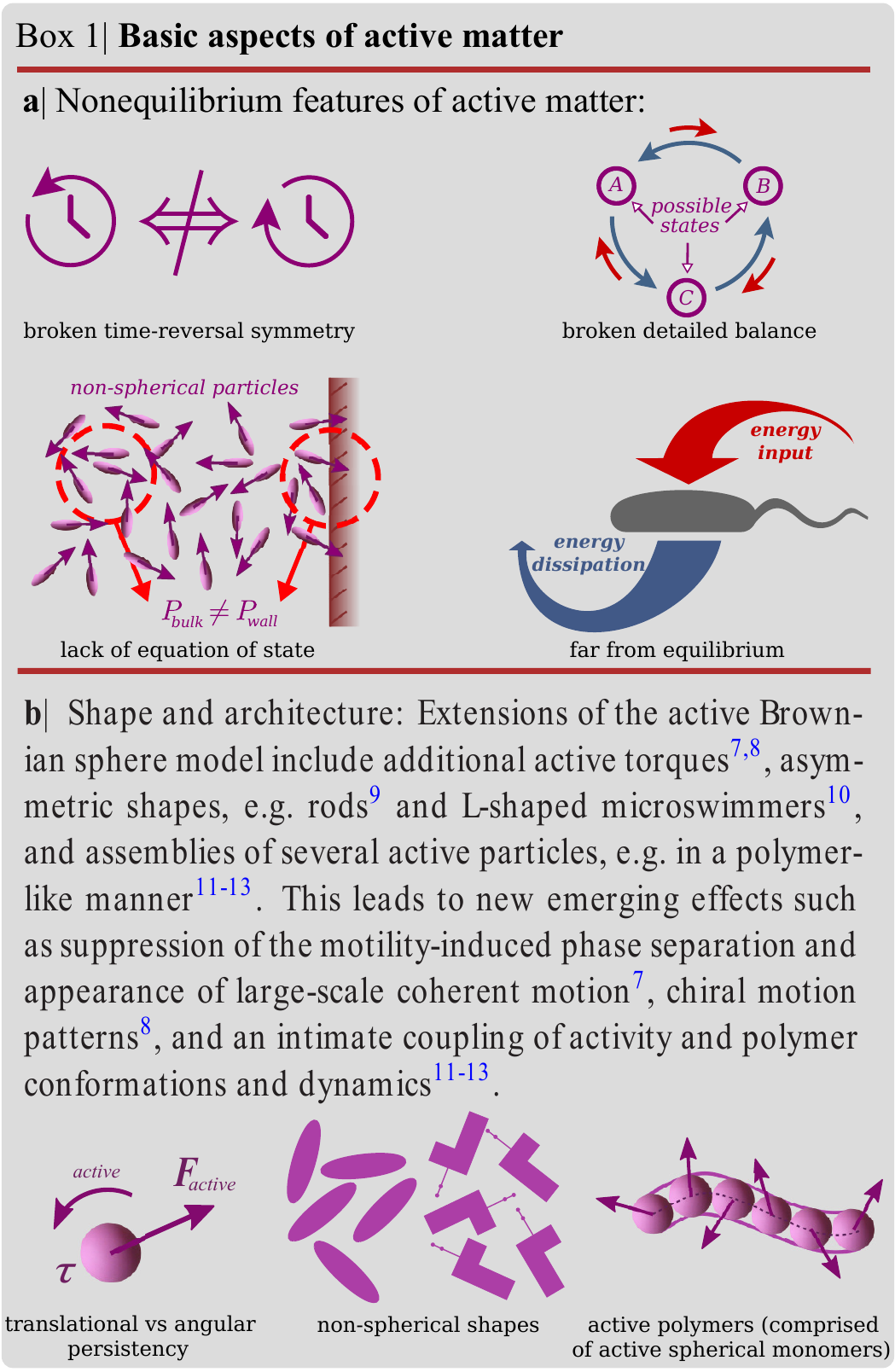}
\end{center}

Active particles, even with purely repulsive interactions, exhibit novel features,
such as motility-induced phase separation (MIPS) [\cite{Bechinger16,Redner13,Cates15,
Wysocki14,Stenhammar14,Elgeti15b,Wysocki16,Stenhammar15,Digregorio18,Fily14b}], 
wall accumulation [\cite{Elgeti13,Fily14,Das19}], \Red{capillary action in spite of 
wall-particle repulsion [\cite{Wysocki19}],} and an active pressure (denoted 
as swim pressure) [\cite{Takatori14,Solon15,Winkler15,Fily18,Das19}]. Intuitively, 
the separation into dense and dilute regions of ABPs is explained by a positive 
feedback between blocking of persistent particle motion by steric interactions, 
and an enhanced probability of collisions with further particles at sufficiently 
large concentrations. \Red{The reduction in speed by collisions implies a local 
increase in density, which further increases the collision frequency in those 
regions. This mechanism eventually leads to particle accumulation and phase
separation. During MIPS, the phase-separated domains grow self-similarly with 
time and their size is only limited by the system dimensions [\cite{Wysocki14,
Stenhammar14}]}. Remarkably, in three dimensions, ABPs exhibit collective motion 
in the high-density phase-separated state without alignment rule [\cite{Wysocki14}]. 
Similarly, wall accumulation emerges by the (slow) orientational diffusion of 
adsorbed ABPs, which are only able to escape when their propulsion direction 
points away from the wall [\cite{Elgeti13,Elgeti15b}]. The strength of the 
effects depends on the rotational diffusion coefficient, the propulsion velocity, 
and the curvature of the surface. Simulation show that ABPs accumulate 
preferentially in regions of highest curvature [\cite{Fily14,Wysocki15}]. 
While non-equilibrium active systems usually lack a free energy and equations 
of state, spherical ABPs are a notable exception. Analytical considerations 
and simulations yield a pressure equation of state in this case 
[\cite{Takatori14,Solon15,Winkler15,Fily18,Das19}]; however, such equation 
does not exist for nonspherical, elongated ABPs [\cite{Solon15}].

\begin{table*}
\centering
\includegraphics[width=0.9\textwidth]{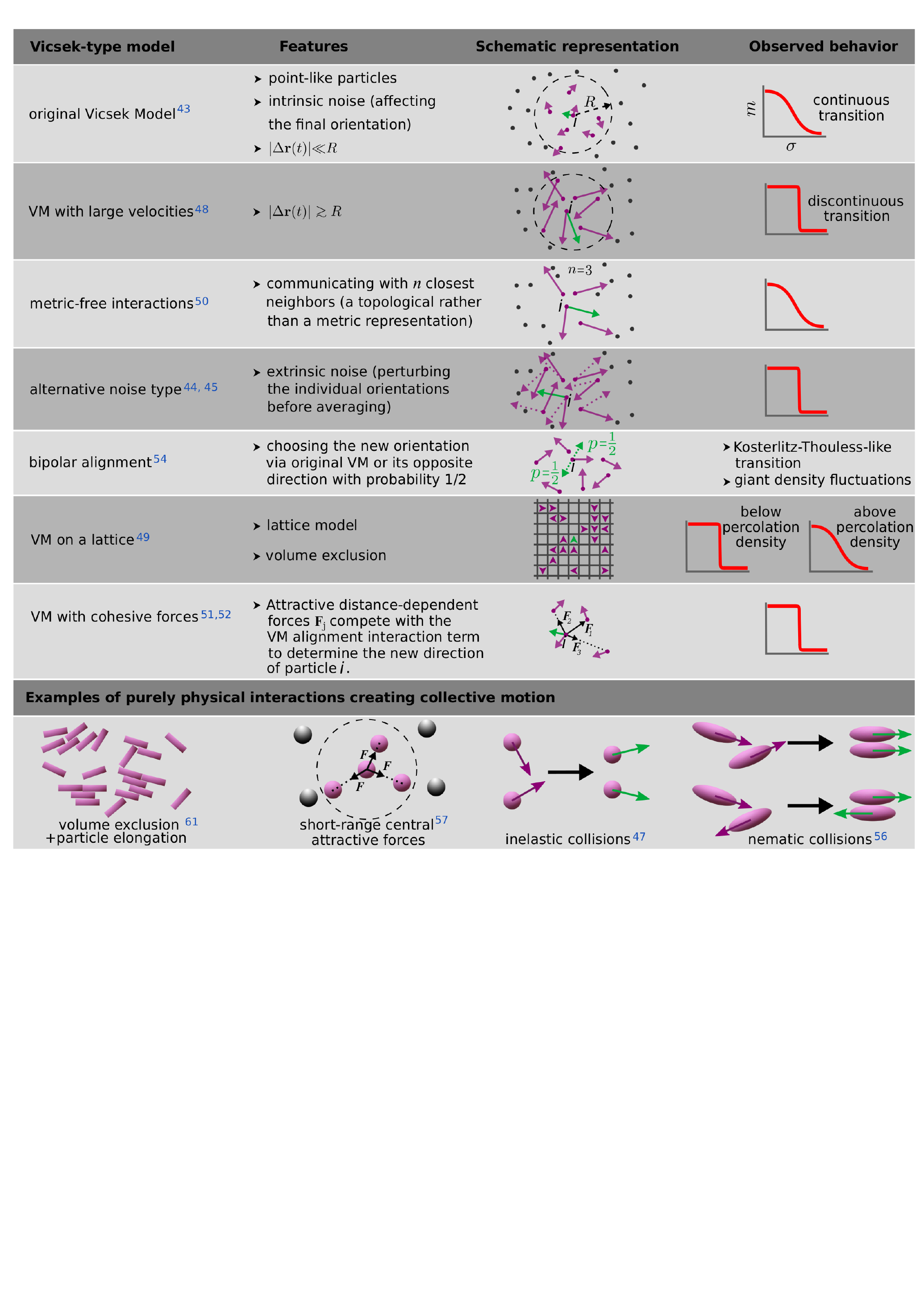}
\caption{{\bf (top) A few variants of Vicsek model.} \Red{The green vectors 
indicate the adapted orientation of the central particle $i$. The purple 
arrows are the directions that enter in the calculation for the new 
direction of particle $i$.} \Red{$\sigma$}, $m$, $R$, and $\Delta{\bm 
r}(t)$ represent, respectively, the noise strength, order parameter, 
radius of interaction, and particle displacement during one time step. 
{\bf (bottom) Examples of models exhibiting collective motion in active 
particle systems solely based on physical interactions without explicit 
alignment rules.}}
\label{Tab1}
\end{table*}

\Blue{\subsection{\it Active motion with alignment interactions}}

Collective motion in various systems of living organisms (e.g., flocks of birds 
and animal herds) and ensembles of synthetic elongated active particles shares 
intriguing common features---such as swirling patterns and swarming---related 
to the alignment of motion with their neighbors [\cite{Vicsek12}].

Vicsek et al.~proposed an agent-based minimal model for flocking, accounting
for the interplay between fluctuations and simultaneous interactions of 
multiple agents [\cite{Vicsek95}]. Particles moving in a plane with constant 
velocity $v_0$ align with their neighbors by updating their direction of 
motion at each time step according to
\begin{equation}
\bm{\theta}_i(t{+}1)=\langle\bm{\theta}_i(t)\rangle_R+\bm{\xi}_i(t),
\label{Eq:Vicsek}
\end{equation}
where $\langle\bm{\theta}_i(t)\rangle_R$ is the average orientation
vector of particles located in a circle of radius $R$ surrounding
particle $i$, and $\bm{\xi}_i(t)$ is a random vector with orientation
obtained from a uniform distribution \Red{$[-\sigma\pi,\sigma\pi]$} along
the current direction of motion. The strength of the noise is tuned
by the parameter \Red{$\sigma\!\in\![0,1]$}. The new position of particle
$i$ is then obtained as $\bm x_i(t{+}1){=}\bm x_i(t){+}v_0\,\hat
{\bm \theta}_i$, where $\hat{\bm \theta}_i$ is the unit vector in the
direction of $\bm{\theta}_i(t{+}1)$ and $v_0$ the velocity. A continuous 
transition from disordered to an ordered state occurs upon increasing 
the particle density or decreasing the noise strength \Red{$\sigma$}. The 
global mean normalized velocity $m =  | \sum_{i{=}1}^N \bm{v_i}|/(N v_0)$
is an appropriate order parameter, characterizing the transition from
random ($m{=}0$) to coherent movement ($m{=}1$). Despite of being a
nonequilibrium phase transition, various notions of equilibrium
statistical mechanics are borrowed here due to spontaneous symmetry
breaking and emergence of well-defined macroscopic states.

According to numerous studies [\cite{Vicsek12}], the way in which noise and
disorder are introduced into the system [\cite{Aldana07,Aldana09,Peruani18}],
boundary conditions (periodic vs reflecting, shape, etc.) [\cite{Grossman08}],
range [\cite{Nagy07}] and type of interactions (including hard core,
metric-free, repulsive or attractive) [\cite{Peruani11,Ginelli10,Gregoire03,
Chate08}], and alignment rules (e.g.,~polar or bipolar) [\cite{Szabo09,
Chate06,Mahault18}] influence the resulting patterns (leading to band 
formation [\cite{Ginelli10b,Chate08}], rotating chains [\cite{Strombom11}],
marching groups, etc.) and the essential characteristics of collective motion
such as the nature of the phase transition. \Red{For example, the order of the
transition depends on: (i) whether the noise is {\it intrinsic} (perturbing 
the final orientation) or {\it extrinsic} (perturbing the individual orientations
before averaging) [\cite{Aldana07,Aldana09}], (ii) whether interactions are metric 
(fixed range) or topological (fixed number of partners) [\cite{Ginelli10}], and 
(iii) the relative magnitude of the particle displacement $v_0 \Delta t$ during 
one time step compared to the radius of interaction $R$ [\cite{Nagy07}]. Recent 
studies indicate that the flocking transition in the Vicsek model is reminiscent 
of a liquid-gas transition, rather than an order-disorder transition, but with 
microphase separation in the coexistence region, where traveling ordered bands 
of finite width coexist with a disordered gas [\cite{Solon15b}]. By contrast, 
the corresponding lattice model, called active Ising model, exhibits full phase 
separation [\cite{Solon13}].} In animal herds, agents sense the motion of their 
neighbors, resulting in polar alignment. However, an explicit (polar) alignment 
rule is not a necessary condition for coherent motion: the alignment can be 
induced implicitly through purely physical local interactions such as inelastic 
[\cite{Grossman08}] or nematic collisions [\cite{Ginelli10b}], short-range 
\Red{interactions} [\cite{Strombom11,Szabo06}], or volume exclusion in combination 
with shape-induced (particle elongation) effects [\cite{Peruani06}]. An important 
question is the minimal requirements for emerging collective motion by solely 
physical interactions. Table~\ref{Tab1} lists some variants of the Vicsek model 
as well as a few systems which show collective motion without explicit alignment 
rules.

\Blue{\subsection{\it Continuum models of dry active matter}}

The large-scale behavior of many-particle systems can be captured by continuum 
models, which describe the evolution of continuous slow variables such as number 
density $n(\bm r,t)$ and velocity $\bm{v}(\bm r,t)$ fields. Such macroscopic 
approaches capture the major features of active matter, e.g., collective 
motion, by considering the conservation laws and broken continuous symmetries. 
Continuum theories can be constructed via coarse graining of a microscopic 
model, adopting symmetry arguments, or using out-of-equilibrium thermodynamics 
close to equilibrium. Representative continuum models of dry active matter, 
which conserve the number of particles while the momentum is not conserved, 
are provided in Table~\ref{Tab2}.

\begin{table*}
\centering
\includegraphics[width=0.88\textwidth]{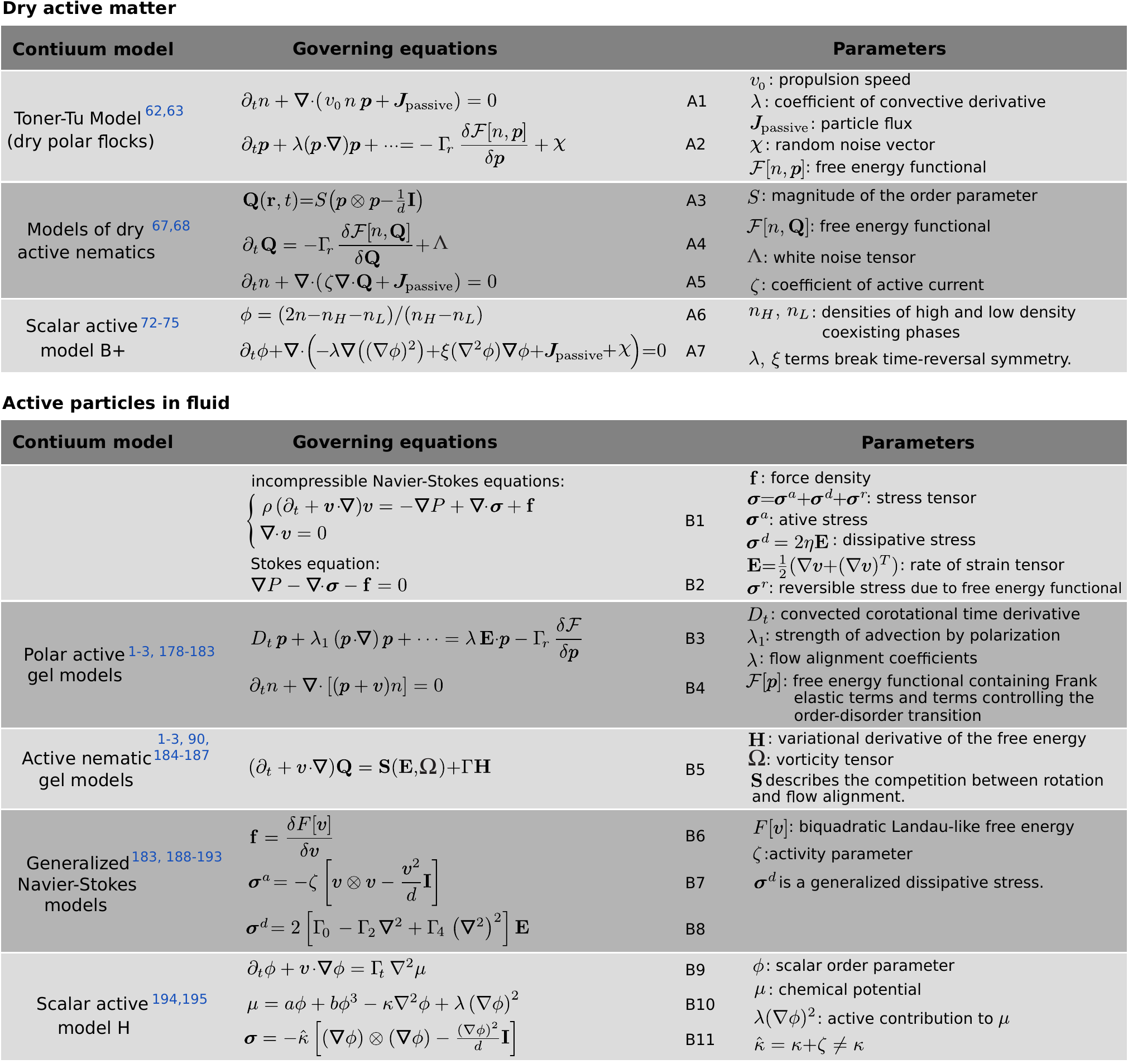}
\caption{{\bf The governing equations of continuum models for dry active
matter (top) and active particles in fluids (bottom)}. $P$ denotes pressure,
$n, \rho$ number and mass density, ${\bm v}$ velocity, ${\bm Q}({\bm r},t)$
nematic alignment tensor, ${\bm p}$ polarization or nematic director, ${\bf
I}$ identity tensor, $T$ temperature, $\Gamma_t, \Gamma_r$ translational
and rotational mobility, and $\eta$ viscosity.}
\label{Tab2}
\end{table*}

{\it Dry polar flocks --} A continuum description of the Vicsek model, 
solely based on symmetry considerations, was first proposed by Toner and 
Tu [\cite{Toner95,Toner98,Toner05,Bertin06,Ihle11}]. The theory describes 
active particles distinguishing front from rear and, therefore, are 
characterized by a polarization field ${\bm p}({\bm r},t)$ corresponding 
to the vectorial orientation of active particles and their direction of 
swimming. The conservation of the number density, $n({\bm r},t)$, results 
in a continuity equation including an active contribution $v_0\,n\,{\bm p}$ 
to the particle flux (see equation A1 in Table~\ref{Tab2}). The simplest 
form of the dynamical equation for ${\bm p}$ assumes a relaxation process 
toward the minimum of an effective free energy $\mathcal{F}[n,{\bm p}]$ 
including, amongst others, terms responsible for the spontaneous 
polarization and the energetic cost due to elastic distortions of the
orientational field. As shown in equation A2, activity enters solely via
an advective term $\lambda({\bm p}{\cdot}\bm\nabla){\bm p}$, which
accounts for the fact that distortions in ${\bm p}$ are advected by itself
because ${\bm p}$ represents both the order parameter and the velocity.
Flocks do not conserve momentum and the system lacks Galilean invariance,
therefore, $\lambda{\neq}v_0$, meaning that density and polarization
inhomogeneities advect at different velocities. The model exhibits giant
number fluctuations and long-range order in 2D (forbidden in thermal 
equilibrium by the Mermin-Wagner theorem) [\cite{Toner05,Toner95}]. 
By relaxing the constraint on number-density conservation, the model 
can be generalized and applied to systems involving birth-division-death 
processes [\cite{Ranft10}] as, e.g., in dense bacterial colonies.

{\it Dry active nematics --} A system of apolar (or head-tail symmetric)
active particles can exhibit a state with long-range directional order,
however, with zero global drift velocity due to the nematic symmetry.
Active nematics can be characterized by a symmetric second-rank tensor
$\bm{Q}({\bm r},t)$ representing the local alignment of the neighboring
particles (see equation A3). The overdamped dynamics of $\bm{Q}$ obeys
equation A4, \Red{reducing to the free energy} $\mathcal{F}[n,\bm{Q}]$ consisting 
of entirely quasi-passive terms. The continuity equation A5 for the 
density $n({\bm r},t)$ contains an active current ${\bm J}_\text{active}
{=}\zeta\bm\nabla{\cdot}\bm{Q}$ [\cite{Narayan07,Ramaswamy03}] originating 
from the \Red{force dipole} of active particles, which accounts for a particle 
flux along or against the curvature $\bm\nabla{\cdot}\bm{Q}$, \Red{thus 
violating the time-reversal symmetry}. Despite the apparent simplicity, 
active nematics exhibit interesting behavior, e.g.,~giant number 
fluctuations and self-propulsion of topological defects [\cite{Marchetti13,
Narayan07,Ramaswamy03}]. \Red{An extension to a more complex environment, 
e.g., viscoelasticity by polymers [\cite{Hemingway15}], implies additional 
effects, such as drag reduction, spontaneous flows by an antagonistic 
coupling between polymer and nematic orientations, and active turbulence 
in a sufficiently soft elastomeric solid [\cite{Schwarz13}].}

{\it Dry scalar active matter --} Anisotropic interactions are responsible
for orientational order-disorder transitions in active nematics or polar
matter. In contrast, active particles with spherical symmetry (i.e., without
alignment interactions) do not display a global directional order (${\bm p}
{=}\bm{Q}{=}0$) and the only remaining slow variable is the scalar number
density $n$. To explore the physics of motility-induced phase separation in 
scalar active matter, the passive Model B [\cite{Hohenberg77}]-- a field-theoretical 
model for diffusive phase separation of the number density (characterized by 
a conserved scalar order parameter $\phi$)-- has been extended by an active
chemical potential $\mu_\text{active}=\lambda(\bm\nabla\phi)^2$ that breaks
the time-reversal symmetry at leading order in the density gradient expansion
[\cite{Wittkowski14,Tjhung18,Stenhammar13,Cates13,tailleur08}]. To allow 
for circulating real-space particle currents in steady state, further terms 
that break the gradient structure of the active current have been added 
[\cite{Tjhung18}] (see equations A6 and A7). The resulting active model 
B+ (AMB+) displays complex dynamics such as microphase separation (bubbles 
of a finite length-scale) and reverse Ostwald process (increase of the 
number of small bubbles while bigger bubbles evaporate) [\cite{Tjhung18}].

\Blue{\section{Active particles in fluids}}
\label{Sec:Fluids}

Hydrodynamic interactions are fundamental for active particles immersed
in a fluid, and  determine their behavior \Red{in various respects}. On the one
hand, they are an integral part of \Red{the propulsion system} of most of the
biological and various synthetic microswimmers---without hydrodynamic 
interactions, no propulsion [\cite{Elgeti15b}]. On the other hand, 
hydrodynamics determines the behavior of microswimmers at walls, in 
channels, as well as their collective behavior. Microscale modeling 
approaches provide insight into the underlying physical mechanism from 
the level of individual microswimmers up to the emergent collective 
behaviors on \Red{large length scales}.

\Blue{\subsection{\it Universal features}}

The hydrodynamics of simple {\it Newtonian} fluids is governed by the
Navier-Stokes equation, see equation~{\it B1} in Table~\ref{Tab2}. In
the limit of low Reynolds numbers,
\begin{equation}
Re= \rho v_0 L/\eta \ll 1
\end{equation}
(with $L$, $v_0$, and $t_0{=}L/v_0$ being a characteristic length, 
swim velocity, and time), this equation reduces to the Stokes equation, 
where inertial terms are negligible (equation B2 in Table~\ref{Tab2}). 
Characteristic values for microswimmers are body lengths $L{\sim}O(\mu m)$ 
and swimming velocities $v_0{\sim}O(\mu m/s)$, hence, $Re{\lesssim}10^{-3}$
for water. In this limit, hydrodynamics becomes time independent and
the dynamics is reversible. This has the important consequence that 
a microswimmer with a time reversible stroke cannot propel (scallop
theorem), as first pointed out in a seminal paper by Purcell 
[\cite{Purcell77}].

The solution of the Stokes equation is determined by the Oseen tensor
\begin{equation} \label{eq:oseen}
H_{\alpha\beta}(\bm{r}) = \frac{1}{8\pi \eta r} \left[\delta_{\alpha,\beta}
+ r_\alpha r_\beta/r^2 \right], \;\;\;\;\;\;\;\;\;(r{=}|\bm r|)
\end{equation}
the Greens function of the Stokes equation. The flow
field obtained from $\mathbf{H}$ for a point force is typically denoted
as Stokeslet (cf.\ Box 2). It is important to note that the Stokeslet
does {\em not} describe an autonomous microswimmer, because the swimmer
must be force- and torque-free. Instead, a swimmer usually consists of
a {\it motor}, propelling the fluid, and a {\it cargo}, being pushed or
dragged forward. Approximating these two components by point forces with
opposite directions $\hat{\bm e}$ and $-\hat{\bm e}$, and equal magnitude
$f_0$, yields the {\it dipole swimmer} with the flow field of equation~C1,
where $P{=}f_0L$ is the dipole strength (cf.\ Box~2). The sign of $P$
distinguishes a {\it pusher} ($P{>}0$, motor in the back) from a {\it puller}
($P{<}0$, motor in front), with equal flow lines but opposite flow directions
(cf.\ Box 2) [\cite{Elgeti15b}]. The flow field of a microswimmer is typically
more complex and comprises higher order multipoles [\cite{Spagnolie12,Winkler18}].
However, in the far-field (far away from the swimmer compared to its own size)
the dipole contribution C1 dominates.

\begin{widetext}
\centering
\includegraphics[width=0.9\textwidth]{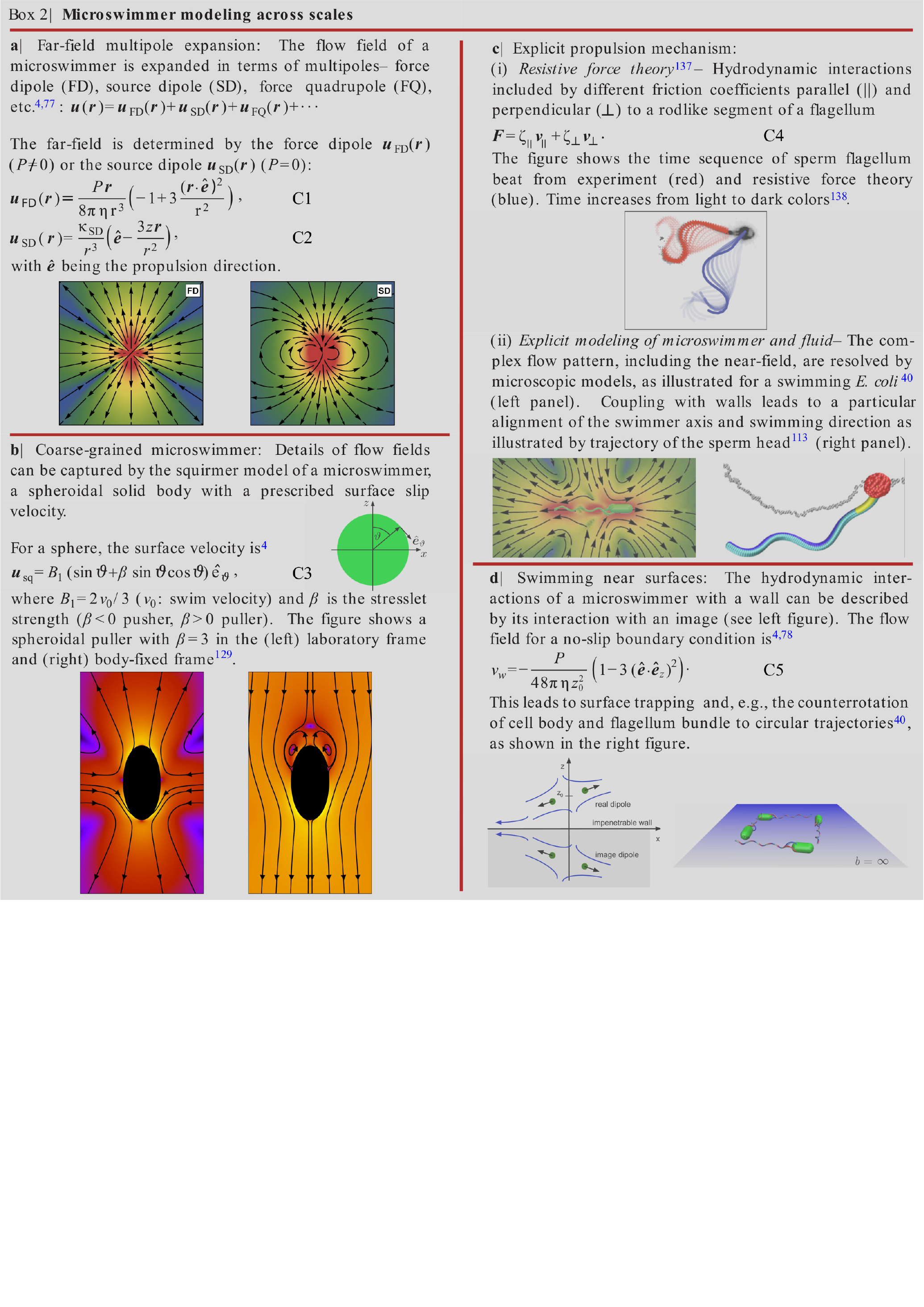}
\end{widetext}

The dipole flow field has important consequences for the interaction of
swimmers with walls [\cite{Berke08,Lauga09,Elgeti16,Spagnolie12,Elgeti15b}]
and other swimmers. Here, inflow/outflow generates an effective attraction/repulsion.
Together with the flow-induced torque, aligning pushers with a wall, pushers
are attracted at walls. It should be noticed that flow-induced interactions
add to an effective attraction emerging by propulsion, slow reorientation,
and steric interactions, which also exists for dry active matter, and is
thus independent of pusher- or puller-type swimming [\cite{Li09,Elgeti09,
Elgeti13,Elgeti15}]. In the latter case, the basic mechanism consists of
arrival at the wall with the propulsion direction toward the wall, a slow
reorientation determined by the rotational diffusion, during which the
swimmer stays at the wall, and finally departure when the orientation
points away from the wall. Hydrodynamics increases the wall detention
time [\cite{Schaar15}].

The hydrodynamic interactions also play an important role in the nematic
arrangement of elongated or rodlike microswimmers [\cite{Saintillan07,
Saintillan08}]. When the aspect ratio exceeds about five, the nematic
phase of passive rods is stable for sufficiently high volume fractions.
Such an arrangement of microswimmers is sensitive to slight perturbations,
e.g., by a sinusoidal reorientation wave. The flow field of pushers
(pullers) at the nodes of this wave  enhances (reduces) the perturbation,
and thus destabilizes (stabilizes) the nematic phase. This hydrodynamic
instability of the nematic phase of extensile active systems lies at the
heart of the intriguing dynamics of active nematics [\cite{Sanchez12,
Thampi13,Giomi13,Keber14}].

\Red{Hydrodynamics can also lead to high-speed and long-range communication 
between freely swimming cells. For example, studies of the protist {\em 
Spirostomum ambiguous} suggest that long-ranged vortex flows can be 
generated by fast cell contraction, which, in turn, trigger contraction 
of neighboring cells and, hence, promote collective behavior [\cite{Mathijssen19}].}

\Blue{\subsection{\it Swimming in viscoelastic fluids}}

Microorganisms often move in complex environments, which are rather viscoelastic
than Newtonian. Such environments break the time-reversal symmetry of Newtonian
fluids, and allow for self-propulsion even for a time-symmetric internal motion,
seemingly violating the scallop theorem [\cite{Qiu14}]. Viscoelastic fluids
are usually polymer solutions with a wide spectrum of  properties, ranging
from shear thinning to viscoelastic (characterized by storage and loss modulus),
and, hence, can affect the swimming  behavior of individual microswimmers
[\cite{Qin15,Patteson15}]  as well as their collective properties [\cite{Li16}].
Theoretical studies reveal both reduced [\cite{Lauga07,Fu09}] and enhanced
[\cite{Spagnolie13,Man15}] swim speeds, with a transition from slow small-amplitude
swimming to fast large-amplitude locomotion [\cite{Spagnolie13}]. Viscoelastic
[\cite{Liu11,Spagnolie13}] and shear thinning [\cite{Gagnon14}] effects are
typically small. However, some unexpected phenomena can appear, which are
related to the complex nature of polymer solutions. Microstructured fluids will
generically phase-separate near surfaces, which can lead to the presence of
low-viscosity fluid layers. Models show that this promotes slip and reduces
viscous friction near the surface of the swimmer, which may increase swim
speeds by orders of magnitude [\cite{Man15}]. Experiments of {\em E.~coli}
in concentrated polymer solutions indicate that peculiarities of flagellated
locomotion are indeed due to the fast-rotating flagellum, giving rise to a
lower local viscosity in its vicinity [\cite{Martinez14}]. A detailed modeling
and simulation of bacteria in dense polymer solutions, with explicit polymers, 
reaches similar conclusions, specifically a depletion of polymers at the 
flagellum is obtained [\cite{Zottl19}]. An increased swim speed with
increasing polymer density is predicted, due to a non-uniform distribution
of polymers in the vicinity of the bacterium, leading to an apparent slip,
in combination with the chirality of the bacterial flagellum [\cite{Zottl19}].

\Blue{\subsection{\it Hydrodynamics: mesoscale simulation techniques}}

Various hydrodynamic simulation approaches have been developed, which facilitate
the study of mesoscopic active-matter agents. Prominent mesoscale simulation
approaches are the {\it Lattice Boltzmann} (LB) method [\cite{McNamara88,Dunweg09}],
{\it Dissipative Particle Dynamics} (DPD) [\cite{Espanol95}], and the {\it
Multiparticle Collision Dynamics} (MPC) approach [\cite{Kapral08,Gompper09}].
All these approaches are essentially alternative ways of solving the
Navier-Stokes equation and its generalizations. The LB method yields an
approximate solution of the Boltzmann equation, i.e., a single particle
phase-space distribution function. An advantage of LB is that thermal
fluctuations can be turned on and off as desired [\cite{Dunweg09}]. DPD
and MPC are particle-based approaches, where the fluid is represented by
point particles. The DPD dynamics proceeds analogous to traditional molecular
dynamics (MD) simulations, however, with pairwise momentum-conserving
stochastic and friction forces. By specific pairwise DPD particle interactions,
compressibility of the fluid can be controlled. MPC consists of alternating
streaming and collision steps, with a ballistic streaming motion and local
momentum-conserving stochastic interactions (collisions), e.g., by
rotation of relative velocities [\cite{Gompper09}]. By the point-particle
nature of the MPC particles, no fluid-induced depletion occurs, and a rather
continuous representation of the fluid is obtained. Other simulation approaches
implicitly take hydrodynamic interactions into account via a hydrodynamic
tensor, e.g., Oseen tensor (equation \ref{eq:oseen}) [\cite{Elgeti15b}] or moment
expansion [\cite{Ishikawa08,Spagnolie12,Mathijssen16}] and mobilities
[\cite{Singh15}].

Coupling of a microswimmer with the fluid can be achieved in various ways,
depending on the nature of the microswimmer model and the desired extent
of coarse-graining. Rather detailed models of, e.g., sperm and {\em E.\ coli}
cells applying no-slip boundary conditions on the flagella and cell body,
combined with a momentum conserving propulsion mechanism, e.g., rotation
of a flagellum in case of a bacteria combined with counter-rotation of the
cell body, leads to swimming motion. This applies to microswimmers in an
explicit [\cite{Elgeti10,Hu15}] and implicit [\cite{Watari10,Shum10,
Pimponi16}] solvent.

Squirmers are a coarse-grained representation of microswimmers, modeled
as a colloid with prescribed fluid velocity at its surface (slip velocity
$\bm{v}_\text{sq}$) [\cite{Lighthill52,Blake71,Ishikawa06,Pedley16}]. This
approach was originally designed for ciliated microswimmers, such as {\em
Paramecia} [\cite{Pedley16}]. Nowadays, it is considered as a generic model
for a broad class of microswimmers, ranging from diffusiophoretic particles
[\cite{Bechinger16}] to biological cells, and has been applied to study 
collective effects in bulk [\cite{Llopis10,Gotze10,Ishikawa06,Evans11,
Alarcon13,Molina13,Yoshinaga17}], at surfaces [\cite{Ishikawa08,Llopis10,
Ishimoto13,Lintuvuori16}], and in narrow  slits [\cite{Theers16,Theers18}]. 
Typically, the slip velocity of a sphere is approximated by equation C3 of 
Box 2 [\cite{Blake71,Pedley16,Theers16}]. Extension to prolate spheroidal 
microswimmers have been proposed [\cite{Keller77,Ishimoto13,Theers16b}]. 
The squirmer model has been applied in combination with the boundary 
element method [\cite{Ishikawa06}], the LB approach [\cite{Llopis10,
Lintuvuori16,Yoshinaga17}], and the MPC [\cite{Theers16b,Theers18}] 
representation of the fluid.

A further level of coarse-graining is obtained by taking only far-field
hydrodynamics into account, and representing a microswimmer as a force
dipole (see above), where one particle is moving in the direction of the
applied force and the total momentum is conserved by imposing the opposite
force on the fluid [\cite{Nash10}]. Extensions to dumbbell-type swimmers
[\cite{Hernandez-Ortiz05}] or even more complex spherical [\cite{deGraaf16,
Menzel16}] and rodlike [\cite{Elgeti09,deGraaf16}] structures have been
proposed. Such an approach allows for the study of a large number of
microswimmers with a minimal numerical effort. However, the near-field is
not adequately accounted for [\cite{deGraaf16}], which becomes relevant for
the swimming behavior near a surface, in thin slits, or even the collective
behavior in dense systems.

\Blue{\subsection{\it Biological swimmers}}

Cell motility is a major achievement of biological evolution and is essential
for a wide spectrum of cellular activities, such as search for food, reproduction,
or escape from predators [\cite{Elgeti15b}]. The spectrum of microswimmers is
wide, ranging from bacteria, e.g., {\em  Escherichia coli}, protozoa, e.g.,
dinoflagellates, algae, e.g., {\em Chlamydomonas reinhardtii}, to spermatozoa.
Unravelling the underlying propulsion mechanisms is essential for the understanding
of their behavior, possible utilization of microswimmers in medicine, ecology,
and technical applications, or for biomimetics by transferring biological concepts
into synthetic swimmers. Biological microswimmers, both prokaryotes and eukaryotes,
exploit flagella for propulsion, although the structure of their flagella differs
[\cite{Lighthill76,Lauga09,Elgeti15b}]. Bacteria typically use one or several
rotating helical flagella for locomotion, whereas a flagellum (or cilium) of an
eukaryote beats in a wave-like fashion [\cite{Saggiorato17}]. Explicit modeling
of these microswimmers requires to account for three components: the cell body,
the flagellum or several flagella, and the embedding fluid (see above). Typically,
such cells are considered as neutrally buoyant objects, with spherical, spheroidal,
or cylindrical cell body and (an) attached flagellum/flagella. Both parts are
either considered as solid bodies [\cite{Lighthill76,Shum15,Pimponi16,Lauga16}]
or are composed of linked discrete points in a {\it crane-like} fashion for
eukaryotic [\cite{Elgeti10,Rode19}] or bacteria [\cite{Reigh12,Hu15,Reichert05,
Vogel12,Watari10,Janssen11}] flagella. For the latter, traditional polymer
models are employed, or a flagellum is described by the helical wormlike chain 
model [\cite{Elgeti15b}]. A bacterium is propelled by independent rotation
of flagella via an applied torque. Assigning the opposite forces and torques
to the cell body ensures a force- and torque-free swimming [\cite{Elgeti15b}].
In any case, propulsion is due to frictional anisotropy of the thin flagellum
[\cite{Lauga09,Elgeti15b,Lauga16}].

Simulation studies emphasize the importance of hydrodynamic interactions for,
e.g., the synchronization of bacteria flagella in the bundling process, e.g.,
for {\em E. coli} [\cite{Reigh12,Janssen11}] or the beating of {\em Chlamydomonas}
flagella. Moreover, they illustrate the complexity of the hydrodynamic flow field 
adjacent to a cell, which is important for cell-cell scattering/interaction
processes [\cite{Hu15}]. Interactions of microswimmers with surfaces are
fundamental in many biological processes, e.g., biofilm formation and egg
fertilization. As pointed out above (see Box 2), surface hydrodynamic interactions
determine the swimmer orientation [\cite{Spagnolie12}]. In the far-field approximation, 
{\em E. coli} and {\em sperm} are pushers and orient preferentially parallel to 
a surface, whereas {\em Chlamydomonas} is a puller and correspondingly aligns 
perpendicular to surfaces. For bacteria, the rotation of the helical bundle 
and the counter-rotation of the cell body lead to circular trajectories with 
a handedness and circle radius depending on the surface-slip length [\cite{Hu15b,
Lemelle13}]---clockwise trajectories follow for no-slip [\cite{Lauga06}] and 
counterclockwise trajectories for perfect-slip boundary conditions 
[\cite{DiLeonardo11}]. Simulations show that a cell is sensitive to nanoscale 
changes in the surface slip length, being itself significantly larger [\cite{Hu15b}].

Bacterial suspension show an intriguing chaotic state of
collective motion called {\it active turbulence} [\cite{Dombrowski04}].
Simulation studies of the collective behavior of spherical squirmers show
cluster formation in thin films with no-slip boundary conditions (quasi-2D)
[\cite{Theers18}]. However, no phase separation is obtained as for ABPs
[\cite{Theers18,Bialke12}]. The formation of small clusters, rather than
the appearance of MIPS for spherical squirmer, is attributed to changes
in the orientational dynamics by interference of the flow fields of the
individual squirmers [\cite{Matas-Navarro14,Theers18}]. In contrast,
spheroidal squirmers exhibit phase separation and swarming already for
rather small activities [\cite{Theers18}]. Hence, shape and hydrodynamics
together govern structure formation of active matter.

\Blue{\subsection{\it Artificial active matter}}

Various artificial active agents (motors) have been synthesized during the
last decades, ranging from tens of nanometers to micrometers, exploiting
diverse propulsion mechanisms [\cite{Bechinger16,Gaspard19,Bayati19}]. Prominent
strategies are based on the slippage of fluid at the surface of the solid
particle due to phoretic effects such as diffusiophoresis, involving
concentration gradients, thermophoresis, involving thermal gradients, and
electrophoresis by inhomogeneities in charge distributions of electrolytes
[\cite{Ruckner07,Yang11,Saha14,Michelin14,Liebchen15,Stark18,Gaspard19,Bayati19,
Bechinger16}]. Modeling of artificial active matter aims at a quantitative
understanding of the underlying propulsion mechanisms and the design of
novel propulsion strategies for practical and technological applications,
such as targeted drug delivery.

In general, phoretic effects appear by molecular interactions of a solute
in the solution together with the solid particle (motor). Surface reactions
generate inhomogeneous concentration fields of reactant and product species
in the vicinity of the motor implying concentration gradients over the
particle surface. These gradients cause a slippage between the fluid and
the particle via diffusiophoresis and induce propulsion of the particle
in the fluid. Directed motion appears by controlling the surface reaction,
e.g., an asymmetric reaction process on the surface of a Janus particle
[\cite{Gaspard19,Bechinger16,Moran17}]. The theoretical description of the active
process requires to solve the diffusion equation for the concentration of
the solute, the hydrodynamic transport problem, and the solid-body equations
of motion of the particle. If advection of the solute is neglected, the
fluid-chemical transport problem is decoupled an the solute diffusion equation
can be solved first. This can then be exploited in the fluid problem to
compute the swimming speed and the flow field [\cite{Michelin14}]. The
assumption applies at small P\'eclet numbers, i.e., at small activities
and particle sizes, and at large solute diffusion. Contrary, solute
advection becomes important at large P\'eclet numbers and larger particles
[\cite{Howse07}], and significantly impacts the solute velocity
[\cite{Michelin14}]. An adequate account of the interactions of the
various components involved in phoresis successfully describes the
salient features of artificial active particles in solution, e.g.,
swimming velocity and surface effects, as for Janus particles hovering
over or swimming parallel to a surface [\cite{Uspal15,Bechinger16}].

\Blue{\subsection{\it Response to external fields -- Taxis}}

Biological microswimmers respond to a large variety of external fields by
redirecting their motion, in order to locate a target or avoid unfavorable
environmental conditions. This directed motion is called taxis. Prominent
examples are chemo-, photo-, gravi-, magneto-, and rheo-taxis, which describe 
the response to chemical gradients, light, gravitational, magnetic, and 
flow fields, respectively. Chemotaxis is used by sperm to find the egg, 
phototaxis by {\em Chlamydomonas} algae to swim toward the light, and 
rheotaxis by sperm and bacteria to move upstream in flow [\cite{Ishimoto15,
Koh16,Uspal15b,Mathijssen19b}]. Modelling and simulation have been employed to understand these 
phenomena. For example, one mechanism of sperm chemotaxis is the adjustment 
of the trajectory curvature with a time delay in response to changes in the
chemoattractant concentration [\cite{Friedrich07}]. It is important to note 
that chemo- and phototaxis in biological system often rely on internal 
biochemical signalling processes [\cite{Tu13,Camley16}].

Redirection of motion in external fields also exists for artificial 
microswimmers. The mechanisms are usually different, and rely more on 
direct physical effects. Examples are gravitaxis of chiral (in two dimensions)
$L$-shaped microswimmers, which can balance the hydrodynamic and gravitational
torques due to shape asymmetry by swimming against gravity [\cite{tenHagen14}],
collective gravitaxis of bottom-heavy swimmers which form convective swirls in
films of finite thickness [\cite{Kuhr17}], and phototaxis of thermophoretic
colloids which reduce the activity of neighboring colloids by casting a shadow
on them [\cite{Cohen14}].

\Blue{\subsection{\it Continuum models for active motion in fluids}}

To investigate phenomena occurring at large time and length scales,
hydrodynamic theories based on conserved quantities (slow variables)
and broken continuous symmetries (order parameters) have been developed,
which describe a broad class of systems [\cite{Martin72}]. Here we
discuss a few continuum models of wet active matter, i.e.\ suspensions
of active particles with momentum conservation.

{\it Wet active liquid crystals --} Of particular interest are suspensions
of active rodlike or elongated objects (e.g.\ swimming organisms, cytoskeleton,
or tissues) embedded in a momentum-conserving solvent and generating
active stresses [\cite{Toner05,Ramaswamy10,Marchetti13,Prost15,Julicher18,
Carenza19}]. The term {\it active gel} is also used in the context of the 
cytoskeleton and tissues, referring to their viscoelastic nature 
[\cite{Prost15,Julicher18}]. Slow variables are the number density
$n({\bm r},t)$ of active particles and the total momentum density
$\rho\,{\bm v}({\bm r},t)$ of the suspension with mass density $\rho$.

Polar active gels consist of particles distinguishing front from rear
and are characterized by a polarization field ${\bm p}({\bm r},t)$
corresponding to the vectorial orientation of active particles. The
equations of motion for active gels are derived on the basis of
{\it (i)} symmetry [\cite{Simha02,Hatwalne04}], {\it (ii)} irreversible
thermodynamics [\cite{Kruse04,Prost15,Julicher18}], or {\it (iii)}
coarse-graining a microscopic theory [\cite{Giomi08,Baskaran09}]. The
earliest phenomenological description of wet polar gel [\cite{Simha02,
Ramaswamy10}] extends the Toner-Tu model of dry active matter, equation~A2,
by terms coupling the orientation ${\bm p}$ to the flow, see equation~B3
in Table~\ref{Tab2}. The fluid velocity obeys the incompressible
Navier-Stokes equations~B1 (NSE), with passive (viscous, elastic and
interface) and active contributions to the stress tensor $\bm{\sigma}$,
where the latter is responsible for a spontaneous shear flow without
an external stress. The assumption that every swimmer (or molecular
motor) exerts an active dipolar force on the solvent (or filament
network) yields, to leading order in a gradient expansion, an active
stress
\begin{equation}
\bm{\sigma}^{a}{=}-\zeta n({\bm r},t)\left({\bm p}\otimes{\bm p}
-\frac{{\bm p}^2}{d}\bm{I}\right)+O(\nabla),
\label{active_stress}
\end{equation}
where $d$ is the dimension, $\bm{I}$ the unit tensor and $\zeta$ the
activity strength, that is $\zeta{>}0$ for extensile (pushers) and
$\zeta{<}0$ for contractile particles (pullers). Note that $\bm{
\sigma}^{a}$ in equation~\ref{active_stress} has nematic symmetry,
${\bm p}{\rightarrow}-{\bm p}$, and active stress with purely polar
symmetry arises first in terms containing gradients of ${\bm p}$
[\cite{Giomi08,Linkmann19}]. The concentration $n$ of active particles
evolves via the continuity equation~B4.

Active gels of apolar (or head-tail symmetric) particles, called wet
active nematics, are described by the nematic alignment tensor field
$\bm{Q}$ introduced in equation~A3, where ${\bm p}$ now denotes the 
nematic director. The evolution of $\bm{Q}$ in active nematics at 
constant concentration $n$ is governed by the nematodynamic equation~B5
[\cite{Marenduzzo07,Marenduzzo07b,Giomi11,Giomi13,Doostmohammadi18}]
accompanied by the incompressible NSE, equation~B1, with an active
contribution $\bm{\sigma}^{a}{=}-\zeta\bm{Q}$ to the total stress
tensor $\bm{\sigma}$, cf.\ equation~\ref{active_stress}. Note that
dynamical equations for active nematics can be obtained from that
of polar active gels by interpreting ${\bm p}$ as nematic director
and dropping terms which violate the invariance of the nematic.

{\it Generalized Navier-Stokes equations (GNSE) --} Two-fluid models
consists of equations for the fluid velocity ${\bm v}$, the concentration
$n$ and the order parameter characterizing the active constituents,
like the polarization ${\bm p}$. For dense suspensions (with constant $n$)
it is possible to enslave ${\bm v}$ to ${\bm p}$ or vice versa (to
eliminate one of the vector fields) leading to a simpler one-fluid
descriptions of the system [\cite{Linkmann19}]. GNSE is a class of
minimal, single vector-field models for active fluids like dense
microbial suspensions (${\bm v}$ enslaved to ${\bm p}$, bacterial flow
model) [\cite{Dunkel13}] or sole passive solvent driven by active components
(${\bm p}$ enslaved to ${\bm v}$, solvent flow model) [\cite{Slomka15}].

In the bacterial flow model, ${\bm v}$ denotes the velocity of the active
subcomponents and GNSE is an extension of the Toner-Tu model: NSE plus a
biquadratic Landau velocity potential $F[{\bm v}]$ (see equations~B1 and
B6), an active nematic stress contribution via equation~B7, and higher
order terms in the Fourier expansion of the stress tensor according to
equation~B8 with $\Gamma_0{<}0$, $\Gamma_2{>}0$ and $\Gamma_4{=}0$ needed
to account for non-local interactions and to reproduce local polar order
observed in mesoscale turbulence [\cite{Dunkel13,Dunkel13b,Wensink12}].

GNSE without a velocity potential and the active stress but with additional
higher-order stresses via equation~B8 with $\Gamma_0,\Gamma_4{>}0$
and $\Gamma_2{<}0$, accounting for non-Newtonian effects [\cite{Slomka15}]
and originating from active stress with purely polar symmetry [\cite{Linkmann19}],
describes the solvent dynamics in the presence of active components. Solvent
flow model was used to study the rheology of active fluids [\cite{Slomka17}]
and active turbulence [\cite{Slomka17b,Linkmann19}].

{\it Active Model H (AMH)--} Active gel models and GNSE consider explicitly
or implicitly alignment interactions between active particles. In contrast,
AMH describes scalar active matter ($\bm{p}{=}\mathbf{Q}{=}0$) and is an
extension of the active model B (see equations~A6, A7 in Table~\ref{Tab2})
to account for the momentum-conserving solvent [\cite{Tiribocchi15,Cates19}].
Here, the dynamics of the order-parameter field $\phi({\bm r},t)$ obeys
the diffusive equations of AMB [\cite{Wittkowski14,Cates19}] together
with an additional advective term ${\bm v}{\cdot}\bm\nabla\phi$ coupling
$\phi$ to the velocity ${\bm v}$ of the fluid, whose dynamics is governed
by the NSE equations~B1. The governing equations of the model are given
in equations~B9 and B10. The violation of the thermodynamic relation
between stress $\bm{\sigma}$ and chemical potential $\mu$ leads to an active
contribution to the deviatoric stress (see equation~B11); at interfaces
the polarization is large (${\bm p}{\sim}\nabla\phi$) and equation~\ref{active_stress}
justifies equation~B11. The active stress contribution is positive
for extensile and negative for contractile swimmers and in the latter
case results in an unusual arrested motility-induced phase separation
[\cite{Tiribocchi15}].

\Blue{\section{Cells and tissues}}

\label{Sec:Cells}

Living matter is active matter on all scales: on the protein scale
with molecular ATP-consuming machines like motor proteins, ATPase
pumps and protein factories, the ribosome; on the cellular scale
with cell shape transformation, polarization, migration and division;
on the multi-cellular scale with growing biofilms and tissues, tumors,
and developing organs; and on the macro-scale with groups, swarms, and
herds of animals and humans. The different scales necessitate
different model approaches, ranging from all-atom molecular dynamics
(MD) simulations for force generating protein machines, over
coarse-grained particle-based models and continuum models, to
agent-based models with phenomenological interaction rules.

\Blue{\subsection{\it Cytoskeltal filaments and molecular motors}}

The ubiquitous microscopic origin of the activity in living matter
is the biochemical force generation via energy (ATP) consuming
polymerization and depolymerization of cytoskeleton filaments (actin
and microtubules) and the collective action of molecular motors.
Force generation during polymerization of actin filaments and
microtubules is based upon a ratchet mechanism (Box~3): Thermal
fluctuations of the target against which the filament polymerizes,
e.g., when pushing against the plasma membrane during the formation
of filopodia or lemellopodia, allow for the occasional, ATP-dependent,
insertion of a new subunit, even when an external force opposes the
motion of the target [\cite{Mogilner96,Dogterom05}]. Asymmetric
ATP-dependent polymerization and depolymerization rates at the two
ends of cytoskeletal filaments also lead to their effective forward
motion called treadmilling [\cite{Mogilner08,Erlenkamper09}] (Box~3).
Both processes are far from equilibrium and at the heart of cell motility.

Force generation of molecular motors [\cite{Howard01}] has its origin
in the massive, ATP-dependent conformational changes that span spatially
from the atomic to the molecular level and temporarily up to milli-seconds.
All-atom MD simulations are in principle capable of elucidating the
underlying molecular processes, while the study of the necessary long
time-scale dynamics require efficient sampling techniques and
coarse-grained approaches coupled with all-atom MD simulations,
termed multiscale MD simulations [\cite{Nielsen10,Ekimoto18}] (note
that stability requirements demand time steps of the order of
femto-seconds).

To understand the collective behavior of whole ensembles of molecular
motors and their interaction with the filament network of the
cytoskeleton and the membrane, the number of degrees of freedom has
to be reduced drastically. Discrete kinetic and stochastic models for
individual motors, as exemplarily sketched in Box~3, yield predictions
for the mean velocity and other observables as a function of an
imposed load force, the ATP concentration, and other variables
[\cite{Kolomeisky07,Chowdhury13}]. The effect of the attachment of
opposing motors to one cargo are described by a tug-of-war model
[\cite{Klumpp05}]. Many motors on a single track can lead to
molecular motor traffic jams and are described by asymmetric
exclusion process (ASEP) models [\cite{ Appert-Rolland15}] void
of any mechano-chemistry.

Particle-based models for cytoskeleton filaments and networks of
crosslinked filaments are based on the wormlike chain (WLC) model for
semiflexible polymers [\cite{Bausch06,Huber13,
Broedersz14}]. Computational studies of elastic and collective properties
of semiflexible filaments are commonly based on a discretization of the
filament into a finite number of segments and subsequent simulation of
the Langevin dynamics. The latter is difficult to perform for microtubules
since they are  a) nearly incompressible in the longitudinal direction,
which necessitates the use of constrained Langevin dynamics (fixed segment
length) and b) are extremely stiff, which necessitates an extremely small
time step in the presence of noise (and which is therefore often neglected).
\Red{There are several powerful software packages for simulations of biological 
environments and particle-based modelling of active systems, see list and 
short description in Supplementary Information, Table S1} [\cite{Nedelec07}]. 
Specific active aspects of microtubule dynamics comprise their length regulation 
[\cite{Erlenkamper09,Mohapatra16}] and the spindle dynamics during mitosis 
[\cite{Mogilner10,Pavin16}].

\begin{widetext}
\centering
\includegraphics[width=0.95\textwidth]{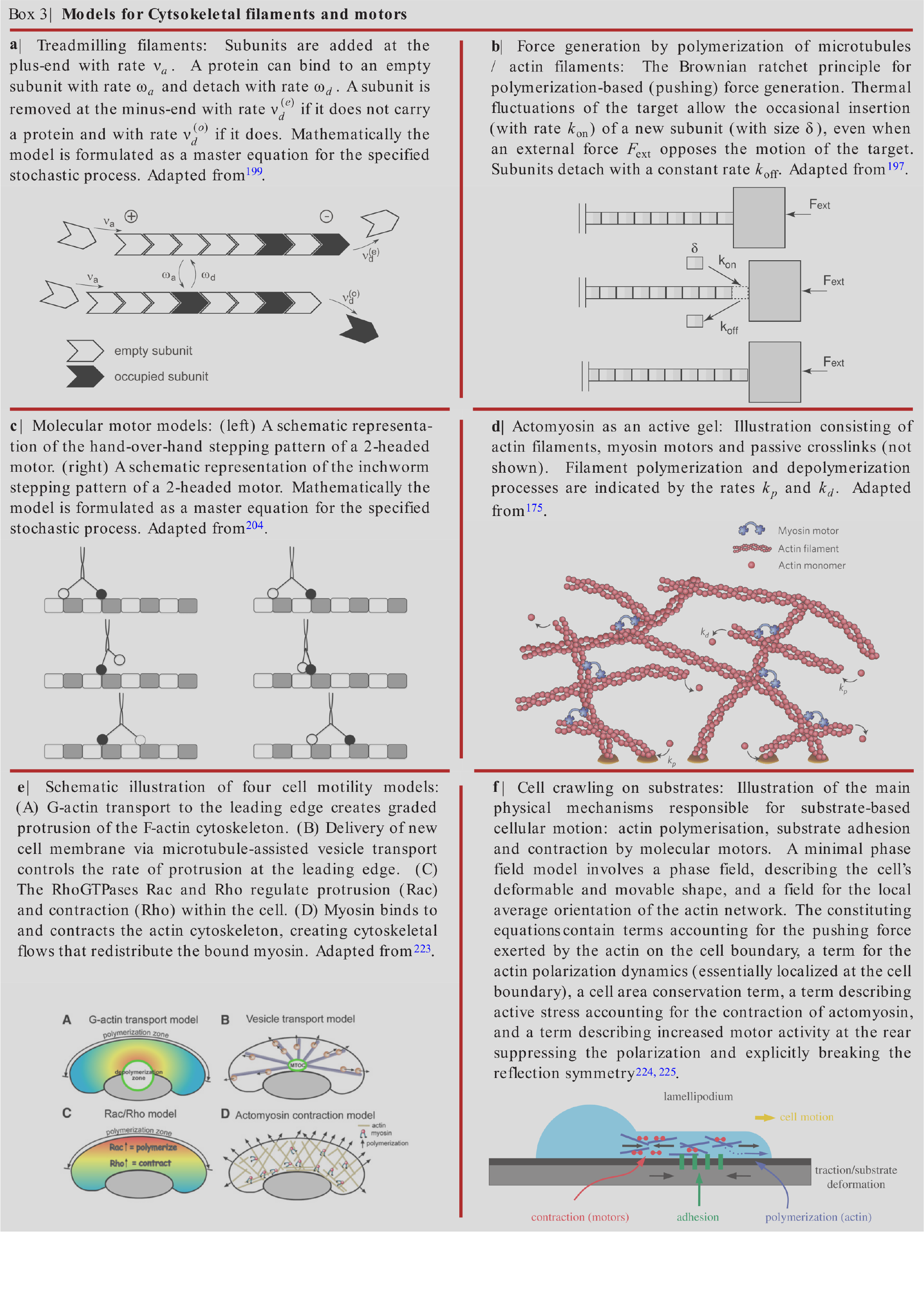}
\end{widetext}

The addition of molecular motors to filament networks (like myosin
to actin networks) generates forces that drive the network far from
equilibrium and can dramatically alter its stiffness, amplify stress,
or lead to network contractility.
These effects were studied in an extensible WLC model for semiflexible
polymer in which force dipoles were introduced into the network at
neighboring crosslinks [\cite{Broedersz11, Ronceray16}], cf.\ Box~3.
On large length and time scales, polymerizing and depolymerizing
(treadmilling) actin filaments interconnected by active myosin motors
and passive crosslinkers, so called acto-myosin, can be understood
as an active gel for which continuum models have been developed,
cf.\ the previous section and Refs.~[\cite{Julicher07,Joanny09,Prost15,
Julicher18}]. From this perspective acto-myosin is an active, nematic
liquid described by appropriately modified Navier-Stokes equations:
the central quantity is the stress tensor which depends on the
velocity gradient, the orientational field of the filaments, and
an active stress generated by the motors and actin polymerization. 
Since cytoskeletal filaments can act as tracks to motors and in turn 
motors can move filaments, the active stress is coupled to the orientational 
field. Within this active hydrodynamics framework cell locomotion and
cellular shape changes and many experimental situations have been
successfully discussed [\cite{Prost15,Julicher18}].

\Blue{\subsection{\it Cell motility models}}

The central molecular machinery that enables eukaryotic cells
to spontaneously move is the actin cytoskeleton responsible for
cellular shape changes, like the formation of thin membrane-bound
protrusions. Thus, in addition to a mathematical representation
of the acto-myosin machinery, cell-motility models must involve a
representation of the cell membrane and a description of
adhesion and force transmission to a substrate or a three-dimensional
environment [\cite{Mogilner08,Jilkine11,Holmes12,
Danuser13,teBoekhorst16}]. In the paradigmatic model
of cell crawling on substrates, migration is divided into discrete
steps: (a) protrusion based on actin growth and polymerization force;
(b) formation of new adhesions at the front; (c) release and recycling
of adhesions at the rear; and finally, (d) actin-myosin-powered
contraction of the cytoplasm, resulting in forward translocation
of the cell body [\cite{Danuser13}], cf.\ Box~3. In a model that
relies mainly on actin treadmilling and diffusing actin nucleators,
cell crawling is driven by actin polymerization waves without motors
[\cite{Doubrovinski11}].

The major technical challenge for a continuum modeling of cell migration
is the presence of a moving boundary and the nonlinear and nonlocal
coupling of cytoskeletal dynamics to a moving and deformable membrane. 
Various continuum models for cell motility on substrates have been 
employed [\cite{Doubrovinski11,Wolgemuth11,Holmes12,Ziebert12,Ziebert16,
Linsmeier16}]. Commonly, three different modeling approaches have been 
used: sharp-interface models, in which the interface is represented by 
a curve that moves with some velocity, level set methods, and 
diffuse-interface models. In the latter two descriptions, a phase field 
distinguishes the two phases (the interior and the exterior of the cell), 
where either the zero contour of the phase field determines the position 
of the membrane, or there is a gradual variation of the different 
physical quantities across the interface [\cite{Singer-Loginova08}]. 
Phase-field models have also been used for multi-cellular systems 
[\cite{Nonomura12,Camley14}], \Red{as for collective cell migration 
[\cite{Najem16}] (for a review about physical models for collective 
cell motility see [\cite{Camley17}]) and tissues [\cite{Mueller19,
Wenzel19}], see also Box 4.}

Alternatively, microscopic models with explicit membrane and self-propelled 
(pulling or pushing) filaments can be employed [\cite{Abaurrea17,Abaurrea19}]. 
This approach incorporates fluctuations of the internal structure, persistent 
and random-walk-like motion, and shape changes in response to external conditions.

\Blue{\subsection{\it Tissues}}

Tissues are aggregates of adherent cells, sometimes organized in layers
(e.g., epithelia). In addition to constitute a viscoelasto-plastic
material [\cite{Preziosi10}], tissues generate actively internal
tension via cell proliferation and death, as for instance during
growth, and generate active stress by cellular force generation,
as during muscle contraction. The technical challenge
for a continuum formulation of volumetric growth in soft elastic
tissues is the persistent change of the equilibrium configurations
against which small deformations must be defined [\cite{Rodriguez94}].

Particle-based models for tissue growth represent cells as spheres
that continuously deform into dumbbells until division occurs [\cite{Drasdo05,
Basan11}], cf.\ Box~4. The particles representing cells can adhere to
each other, maintain volume exclusion, exert an active growth pressure
on their surrounding, expand in size until reaching a size checkpoint,
divide when reaching this checkpoint size, undergo apoptosis, exert
random forces on neighboring cells, regulate to their homeostatic
state via cell division and apoptosis in a confined volume, and comply
with force balance and momentum conservation. Either Langevin dynamics
[\cite{Drasdo05}] or dissipative particle dynamics [\cite{Drasdo05,
Basan11}] have been employed as constitutive dynamics. Simulations
show that stress-induced growth inhibition is responsible for the
transition from exponential to sub-exponential growth experimentally
observed in tumor spheroids, and lack of nutrients determines the
size of the necrotic core but not the size of the tumor [\cite{Drasdo05}].
Moreover, cell division and apoptosis also lead to a fluidization of
the tissue [\cite{Basan11,MalmiKakkada18,MatozFernandez17}] as 
further analyzed in a continuum theory described below.

\begin{widetext}
\centering
\includegraphics[width=0.97\textwidth]{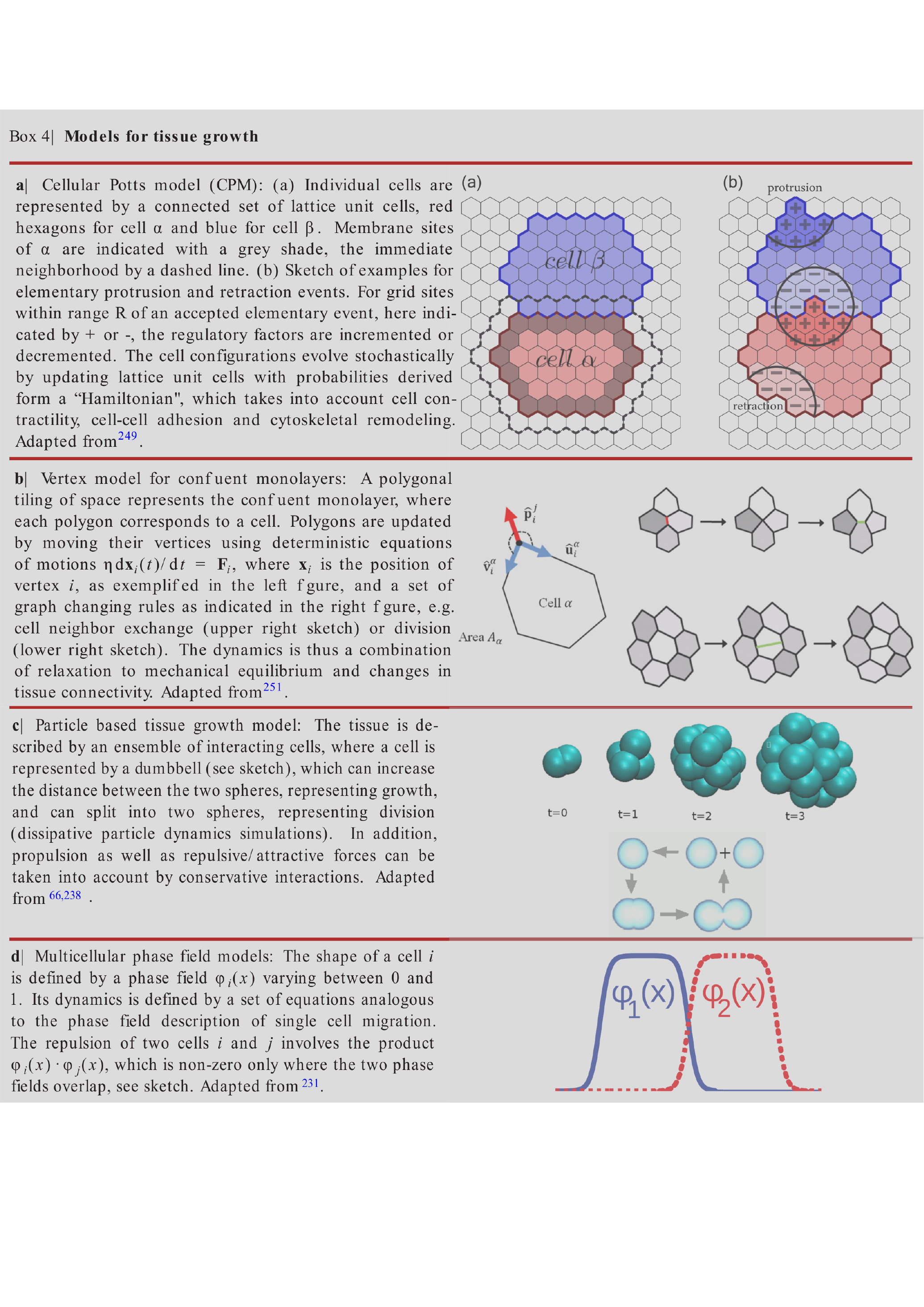}
\end{widetext}

Alternatively, lattice-based models for tissue growth and morphogenesis
have been extensively employed. The cellular Potts model (CPM)
[\cite{Graner92,Glazier93,Chen07}] defines the cell shape with the
help of discrete variables assigned to each site of a regular lattice,
cf.\ Box~4. An energy function regulates cell volume, cell surface
area, cell adhesion, etc. The CPM has been applied to development
[\cite{Mare01}] and vasculogenesis [\cite{Merks08}], but also to
cell migration [\cite{Maree06}] and cell shape dynamics on
micropatterned surfaces [\cite{Albert14,Segerer15}]. Another lattice
model is the confluent tissue vertex or Voronoi model [\cite{Hufnagel07,
Fletcher14,Sussman18,Barton17}], where confluent monolayers are 
represented as a polygonal tiling of space and each polygon corresponds 
to a cell, cf.\ Box~4. \Red{These models have been used to study the 
jamming transition in tissues [\cite{Oswald17}] with a CPM [\cite{Chiang16}] 
and a Voronoi [\cite{Bi16}] model.}

Continuum models for tissue mechanics have a long tradition in the
field of Biomechanics and Biomedical Engineering [\cite{Fung10}].
A major challenge in continuum models for tissue growth is the
coupling between growth rate and local stress, which is modeled
by a dependence of the growth-rate tensor on the stress tensor
[\cite{Rodriguez94,Dunlop10,Ambrosi12}]. A simple low-dimensional
example is the mathematical model of nonuniform growth in a
monolayer [\cite{Shraiman05}], which incorporates a mechanical
feedback mechanism via an explicit dependence of the local tissue
growth rate on the degree of local compression (or stretching)
of the tissue. In Ref.~[\cite{Ranft10}], three-dimensional tissues
were considered as elastic media and it was shown that the coupling
of cell division and cell death to the local stresses effectively
leads to viscoelastic behavior with a relaxation time set by the
rate of cell division.

It should also be mentioned that various
continuum models for tumor growth are based on the theory of mixtures
[\cite{Byrne03}] as reviewed in Ref.~[\cite{Tracqui09}]. Since
tissue growth in a living organism requires nutrition supply and
oxygen, a blood vessel network should be included: Hybrid models
describe the tumor mass with a continuum model and the dynamically
changing vascularization with a discrete pipe network [\cite{Rieger15,Fredrich19}].
Simulations show that the incorporation of a blood vessel network leads
to a characteristic compartmentalization of the tumor into several regions
differing in vessel density and diameter, and in necrosis.

\begin{table*}
\centering
\includegraphics[width=0.99\textwidth]{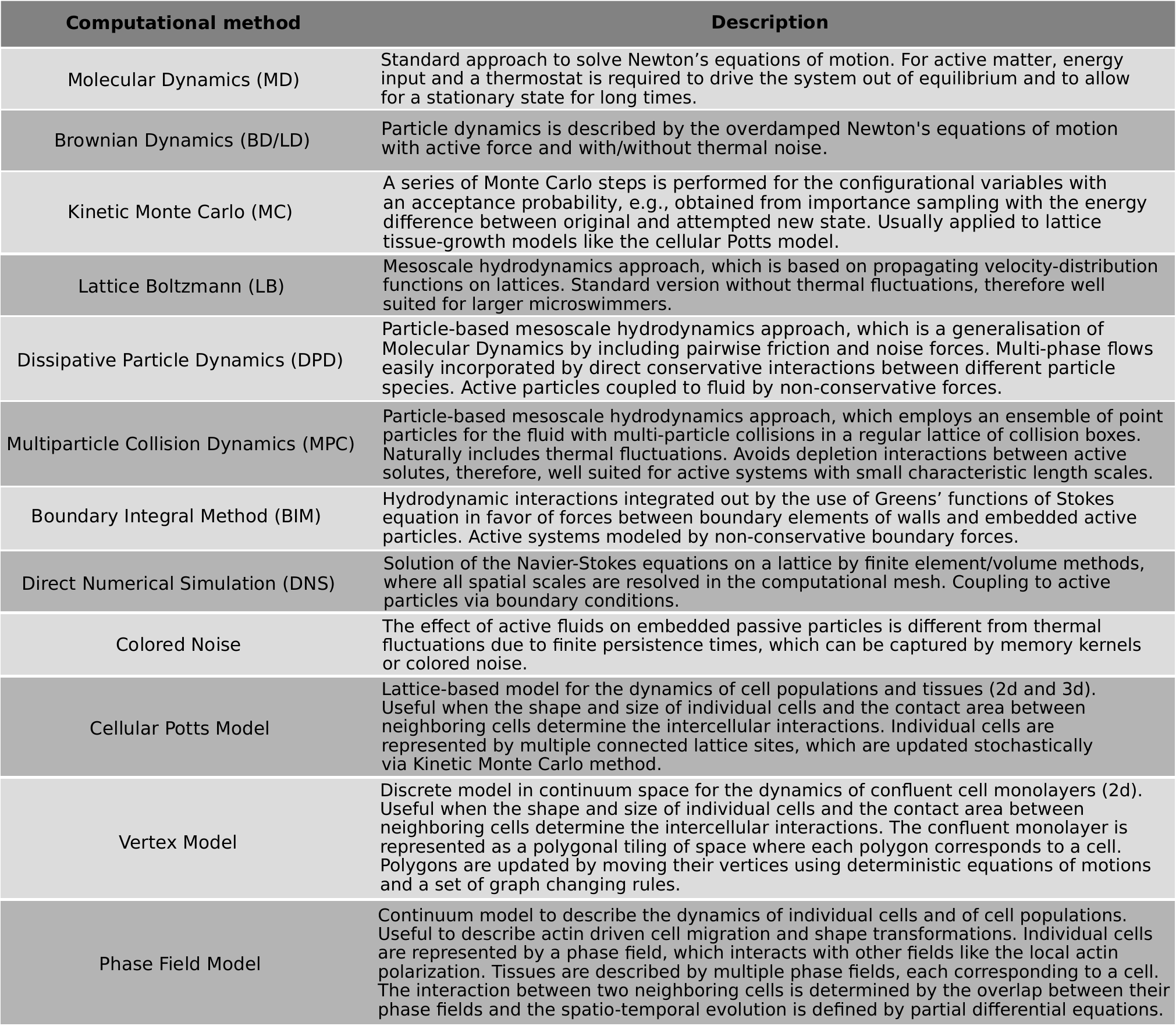}
\caption{{\Red{\bf Computational methods for simulation of active 
matter.}}}
\label{Tab3}
\end{table*}

\Blue{\section{Animal groups}}

\label{Sec:AnimalGroups}

\Red{Collective migration is of paramount importance for a wide variety of animals, 
such as swarms of insects, fish schools, bird flocks, herds of game, and human crowds. 
The models of dry active matter (cf.\ Table~\ref{Tab1}), in particular the Vicsek model 
and its extensions [\cite{Chate08}], capture prototypical aspects of the collective 
behavior of such animal groups. Despite of sharing some universal features, the 
observed motion patterns differ substantially among animal groups due to differences 
in the nature of interactions between individuals, which requires adaptation of basic 
models. Important characteristics include social behavior mediated by chemical,
acoustic, or optical signals. Chemical signaling comprises non-reciprocal 
attractive-repulsive interactions, which imply specific actions such as 
pursuit-escape behavior [\cite{Romanczuk09,Simpson06,Agudo19}]. Particular 
attention has been paid to modeling vision-based interactions, where the emerging 
motion patterns depend on the field of view [\cite{Pearce14,Lavergne19,Ballerini08}].}

\Red{Various strategies are applied to unravel the mechanisms underlying the 
collective behavior of animal groups. In a phenomenological top-down approach, 
(physical) interactions are deduced from observations [\cite{Bajec09,Ballerini08}]. 
Alternatively, in a bottom-up approach, the effect of complex information processing 
strategies of individual agents, e.g., delayed signal processing, is studied 
[\cite{Mijalkov16,Charlesworth19,Khadka18}]. Collective behavior can even emerge 
from a purely probabilistic approach by considering intrinsic motivation and 
maximization of future options via processing of sensed information, without 
any a priori specification of social forces or individual interaction rules 
[\cite{Charlesworth19,Mann15}].}

\Red{Living active agents travelling together have evolutionary developed 
efficient decision-making mechanisms to search, forage, escape, etc. For 
instance quorum sensing, i.e., the ability of individuals to coordinate 
their activities according to the local population  density, facilitates 
information transfer in these systems [\cite{Ward08,Lavergne19,Abaurrea18}]. 
While such {\it cooperative} interactions are powerful tools to reach 
collective decisions [\cite{Castellano09}], in some cases only a few 
individuals possess the required knowledge about the migration route, 
location of food sources, etc.\ [\cite{King08}]. Accordingly, numerical 
models have been studied to understand the mechanisms of effective leadership 
in biological systems, and the impact of the density of informed individuals 
and their communications on the information transfer in animal groups 
[\cite{Couzin05,Freeman09,Baeuerle18}]. A realistic description of 
collective human motion is crucial to prevent crowd disasters 
[\cite{Moussaid11,Faria10,Bain19}].}

\Blue{\section{Outlook and future directions}}

A powerful pool of computational models for active matter, particle-based 
or field-theoretical, have been developed in the recent years \Red{(see a 
list of computational methods for active matter in Table~\ref{Tab3}). This 
has lead to an enormous knowledge gain for the non-equilibrium behavior 
of active systems.} Yet, many challenges remain.

The statistical-physics description of active systems needs to be extended
in several directions, like the incorporation of an activity-dependent noise
to account for the stochastic nature of active forces, or the generalization 
of field-theoretical models to systems far from equilibrium. Current field 
theories are based on a linear expansion of fluxes in terms of forces and 
can therefore in principle describe only systems close to equilibrium; 
however, biological systems are typically far from equilibrium and there 
is no comprehensive \Red{continuum} theory for systems far from equilibrium yet.

The investigation of the collective behavior of self-propelled particles
is limited so far to relatively simple systems. In particular, the interplay
of different interactions, such as shape and hydrodynamics, of complex 
environments, such as viscoelastic fluids and intricate confining geometries, 
external fields, such as gravity and turbulent flows, mixtures of active 
and passive particles, as well as information exchange by vision-like 
interactions, holds many unexpected surprises to be discovered.

\Red{A plethora of models for cell motility has been employed in studies of
migration on two-dimensional substrates. However, many cells in
living organisms move in three-dimensional (3D) environments, like the
extra-cellular matrix (ECM), and there is a lack of continuum models 
for motility in 3D. First studies have focused on rigidity sensing
and subsequent polarization of the cell toward stiffer ECM [\cite{Kim18}].
Furthermore, for migration in 3D environments, cells can utilize other
modes of motility in addition to adhesion-based lamellipodial propulsion, 
in particular blebbing motility, which is based on the formation of blebs 
at the leading edge (blebs emerge when the plasma membrane of a cell 
delaminates locally from its actin cortex, leading to pressure-driven 
protrusions, followed by membrane re-attachment) [\cite{Paluch13}]. 
Again continuum modeling has received little attention so far. A first 
modeling study for blebbing-induced migration in 2D still crucially 
depends on protrusions adhering to a substrate [\cite{Tozluoglu13}].}

\Red{Finally, signalling is a very important aspect of cell motility,
tissue growth, the development of bacterial colonies, and the collective
motion of animal herds, which has been neglected in most computational 
models for active matter so far. Challenges include the modeling of 
the interplay of biochemistry and mechanics on a single cell level 
[\cite{Besser07, Nishikawa17,Gross19}]), and the reaction to external 
factors like food, smell, or light (sun) in flocks, schools, and herds.
Quantitative models for tissue growth, morphogenesis, biofilm formation, 
wound healing, or the cancer growth and metastastisis need to include 
biochemical signaling coupled to mechanical stress. Similarly, realistic 
models for human groups, car or pedestrian traffic need to take into account
heterogeneous behavior and varying reactions of individuals to external 
cues, which also asks for the investigation of a putative leader role.}

\Red{An overarching challenge for the whole field is the distinction
between the generic and specific properties and behavior of a particular
active-matter system. What are the universal properties shared by a
large class of systems, and when do specific mechanisms of propulsion or
interactions come into play and dominate the behavior?}\\

\noindent{\bf Acknowledgements} \\
M.\,R.\,S., A.\,W., and H.\,R.\ acknowledge support by DFG within SFB
1027 (A3, A7). R.\,G.\,W.\ and G.\,G.\ acknowledge funding by DFG
within the priority program SPP 1726 ``Microswimmers -- from Single
Particle Motion to Collective Behavior''. \\

\noindent{\bf Author contributions} \\
All authors contributed to all aspects of manuscript preparation,
revision and editing. Request for materials: h.rieger@mx.uni-saarland.de
or shaebani@lusi.uni-sb.de. The authors declare no competing interests. \\

\bibliography{Refs}

\begin{thebibliography}{100}
\expandafter\ifx\csname url\endcsname\relax
  \def\url#1{\texttt{#1}}\fi
\expandafter\ifx\csname urlprefix\endcsname\relax\def\urlprefix{URL }\fi
\providecommand{\bibinfo}[2]{#2}
\providecommand{\eprint}[2][]{\url{#2}}

\bibitem{Marchetti13}
\bibinfo{author}{Marchetti, M.~C.}, \bibinfo{author}{Joanny, J.~F.},
  \bibinfo{author}{Ramaswamy, S.}, \bibinfo{author}{Liverpool, T.~B.},
  \bibinfo{author}{Prost, J.}, \bibinfo{author}{Rao, M.} \&
  \bibinfo{author}{Simha, R.~A.}
\newblock \bibinfo{title}{Hydrodynamics of soft active matter}.
\newblock \href{http://dx.doi.org/10.1103/RevModPhys.85.1143}{\emph{Rev. Mod.
  Phys.}}
  \href{http://dx.doi.org/10.1103/RevModPhys.85.1143}{\textbf{\bibinfo{volume}{85}},
  \bibinfo{pages}{1143--1189}}
  \href{http://dx.doi.org/10.1103/RevModPhys.85.1143}{ (\bibinfo{year}{2013})}.

\bibitem{Ramaswamy10}
\bibinfo{author}{Ramaswamy, S.}
\newblock \bibinfo{title}{The mechanics and statistics of active matter}.
\newblock
  \href{http://dx.doi.org/10.1146/annurev-conmatphys-070909-104101}{\emph{Annu.
  Rev. Condens. Matter Phys.}}
  \href{http://dx.doi.org/10.1146/annurev-conmatphys-070909-104101}{\textbf{\bibinfo{volume}{1}},
  \bibinfo{pages}{323--345}}
  \href{http://dx.doi.org/10.1146/annurev-conmatphys-070909-104101}{
  (\bibinfo{year}{2010})}.

\bibitem{Toner05}
\bibinfo{author}{Toner, J.}, \bibinfo{author}{Tu, Y.} \&
  \bibinfo{author}{Ramaswamy, S.}
\newblock \bibinfo{title}{Hydrodynamics and phases of flocks}.
\newblock \href{http://dx.doi.org/10.1016/j.aop.2005.04.011}{\emph{Ann. Phys.}}
  \href{http://dx.doi.org/10.1016/j.aop.2005.04.011}{\textbf{\bibinfo{volume}{318}},
  \bibinfo{pages}{170--244}}
  \href{http://dx.doi.org/10.1016/j.aop.2005.04.011}{ (\bibinfo{year}{2005})}.

\bibitem{Elgeti15b}
\bibinfo{author}{Elgeti, J.}, \bibinfo{author}{Winkler, R.~G.} \&
  \bibinfo{author}{Gompper, G.}
\newblock \bibinfo{title}{Physics of microswimmers{\textemdash}single particle
  motion and collective behavior: a review}.
\newblock \href{http://dx.doi.org/10.1088/0034-4885/78/5/056601}{\emph{Rep.
  Prog. Phys.}}
  \href{http://dx.doi.org/10.1088/0034-4885/78/5/056601}{\textbf{\bibinfo{volume}{78}},
  \bibinfo{pages}{056601}}
  \href{http://dx.doi.org/10.1088/0034-4885/78/5/056601}{
  (\bibinfo{year}{2015})}.

\bibitem{Bechinger16}
\bibinfo{author}{Bechinger, C.}, \bibinfo{author}{Di~Leonardo, R.},
  \bibinfo{author}{L\"owen, H.}, \bibinfo{author}{Reichhardt, C.},
  \bibinfo{author}{Volpe, G.} \& \bibinfo{author}{Volpe, G.}
\newblock \bibinfo{title}{Active particles in complex and crowded
  environments}.
\newblock \href{http://dx.doi.org/10.1103/RevModPhys.88.045006}{\emph{Rev. Mod.
  Phys.}}
  \href{http://dx.doi.org/10.1103/RevModPhys.88.045006}{\textbf{\bibinfo{volume}{88}},
  \bibinfo{pages}{045006}}
  \href{http://dx.doi.org/10.1103/RevModPhys.88.045006}{
  (\bibinfo{year}{2016})}.

\bibitem{Romanczuk12}
\bibinfo{author}{Romanczuk, P.}, \bibinfo{author}{B{\"a}r, M.},
  \bibinfo{author}{Ebeling, W.}, \bibinfo{author}{Lindner, B.} \&
  \bibinfo{author}{Schimansky-Geier, L.}
\newblock \bibinfo{title}{Active {B}rownian particles}.
\newblock \href{http://dx.doi.org/10.1140/epjst/e2012-01529-y}{\emph{Eur. Phys.
  J. Spec. Top.}}
  \href{http://dx.doi.org/10.1140/epjst/e2012-01529-y}{\textbf{\bibinfo{volume}{202}},
  \bibinfo{pages}{1--162}}
  \href{http://dx.doi.org/10.1140/epjst/e2012-01529-y}{
  (\bibinfo{year}{2012})}.

\bibitem{Nguyen14}
\bibinfo{author}{Nguyen, N. H.~P.}, \bibinfo{author}{Klotsa, D.},
  \bibinfo{author}{Engel, M.} \& \bibinfo{author}{Glotzer, S.~C.}
\newblock \bibinfo{title}{Emergent collective phenomena in a mixture of hard
  shapes through active rotation}.
\newblock \href{http://dx.doi.org/10.1103/PhysRevLett.112.075701}{\emph{Phys.
  Rev. Lett.}}
  \href{http://dx.doi.org/10.1103/PhysRevLett.112.075701}{\textbf{\bibinfo{volume}{112}},
  \bibinfo{pages}{075701}}
  \href{http://dx.doi.org/10.1103/PhysRevLett.112.075701}{
  (\bibinfo{year}{2014})}.

\bibitem{Lowen16}
\bibinfo{author}{L{\"o}wen, H.}
\newblock \bibinfo{title}{Chirality in microswimmer motion: From circle
  swimmers to active turbulence}.
\newblock \href{http://dx.doi.org/10.1140/epjst/e2016-60054-6}{\emph{Eur. Phys.
  J. Spec. Top.}}
  \href{http://dx.doi.org/10.1140/epjst/e2016-60054-6}{\textbf{\bibinfo{volume}{225}},
  \bibinfo{pages}{2319--2331}}
  \href{http://dx.doi.org/10.1140/epjst/e2016-60054-6}{
  (\bibinfo{year}{2016})}.

\bibitem{Peruani16}
\bibinfo{author}{Peruani, F.}
\newblock \bibinfo{title}{Active {B}rownian rods}.
\newblock \href{http://dx.doi.org/10.1140/epjst/e2016-60062-0}{\emph{Eur. Phys.
  J. Spec. Top.}}
  \href{http://dx.doi.org/10.1140/epjst/e2016-60062-0}{\textbf{\bibinfo{volume}{225}},
  \bibinfo{pages}{2301--2317}}
  \href{http://dx.doi.org/10.1140/epjst/e2016-60062-0}{
  (\bibinfo{year}{2016})}.

\bibitem{tenHagen15}
\bibinfo{author}{ten Hagen, B.}, \bibinfo{author}{Wittkowski, R.},
  \bibinfo{author}{Takagi, D.}, \bibinfo{author}{K\"ummel, F.},
  \bibinfo{author}{Bechinger, C.} \& \bibinfo{author}{L\"owen, H.}
\newblock \bibinfo{title}{Can the self-propulsion of anisotropic microswimmers
  be described by using forces and torques?}
\newblock \href{http://dx.doi.org/10.1088/0953-8984/27/19/194110}{\emph{J.
  Phys. Condens. Matter}}
  \href{http://dx.doi.org/10.1088/0953-8984/27/19/194110}{\textbf{\bibinfo{volume}{27}},
  \bibinfo{pages}{194110}}
  \href{http://dx.doi.org/10.1088/0953-8984/27/19/194110}{
  (\bibinfo{year}{2015})}.

\bibitem{Kaiser15}
\bibinfo{author}{Kaiser, A.}, \bibinfo{author}{Babel, S.}, \bibinfo{author}{ten
  Hagen, B.}, \bibinfo{author}{von Ferber, C.} \& \bibinfo{author}{L{\"o}wen,
  H.}
\newblock \bibinfo{title}{How does a flexible chain of active particles swell?}
\newblock \href{http://dx.doi.org/10.1063/1.4916134}{\emph{J. Chem. Phys.}}
  \href{http://dx.doi.org/10.1063/1.4916134}{\textbf{\bibinfo{volume}{142}},
  \bibinfo{pages}{124905}} \href{http://dx.doi.org/10.1063/1.4916134}{
  (\bibinfo{year}{2015})}.

\bibitem{Eisenstecken16}
\bibinfo{author}{Eisenstecken, T.}, \bibinfo{author}{Gompper, G.} \&
  \bibinfo{author}{Winkler, R.~G.}
\newblock \bibinfo{title}{Conformational properties of active semiflexible
  polymers}.
\newblock \href{http://dx.doi.org/10.3390/polym8080304}{\emph{Polymers}}
  \href{http://dx.doi.org/10.3390/polym8080304}{\textbf{\bibinfo{volume}{8}},
  \bibinfo{pages}{304}} \href{http://dx.doi.org/10.3390/polym8080304}{
  (\bibinfo{year}{2016})}.

\bibitem{Eisenstecken17}
\bibinfo{author}{Eisenstecken, T.}, \bibinfo{author}{Gompper, G.} \&
  \bibinfo{author}{Winkler, R.~G.}
\newblock \bibinfo{title}{Internal dynamics of semiflexible polymers with
  active noise}.
\newblock \href{http://dx.doi.org/10.1063/1.4981012}{\emph{J. Chem. Phys.}}
  \href{http://dx.doi.org/10.1063/1.4981012}{\textbf{\bibinfo{volume}{146}},
  \bibinfo{pages}{154903}} \href{http://dx.doi.org/10.1063/1.4981012}{
  (\bibinfo{year}{2017})}.

\bibitem{Kourbane-Houssene18}
\bibinfo{author}{Kourbane-Houssene, M.}, \bibinfo{author}{Erignoux, C.},
  \bibinfo{author}{Bodineau, T.} \& \bibinfo{author}{Tailleur, J.}
\newblock \bibinfo{title}{Exact hydrodynamic description of active lattice
  gases}.
\newblock \href{http://dx.doi.org/10.1103/PhysRevLett.120.268003}{\emph{Phys.
  Rev. Lett.}}
  \href{http://dx.doi.org/10.1103/PhysRevLett.120.268003}{\textbf{\bibinfo{volume}{120}},
  \bibinfo{pages}{268003}}
  \href{http://dx.doi.org/10.1103/PhysRevLett.120.268003}{
  (\bibinfo{year}{2018})}.

\bibitem{Klamser18}
\bibinfo{author}{Klamser, J.~U.}, \bibinfo{author}{Kapfer, S.~C.} \&
  \bibinfo{author}{Krauth, W.}
\newblock \bibinfo{title}{Thermodynamic phases in two-dimensional active
  matter}.
\newblock \href{http://dx.doi.org/10.1038/s41467-018-07491-5}{\emph{Nat.
  Commun.}}
  \href{http://dx.doi.org/10.1038/s41467-018-07491-5}{\textbf{\bibinfo{volume}{9}},
  \bibinfo{pages}{5045}} \href{http://dx.doi.org/10.1038/s41467-018-07491-5}{
  (\bibinfo{year}{2018})}.

\bibitem{Sadjadi15}
\bibinfo{author}{Sadjadi, Z.}, \bibinfo{author}{Shaebani, M.~R.},
  \bibinfo{author}{Rieger, H.} \& \bibinfo{author}{Santen, L.}
\newblock \bibinfo{title}{Persistent-random-walk approach to anomalous
  transport of self-propelled particles}.
\newblock \href{http://dx.doi.org/10.1103/PhysRevE.91.062715}{\emph{Phys. Rev.
  E}}
  \href{http://dx.doi.org/10.1103/PhysRevE.91.062715}{\textbf{\bibinfo{volume}{91}},
  \bibinfo{pages}{062715}} \href{http://dx.doi.org/10.1103/PhysRevE.91.062715}{
  (\bibinfo{year}{2015})}.

\bibitem{Shaebani14}
\bibinfo{author}{Shaebani, M.~R.}, \bibinfo{author}{Sadjadi, Z.},
  \bibinfo{author}{Sokolov, I.~M.}, \bibinfo{author}{Rieger, H.} \&
  \bibinfo{author}{Santen, L.}
\newblock \bibinfo{title}{Anomalous diffusion of self-propelled particles in
  directed random environments}.
\newblock \href{http://dx.doi.org/10.1103/PhysRevE.90.030701}{\emph{Phys. Rev.
  E}}
  \href{http://dx.doi.org/10.1103/PhysRevE.90.030701}{\textbf{\bibinfo{volume}{90}},
  \bibinfo{pages}{030701}} \href{http://dx.doi.org/10.1103/PhysRevE.90.030701}{
  (\bibinfo{year}{2014})}.

\bibitem{Levis14}
\bibinfo{author}{Levis, D.} \& \bibinfo{author}{Berthier, L.}
\newblock \bibinfo{title}{Clustering and heterogeneous dynamics in a kinetic
  {Monte} {Carlo} model of self-propelled hard disks}.
\newblock \href{http://dx.doi.org/10.1103/PhysRevE.89.062301}{\emph{Phys. Rev.
  E}}
  \href{http://dx.doi.org/10.1103/PhysRevE.89.062301}{\textbf{\bibinfo{volume}{89}},
  \bibinfo{pages}{062301}} \href{http://dx.doi.org/10.1103/PhysRevE.89.062301}{
  (\bibinfo{year}{2014})}.

\bibitem{Najafi18}
\bibinfo{author}{Najafi, J.}, \bibinfo{author}{Shaebani, M.~R.},
  \bibinfo{author}{John, T.}, \bibinfo{author}{Altegoer, F.},
  \bibinfo{author}{Bange, G.} \& \bibinfo{author}{Wagner, C.}
\newblock \bibinfo{title}{Flagellar number governs bacterial spreading and
  transport efficiency}.
\newblock \href{http://dx.doi.org/10.1126/sciadv.aar6425}{\emph{Science Adv.}}
  \href{http://dx.doi.org/10.1126/sciadv.aar6425}{\textbf{\bibinfo{volume}{4}}}
  \href{http://dx.doi.org/10.1126/sciadv.aar6425}{ (\bibinfo{year}{2018})}.

\bibitem{Hafner16}
\bibinfo{author}{Hafner, A.~E.}, \bibinfo{author}{Santen, L.},
  \bibinfo{author}{Rieger, H.} \& \bibinfo{author}{Shaebani, M.~R.}
\newblock \bibinfo{title}{Run-and-pause dynamics of cytoskeletal motor
  proteins}.
\newblock \href{http://dx.doi.org/10.1038/srep37162}{\emph{Sci. Rep.}}
  \href{http://dx.doi.org/10.1038/srep37162}{\textbf{\bibinfo{volume}{6}},
  \bibinfo{pages}{37162}} \href{http://dx.doi.org/10.1038/srep37162}{
  (\bibinfo{year}{2016})}.

\bibitem{Redner13}
\bibinfo{author}{Redner, G.~S.}, \bibinfo{author}{Hagan, M.~F.} \&
  \bibinfo{author}{Baskaran, A.}
\newblock \bibinfo{title}{Structure and dynamics of a phase-separating active
  colloidal fluid}.
\newblock \href{http://dx.doi.org/10.1103/PhysRevLett.110.055701}{\emph{Phys.
  Rev. Lett.}}
  \href{http://dx.doi.org/10.1103/PhysRevLett.110.055701}{\textbf{\bibinfo{volume}{110}},
  \bibinfo{pages}{055701}}
  \href{http://dx.doi.org/10.1103/PhysRevLett.110.055701}{
  (\bibinfo{year}{2013})}.

\bibitem{Cates15}
\bibinfo{author}{Cates, M.~E.} \& \bibinfo{author}{Tailleur, J.}
\newblock \bibinfo{title}{Motility-induced phase separation}.
\newblock
  \href{http://dx.doi.org/10.1146/annurev-conmatphys-031214-014710}{\emph{Annu.
  Rev. Condens. Matter Phys.}}
  \href{http://dx.doi.org/10.1146/annurev-conmatphys-031214-014710}{\textbf{\bibinfo{volume}{6}},
  \bibinfo{pages}{219--244}}
  \href{http://dx.doi.org/10.1146/annurev-conmatphys-031214-014710}{
  (\bibinfo{year}{2015})}.

\bibitem{Wysocki14}
\bibinfo{author}{Wysocki, A.}, \bibinfo{author}{Winkler, R.~G.} \&
  \bibinfo{author}{Gompper, G.}
\newblock \bibinfo{title}{Cooperative motion of active {Brownian} spheres in
  three-dimensional dense suspensions}.
\newblock \href{http://dx.doi.org/10.1209/0295-5075/105/48004}{\emph{Europhys.
  Lett.}}
  \href{http://dx.doi.org/10.1209/0295-5075/105/48004}{\textbf{\bibinfo{volume}{105}},
  \bibinfo{pages}{48004}} \href{http://dx.doi.org/10.1209/0295-5075/105/48004}{
  (\bibinfo{year}{2014})}.

\bibitem{Stenhammar14}
\bibinfo{author}{Stenhammar, J.}, \bibinfo{author}{Marenduzzo, D.},
  \bibinfo{author}{Allen, R.~J.} \& \bibinfo{author}{Cates, M.~E.}
\newblock \bibinfo{title}{Phase behaviour of active {Brownian} particles: the
  role of dimensionality}.
\newblock \href{http://dx.doi.org/10.1039/C3SM52813H}{\emph{Soft Matter}}
  \href{http://dx.doi.org/10.1039/C3SM52813H}{\textbf{\bibinfo{volume}{10}},
  \bibinfo{pages}{1489--1499}} \href{http://dx.doi.org/10.1039/C3SM52813H}{
  (\bibinfo{year}{2014})}.

\bibitem{Wysocki16}
\bibinfo{author}{Wysocki, A.}, \bibinfo{author}{Winkler, R.~G.} \&
  \bibinfo{author}{Gompper, G.}
\newblock \bibinfo{title}{Propagating interfaces in mixtures of active and
  passive {Brownian} particles}.
\newblock \href{http://dx.doi.org/10.1088/1367-2630/aa529d}{\emph{New J.
  Phys.}}
  \href{http://dx.doi.org/10.1088/1367-2630/aa529d}{\textbf{\bibinfo{volume}{18}},
  \bibinfo{pages}{123030}} \href{http://dx.doi.org/10.1088/1367-2630/aa529d}{
  (\bibinfo{year}{2016})}.

\bibitem{Stenhammar15}
\bibinfo{author}{Stenhammar, J.}, \bibinfo{author}{Wittkowski, R.},
  \bibinfo{author}{Marenduzzo, D.} \& \bibinfo{author}{Cates, M.~E.}
\newblock \bibinfo{title}{Activity-induced phase separation and self-assembly
  in mixtures of active and passive particles}.
\newblock \href{http://dx.doi.org/10.1103/PhysRevLett.114.018301}{\emph{Phys.
  Rev. Lett.}}
  \href{http://dx.doi.org/10.1103/PhysRevLett.114.018301}{\textbf{\bibinfo{volume}{114}},
  \bibinfo{pages}{018301}}
  \href{http://dx.doi.org/10.1103/PhysRevLett.114.018301}{
  (\bibinfo{year}{2015})}.

\bibitem{Digregorio18}
\bibinfo{author}{Digregorio, P.}, \bibinfo{author}{Levis, D.},
  \bibinfo{author}{Suma, A.}, \bibinfo{author}{Cugliandolo, L.~F.},
  \bibinfo{author}{Gonnella, G.} \& \bibinfo{author}{Pagonabarraga, I.}
\newblock \bibinfo{title}{Full phase diagram of active {B}rownian disks: From
  melting to motility-induced phase separation}.
\newblock \href{http://dx.doi.org/10.1103/PhysRevLett.121.098003}{\emph{Phys.
  Rev. Lett.}}
  \href{http://dx.doi.org/10.1103/PhysRevLett.121.098003}{\textbf{\bibinfo{volume}{121}},
  \bibinfo{pages}{098003}}
  \href{http://dx.doi.org/10.1103/PhysRevLett.121.098003}{
  (\bibinfo{year}{2018})}.

\bibitem{Fily14b}
\bibinfo{author}{Fily, Y.}, \bibinfo{author}{Henkes, S.} \&
  \bibinfo{author}{Marchetti, M.~C.}
\newblock \bibinfo{title}{Freezing and phase separation of self-propelled
  disks}.
\newblock \href{http://dx.doi.org/10.1039/C3SM52469H}{\emph{Soft Matter}}
  \href{http://dx.doi.org/10.1039/C3SM52469H}{\textbf{\bibinfo{volume}{10}},
  \bibinfo{pages}{2132--2140}} \href{http://dx.doi.org/10.1039/C3SM52469H}{
  (\bibinfo{year}{2014})}.

\bibitem{Elgeti13}
\bibinfo{author}{Elgeti, J.} \& \bibinfo{author}{Gompper, G.}
\newblock \bibinfo{title}{Wall accumulation of self-propelled spheres}.
\newblock \href{http://dx.doi.org/10.1209/0295-5075/101/48003}{\emph{EPL}}
  \href{http://dx.doi.org/10.1209/0295-5075/101/48003}{\textbf{\bibinfo{volume}{101}},
  \bibinfo{pages}{48003}} \href{http://dx.doi.org/10.1209/0295-5075/101/48003}{
  (\bibinfo{year}{2013})}.

\bibitem{Fily14}
\bibinfo{author}{Fily, Y.}, \bibinfo{author}{Baskaran, A.} \&
  \bibinfo{author}{Hagan, M.}
\newblock \bibinfo{title}{Dynamics of self-propelled particles under strong
  confinement}.
\newblock \href{http://dx.doi.org/10.1039/C4SM00975D}{\emph{Soft Matter}}
  \href{http://dx.doi.org/10.1039/C4SM00975D}{\textbf{\bibinfo{volume}{10}},
  \bibinfo{pages}{5609--5617}} \href{http://dx.doi.org/10.1039/C4SM00975D}{
  (\bibinfo{year}{2014})}.

\bibitem{Das19}
\bibinfo{author}{Das, S.}, \bibinfo{author}{Gompper, G.} \&
  \bibinfo{author}{Winkler, R.~G.}
\newblock \bibinfo{title}{Local stress and pressure in an inhomogeneous system
  of spherical active {Brownian} particles}.
\newblock \href{http://dx.doi.org/10.1038/s41598-019-43077-x}{\emph{Sci. Rep.}}
  \href{http://dx.doi.org/10.1038/s41598-019-43077-x}{\textbf{\bibinfo{volume}{9}},
  \bibinfo{pages}{6608}} \href{http://dx.doi.org/10.1038/s41598-019-43077-x}{
  (\bibinfo{year}{2019})}.

\bibitem{Wysocki19}
\bibinfo{author}{Wysocki, A.} \& \bibinfo{author}{Rieger, H.}
\newblock \bibinfo{title}{{Capillary action in scalar active matter}}.
\newblock \emph{arXiv e-prints} \bibinfo{pages}{arXiv:1908.03368}
  (\bibinfo{year}{2019}).

\bibitem{Takatori14}
\bibinfo{author}{Takatori, S.~C.}, \bibinfo{author}{Yan, W.} \&
  \bibinfo{author}{Brady, J.~F.}
\newblock \bibinfo{title}{Swim pressure: Stress generation in active matter}.
\newblock \href{http://dx.doi.org/10.1103/PhysRevLett.113.028103}{\emph{Phys.
  Rev. Lett.}}
  \href{http://dx.doi.org/10.1103/PhysRevLett.113.028103}{\textbf{\bibinfo{volume}{113}},
  \bibinfo{pages}{028103}}
  \href{http://dx.doi.org/10.1103/PhysRevLett.113.028103}{
  (\bibinfo{year}{2014})}.

\bibitem{Solon15}
\bibinfo{author}{Solon, A.~P.}, \bibinfo{author}{Fily, Y.},
  \bibinfo{author}{Baskaran, A.}, \bibinfo{author}{Cates, M.~E.},
  \bibinfo{author}{Kafri, Y.}, \bibinfo{author}{Kardar, M.} \&
  \bibinfo{author}{Tailleur, J.}
\newblock \bibinfo{title}{Pressure is not a state function for generic active
  fluids}.
\newblock \href{http://dx.doi.org/10.1038/nphys3377}{\emph{Nat. Phys.}}
  \href{http://dx.doi.org/10.1038/nphys3377}{\textbf{\bibinfo{volume}{11}},
  \bibinfo{pages}{673}} \href{http://dx.doi.org/10.1038/nphys3377}{
  (\bibinfo{year}{2015})}.

\bibitem{Winkler15}
\bibinfo{author}{Winkler, R.~G.}, \bibinfo{author}{Wysocki, A.} \&
  \bibinfo{author}{Gompper, G.}
\newblock \bibinfo{title}{Virial pressure in systems of active {Brownian}
  particles}.
\newblock \href{http://dx.doi.org/10.1039/C5SM01412C}{\emph{Soft Matter}}
  \href{http://dx.doi.org/10.1039/C5SM01412C}{\textbf{\bibinfo{volume}{11}},
  \bibinfo{pages}{6680--6691}} \href{http://dx.doi.org/10.1039/C5SM01412C}{
  (\bibinfo{year}{2015})}.

\bibitem{Fily18}
\bibinfo{author}{Fily, Y.}, \bibinfo{author}{Kafri, Y.},
  \bibinfo{author}{Solon, A.~P.}, \bibinfo{author}{Tailleur, J.} \&
  \bibinfo{author}{Turner, A.}
\newblock \bibinfo{title}{Mechanical pressure and momentum conservation in dry
  active matter}.
\newblock \href{http://dx.doi.org/10.1088/1751-8121/aa99b6}{\emph{J. Phys. A:
  Math. Theor.}}
  \href{http://dx.doi.org/10.1088/1751-8121/aa99b6}{\textbf{\bibinfo{volume}{51}},
  \bibinfo{pages}{044003}} \href{http://dx.doi.org/10.1088/1751-8121/aa99b6}{
  (\bibinfo{year}{2018})}.

\bibitem{Wysocki15}
\bibinfo{author}{Wysocki, A.}, \bibinfo{author}{Elgeti, J.} \&
  \bibinfo{author}{Gompper, G.}
\newblock \bibinfo{title}{Giant adsorption of microswimmers: Duality of shape
  asymmetry and wall curvature}.
\newblock \href{http://dx.doi.org/10.1103/PhysRevE.91.050302}{\emph{Phys. Rev.
  E}}
  \href{http://dx.doi.org/10.1103/PhysRevE.91.050302}{\textbf{\bibinfo{volume}{91}},
  \bibinfo{pages}{050302(R)}}
  \href{http://dx.doi.org/10.1103/PhysRevE.91.050302}{ (\bibinfo{year}{2015})}.

\bibitem{Tjhung18}
\bibinfo{author}{Tjhung, E.}, \bibinfo{author}{Nardini, C.} \&
  \bibinfo{author}{Cates, M.~E.}
\newblock \bibinfo{title}{Cluster phases and bubbly phase separation in active
  fluids: Reversal of the ostwald process}.
\newblock \href{http://dx.doi.org/10.1103/PhysRevX.8.031080}{\emph{Phys. Rev.
  X}}
  \href{http://dx.doi.org/10.1103/PhysRevX.8.031080}{\textbf{\bibinfo{volume}{8}},
  \bibinfo{pages}{031080}} \href{http://dx.doi.org/10.1103/PhysRevX.8.031080}{
  (\bibinfo{year}{2018})}.

\bibitem{Bratanov15}
\bibinfo{author}{Bratanov, V.}, \bibinfo{author}{Jenko, F.} \&
  \bibinfo{author}{Frey, E.}
\newblock \bibinfo{title}{New class of turbulence in active fluids}.
\newblock \href{http://dx.doi.org/10.1073/pnas.1509304112}{\emph{Proc. Natl.
  Acad. Sci. USA}}
  \href{http://dx.doi.org/10.1073/pnas.1509304112}{\textbf{\bibinfo{volume}{112}},
  \bibinfo{pages}{15048--15053}}
  \href{http://dx.doi.org/10.1073/pnas.1509304112}{ (\bibinfo{year}{2015})}.

\bibitem{Weber13}
\bibinfo{author}{Weber, C.~A.}, \bibinfo{author}{Hanke, T.},
  \bibinfo{author}{Deseigne, J.}, \bibinfo{author}{L\'eonard, S.},
  \bibinfo{author}{Dauchot, O.}, \bibinfo{author}{Frey, E.} \&
  \bibinfo{author}{Chat\'e, H.}
\newblock \bibinfo{title}{Long-range ordering of vibrated polar disks}.
\newblock \href{http://dx.doi.org/10.1103/PhysRevLett.110.208001}{\emph{Phys.
  Rev. Lett.}}
  \href{http://dx.doi.org/10.1103/PhysRevLett.110.208001}{\textbf{\bibinfo{volume}{110}},
  \bibinfo{pages}{208001}}
  \href{http://dx.doi.org/10.1103/PhysRevLett.110.208001}{
  (\bibinfo{year}{2013})}.

\bibitem{Hu15}
\bibinfo{author}{Hu, J.}, \bibinfo{author}{Yang, M.}, \bibinfo{author}{Gompper,
  G.} \& \bibinfo{author}{Winkler, R.~G.}
\newblock \bibinfo{title}{Modelling the mechanics and hydrodynamics of swimming
  e. coli}.
\newblock \href{http://dx.doi.org/10.1039/C5SM01678A}{\emph{Soft Matter}}
  \href{http://dx.doi.org/10.1039/C5SM01678A}{\textbf{\bibinfo{volume}{11}},
  \bibinfo{pages}{7867--7876}} \href{http://dx.doi.org/10.1039/C5SM01678A}{
  (\bibinfo{year}{2015})}.

\bibitem{Vicsek12}
\bibinfo{author}{Vicsek, T.} \& \bibinfo{author}{Zafeiris, A.}
\newblock \bibinfo{title}{Collective motion}.
\newblock \href{http://dx.doi.org/10.1016/j.physrep.2012.03.004}{\emph{Phys.
  Rep.}}
  \href{http://dx.doi.org/10.1016/j.physrep.2012.03.004}{\textbf{\bibinfo{volume}{517}},
  \bibinfo{pages}{71--140}}
  \href{http://dx.doi.org/10.1016/j.physrep.2012.03.004}{
  (\bibinfo{year}{2012})}.

\bibitem{Vicsek95}
\bibinfo{author}{Vicsek, T.}, \bibinfo{author}{Czir\'ok, A.},
  \bibinfo{author}{Ben-Jacob, E.}, \bibinfo{author}{Cohen, I.} \&
  \bibinfo{author}{Shochet, O.}
\newblock \bibinfo{title}{Novel type of phase transition in a system of
  self-driven particles}.
\newblock \href{http://dx.doi.org/10.1103/PhysRevLett.75.1226}{\emph{Phys. Rev.
  Lett.}}
  \href{http://dx.doi.org/10.1103/PhysRevLett.75.1226}{\textbf{\bibinfo{volume}{75}},
  \bibinfo{pages}{1226--1229}}
  \href{http://dx.doi.org/10.1103/PhysRevLett.75.1226}{
  (\bibinfo{year}{1995})}.

\bibitem{Aldana07}
\bibinfo{author}{Aldana, M.}, \bibinfo{author}{Dossetti, V.},
  \bibinfo{author}{Huepe, C.}, \bibinfo{author}{Kenkre, V.~M.} \&
  \bibinfo{author}{Larralde, H.}
\newblock \bibinfo{title}{Phase transitions in systems of self-propelled agents
  and related network models}.
\newblock \href{http://dx.doi.org/10.1103/PhysRevLett.98.095702}{\emph{Phys.
  Rev. Lett.}}
  \href{http://dx.doi.org/10.1103/PhysRevLett.98.095702}{\textbf{\bibinfo{volume}{98}},
  \bibinfo{pages}{095702}}
  \href{http://dx.doi.org/10.1103/PhysRevLett.98.095702}{
  (\bibinfo{year}{2007})}.

\bibitem{Aldana09}
\bibinfo{author}{Aldana, M.}, \bibinfo{author}{Larralde, H.} \&
  \bibinfo{author}{Vazquez, B.}
\newblock \bibinfo{title}{On the emergence of collective order in swarming
  systems: A recent debate}.
\newblock \href{http://dx.doi.org/10.1142/S0217979209053552}{\emph{Int. J. Mod.
  Phys. B}}
  \href{http://dx.doi.org/10.1142/S0217979209053552}{\textbf{\bibinfo{volume}{23}},
  \bibinfo{pages}{3661--3685}}
  \href{http://dx.doi.org/10.1142/S0217979209053552}{ (\bibinfo{year}{2009})}.

\bibitem{Peruani18}
\bibinfo{author}{Peruani, F.} \& \bibinfo{author}{Aranson, I.~S.}
\newblock \bibinfo{title}{Cold active motion: How time-independent disorder
  affects the motion of self-propelled agents}.
\newblock \href{http://dx.doi.org/10.1103/PhysRevLett.120.238101}{\emph{Phys.
  Rev. Lett.}}
  \href{http://dx.doi.org/10.1103/PhysRevLett.120.238101}{\textbf{\bibinfo{volume}{120}},
  \bibinfo{pages}{238101}}
  \href{http://dx.doi.org/10.1103/PhysRevLett.120.238101}{
  (\bibinfo{year}{2018})}.

\bibitem{Grossman08}
\bibinfo{author}{Grossman, D.}, \bibinfo{author}{Aranson, I.~S.} \&
  \bibinfo{author}{Jacob, E.~B.}
\newblock \bibinfo{title}{Emergence of agent swarm migration and vortex
  formation through inelastic collisions}.
\newblock \href{http://dx.doi.org/10.1088/1367-2630/10/2/023036}{\emph{New J.
  Phys.}}
  \href{http://dx.doi.org/10.1088/1367-2630/10/2/023036}{\textbf{\bibinfo{volume}{10}},
  \bibinfo{pages}{023036}}
  \href{http://dx.doi.org/10.1088/1367-2630/10/2/023036}{
  (\bibinfo{year}{2008})}.

\bibitem{Nagy07}
\bibinfo{author}{Nagy, M.}, \bibinfo{author}{Daruka, I.} \&
  \bibinfo{author}{Vicsek, T.}
\newblock \bibinfo{title}{New aspects of the continuous phase transition in the
  scalar noise model (snm) of collective motion}.
\newblock \href{http://dx.doi.org/10.1016/j.physa.2006.05.035}{\emph{Physica
  A}}
  \href{http://dx.doi.org/10.1016/j.physa.2006.05.035}{\textbf{\bibinfo{volume}{373}},
  \bibinfo{pages}{445--454}}
  \href{http://dx.doi.org/10.1016/j.physa.2006.05.035}{
  (\bibinfo{year}{2007})}.

\bibitem{Peruani11}
\bibinfo{author}{Peruani, F.}, \bibinfo{author}{Klauss, T.},
  \bibinfo{author}{Deutsch, A.} \& \bibinfo{author}{Voss-Boehme, A.}
\newblock \bibinfo{title}{Traffic jams, gliders, and bands in the quest for
  collective motion of self-propelled particles}.
\newblock \href{http://dx.doi.org/10.1103/PhysRevLett.106.128101}{\emph{Phys.
  Rev. Lett.}}
  \href{http://dx.doi.org/10.1103/PhysRevLett.106.128101}{\textbf{\bibinfo{volume}{106}},
  \bibinfo{pages}{128101}}
  \href{http://dx.doi.org/10.1103/PhysRevLett.106.128101}{
  (\bibinfo{year}{2011})}.

\bibitem{Ginelli10}
\bibinfo{author}{Ginelli, F.} \& \bibinfo{author}{Chat\'e, H.}
\newblock \bibinfo{title}{Relevance of metric-free interactions in flocking
  phenomena}.
\newblock \href{http://dx.doi.org/10.1103/PhysRevLett.105.168103}{\emph{Phys.
  Rev. Lett.}}
  \href{http://dx.doi.org/10.1103/PhysRevLett.105.168103}{\textbf{\bibinfo{volume}{105}},
  \bibinfo{pages}{168103}}
  \href{http://dx.doi.org/10.1103/PhysRevLett.105.168103}{
  (\bibinfo{year}{2010})}.

\bibitem{Gregoire03}
\bibinfo{author}{Gregoire, G.}, \bibinfo{author}{Chat{\'e}, H.} \&
  \bibinfo{author}{Tu, Y.}
\newblock \bibinfo{title}{Moving and staying together without a leader}.
\newblock \href{http://dx.doi.org/10.1016/S0167-2789(03)00102-7}{\emph{Physica
  D}}
  \href{http://dx.doi.org/10.1016/S0167-2789(03)00102-7}{\textbf{\bibinfo{volume}{181}},
  \bibinfo{pages}{157--170}}
  \href{http://dx.doi.org/10.1016/S0167-2789(03)00102-7}{
  (\bibinfo{year}{2003})}.

\bibitem{Chate08}
\bibinfo{author}{Chat{\'e}, H.}, \bibinfo{author}{Ginelli, F.},
  \bibinfo{author}{Gr{\'e}goire, G.}, \bibinfo{author}{Peruani, F.} \&
  \bibinfo{author}{Raynaud, F.}
\newblock \bibinfo{title}{Modeling collective motion: variations on the vicsek
  model}.
\newblock \href{http://dx.doi.org/10.1140/epjb/e2008-00275-9}{\emph{Eur. Phys.
  J. B}}
  \href{http://dx.doi.org/10.1140/epjb/e2008-00275-9}{\textbf{\bibinfo{volume}{64}},
  \bibinfo{pages}{451--456}}
  \href{http://dx.doi.org/10.1140/epjb/e2008-00275-9}{ (\bibinfo{year}{2008})}.

\bibitem{Szabo09}
\bibinfo{author}{Szab\'o, P.}, \bibinfo{author}{Nagy, M.} \&
  \bibinfo{author}{Vicsek, T.}
\newblock \bibinfo{title}{Transitions in a self-propelled-particles model with
  coupling of accelerations}.
\newblock \href{http://dx.doi.org/10.1103/PhysRevE.79.021908}{\emph{Phys. Rev.
  E}}
  \href{http://dx.doi.org/10.1103/PhysRevE.79.021908}{\textbf{\bibinfo{volume}{79}},
  \bibinfo{pages}{021908}} \href{http://dx.doi.org/10.1103/PhysRevE.79.021908}{
  (\bibinfo{year}{2009})}.

\bibitem{Chate06}
\bibinfo{author}{Chat\'e, H.}, \bibinfo{author}{Ginelli, F.} \&
  \bibinfo{author}{Montagne, R.}
\newblock \bibinfo{title}{Simple model for active nematics: Quasi-long-range
  order and giant fluctuations}.
\newblock \href{http://dx.doi.org/10.1103/PhysRevLett.96.180602}{\emph{Phys.
  Rev. Lett.}}
  \href{http://dx.doi.org/10.1103/PhysRevLett.96.180602}{\textbf{\bibinfo{volume}{96}},
  \bibinfo{pages}{180602}}
  \href{http://dx.doi.org/10.1103/PhysRevLett.96.180602}{
  (\bibinfo{year}{2006})}.

\bibitem{Mahault18}
\bibinfo{author}{Mahault, B.}, \bibinfo{author}{Jiang, X.-c.},
  \bibinfo{author}{Bertin, E.}, \bibinfo{author}{Ma, Y.-q.},
  \bibinfo{author}{Patelli, A.}, \bibinfo{author}{Shi, X.-q.} \&
  \bibinfo{author}{Chat\'e, H.}
\newblock \bibinfo{title}{Self-propelled particles with velocity reversals and
  ferromagnetic alignment: Active matter class with second-order transition to
  quasi-long-range polar order}.
\newblock \href{http://dx.doi.org/10.1103/PhysRevLett.120.258002}{\emph{Phys.
  Rev. Lett.}}
  \href{http://dx.doi.org/10.1103/PhysRevLett.120.258002}{\textbf{\bibinfo{volume}{120}},
  \bibinfo{pages}{258002}}
  \href{http://dx.doi.org/10.1103/PhysRevLett.120.258002}{
  (\bibinfo{year}{2018})}.

\bibitem{Ginelli10b}
\bibinfo{author}{Ginelli, F.}, \bibinfo{author}{Peruani, F.},
  \bibinfo{author}{B\"ar, M.} \& \bibinfo{author}{Chat\'e, H.}
\newblock \bibinfo{title}{Large-scale collective properties of self-propelled
  rods}.
\newblock \href{http://dx.doi.org/10.1103/PhysRevLett.104.184502}{\emph{Phys.
  Rev. Lett.}}
  \href{http://dx.doi.org/10.1103/PhysRevLett.104.184502}{\textbf{\bibinfo{volume}{104}},
  \bibinfo{pages}{184502}}
  \href{http://dx.doi.org/10.1103/PhysRevLett.104.184502}{
  (\bibinfo{year}{2010})}.

\bibitem{Strombom11}
\bibinfo{author}{Str{\"o}mbom, D.}
\newblock \bibinfo{title}{Collective motion from local attraction}.
\newblock \href{http://dx.doi.org/10.1016/j.jtbi.2011.05.019}{\emph{J. Theor.
  Biol.}}
  \href{http://dx.doi.org/10.1016/j.jtbi.2011.05.019}{\textbf{\bibinfo{volume}{283}},
  \bibinfo{pages}{145--151}}
  \href{http://dx.doi.org/10.1016/j.jtbi.2011.05.019}{ (\bibinfo{year}{2011})}.

\bibitem{Solon15b}
\bibinfo{author}{Solon, A.~P.}, \bibinfo{author}{Chat\'e, H.} \&
  \bibinfo{author}{Tailleur, J.}
\newblock \bibinfo{title}{From phase to microphase separation in flocking
  models: The essential role of nonequilibrium fluctuations}.
\newblock \href{http://dx.doi.org/10.1103/PhysRevLett.114.068101}{\emph{Phys.
  Rev. Lett.}}
  \href{http://dx.doi.org/10.1103/PhysRevLett.114.068101}{\textbf{\bibinfo{volume}{114}},
  \bibinfo{pages}{068101}}
  \href{http://dx.doi.org/10.1103/PhysRevLett.114.068101}{
  (\bibinfo{year}{2015})}.

\bibitem{Solon13}
\bibinfo{author}{Solon, A.~P.} \& \bibinfo{author}{Tailleur, J.}
\newblock \bibinfo{title}{Revisiting the flocking transition using active
  spins}.
\newblock \href{http://dx.doi.org/10.1103/PhysRevLett.111.078101}{\emph{Phys.
  Rev. Lett.}}
  \href{http://dx.doi.org/10.1103/PhysRevLett.111.078101}{\textbf{\bibinfo{volume}{111}},
  \bibinfo{pages}{078101}}
  \href{http://dx.doi.org/10.1103/PhysRevLett.111.078101}{
  (\bibinfo{year}{2013})}.

\bibitem{Szabo06}
\bibinfo{author}{Szab\'o, B.}, \bibinfo{author}{Sz\"oll\"osi, G.~J.},
  \bibinfo{author}{G\"onci, B.}, \bibinfo{author}{Jur\'anyi, Z.},
  \bibinfo{author}{Selmeczi, D.} \& \bibinfo{author}{Vicsek, T.}
\newblock \bibinfo{title}{Phase transition in the collective migration of
  tissue cells: Experiment and model}.
\newblock \href{http://dx.doi.org/10.1103/PhysRevE.74.061908}{\emph{Phys. Rev.
  E}}
  \href{http://dx.doi.org/10.1103/PhysRevE.74.061908}{\textbf{\bibinfo{volume}{74}},
  \bibinfo{pages}{061908}} \href{http://dx.doi.org/10.1103/PhysRevE.74.061908}{
  (\bibinfo{year}{2006})}.

\bibitem{Peruani06}
\bibinfo{author}{Peruani, F.}, \bibinfo{author}{Deutsch, A.} \&
  \bibinfo{author}{B\"ar, M.}
\newblock \bibinfo{title}{Nonequilibrium clustering of self-propelled rods}.
\newblock \href{http://dx.doi.org/10.1103/PhysRevE.74.030904}{\emph{Phys. Rev.
  E}}
  \href{http://dx.doi.org/10.1103/PhysRevE.74.030904}{\textbf{\bibinfo{volume}{74}},
  \bibinfo{pages}{030904}} \href{http://dx.doi.org/10.1103/PhysRevE.74.030904}{
  (\bibinfo{year}{2006})}.

\bibitem{Toner95}
\bibinfo{author}{Toner, J.} \& \bibinfo{author}{Tu, Y.}
\newblock \bibinfo{title}{Long-range order in a two-dimensional dynamical
  $\mathrm{XY}$ model: How birds fly together}.
\newblock \href{http://dx.doi.org/10.1103/PhysRevLett.75.4326}{\emph{Phys. Rev.
  Lett.}}
  \href{http://dx.doi.org/10.1103/PhysRevLett.75.4326}{\textbf{\bibinfo{volume}{75}},
  \bibinfo{pages}{4326--4329}}
  \href{http://dx.doi.org/10.1103/PhysRevLett.75.4326}{
  (\bibinfo{year}{1995})}.

\bibitem{Toner98}
\bibinfo{author}{Toner, J.} \& \bibinfo{author}{Tu, Y.}
\newblock \bibinfo{title}{Flocks, herds, and schools: A quantitative theory of
  flocking}.
\newblock \href{http://dx.doi.org/10.1103/PhysRevE.58.4828}{\emph{Phys. Rev.
  E}}
  \href{http://dx.doi.org/10.1103/PhysRevE.58.4828}{\textbf{\bibinfo{volume}{58}},
  \bibinfo{pages}{4828--4858}}
  \href{http://dx.doi.org/10.1103/PhysRevE.58.4828}{ (\bibinfo{year}{1998})}.

\bibitem{Bertin06}
\bibinfo{author}{Bertin, E.}, \bibinfo{author}{Droz, M.} \&
  \bibinfo{author}{Gr\'egoire, G.}
\newblock \bibinfo{title}{{B}oltzmann and hydrodynamic description for
  self-propelled particles}.
\newblock \href{http://dx.doi.org/10.1103/PhysRevE.74.022101}{\emph{Phys. Rev.
  E}}
  \href{http://dx.doi.org/10.1103/PhysRevE.74.022101}{\textbf{\bibinfo{volume}{74}},
  \bibinfo{pages}{022101}} \href{http://dx.doi.org/10.1103/PhysRevE.74.022101}{
  (\bibinfo{year}{2006})}.

\bibitem{Ihle11}
\bibinfo{author}{Ihle, T.}
\newblock \bibinfo{title}{Kinetic theory of flocking: Derivation of
  hydrodynamic equations}.
\newblock \href{http://dx.doi.org/10.1103/PhysRevE.83.030901}{\emph{Phys. Rev.
  E}}
  \href{http://dx.doi.org/10.1103/PhysRevE.83.030901}{\textbf{\bibinfo{volume}{83}},
  \bibinfo{pages}{030901}} \href{http://dx.doi.org/10.1103/PhysRevE.83.030901}{
  (\bibinfo{year}{2011})}.

\bibitem{Ranft10}
\bibinfo{author}{Ranft, J.}, \bibinfo{author}{Basan, M.},
  \bibinfo{author}{Elgeti, J.}, \bibinfo{author}{Joanny, J.-F.},
  \bibinfo{author}{Prost, J.} \& \bibinfo{author}{J{\"u}licher, F.}
\newblock \bibinfo{title}{Fluidization of tissues by cell division and
  apoptosis}.
\newblock \href{http://dx.doi.org/10.1073/pnas.1011086107}{\emph{Proc. Natl.
  Acad. Sci. USA}}
  \href{http://dx.doi.org/10.1073/pnas.1011086107}{\textbf{\bibinfo{volume}{107}},
  \bibinfo{pages}{20863--20868}}
  \href{http://dx.doi.org/10.1073/pnas.1011086107}{ (\bibinfo{year}{2010})}.

\bibitem{Narayan07}
\bibinfo{author}{Narayan, V.}, \bibinfo{author}{Ramaswamy, S.} \&
  \bibinfo{author}{Menon, N.}
\newblock \bibinfo{title}{Long-lived giant number fluctuations in a swarming
  granular nematic}.
\newblock \href{http://dx.doi.org/10.1126/science.1140414}{\emph{Science}}
  \href{http://dx.doi.org/10.1126/science.1140414}{\textbf{\bibinfo{volume}{317}},
  \bibinfo{pages}{105--108}} \href{http://dx.doi.org/10.1126/science.1140414}{
  (\bibinfo{year}{2007})}.

\bibitem{Ramaswamy03}
\bibinfo{author}{Ramaswamy, S.}, \bibinfo{author}{Simha, R.~A.} \&
  \bibinfo{author}{Toner, J.}
\newblock \bibinfo{title}{Active nematics on a substrate: Giant number
  fluctuations and long-time tails}.
\newblock \href{http://dx.doi.org/10.1209/epl/i2003-00346-7}{\emph{EPL}}
  \href{http://dx.doi.org/10.1209/epl/i2003-00346-7}{\textbf{\bibinfo{volume}{62}},
  \bibinfo{pages}{196--202}}
  \href{http://dx.doi.org/10.1209/epl/i2003-00346-7}{ (\bibinfo{year}{2003})}.

\bibitem{Hemingway15}
\bibinfo{author}{Hemingway, E.~J.}, \bibinfo{author}{Maitra, A.},
  \bibinfo{author}{Banerjee, S.}, \bibinfo{author}{Marchetti, M.~C.},
  \bibinfo{author}{Ramaswamy, S.}, \bibinfo{author}{Fielding, S.~M.} \&
  \bibinfo{author}{Cates, M.~E.}
\newblock \bibinfo{title}{Active viscoelastic matter: From bacterial drag
  reduction to turbulent solids}.
\newblock \href{http://dx.doi.org/10.1103/PhysRevLett.114.098302}{\emph{Phys.
  Rev. Lett.}}
  \href{http://dx.doi.org/10.1103/PhysRevLett.114.098302}{\textbf{\bibinfo{volume}{114}},
  \bibinfo{pages}{098302}}
  \href{http://dx.doi.org/10.1103/PhysRevLett.114.098302}{
  (\bibinfo{year}{2015})}.

\bibitem{Schwarz13}
\bibinfo{author}{Schwarz, U.~S.} \& \bibinfo{author}{Safran, S.~A.}
\newblock \bibinfo{title}{Physics of adherent cells}.
\newblock \href{http://dx.doi.org/10.1103/RevModPhys.85.1327}{\emph{Rev. Mod.
  Phys.}}
  \href{http://dx.doi.org/10.1103/RevModPhys.85.1327}{\textbf{\bibinfo{volume}{85}},
  \bibinfo{pages}{1327--1381}}
  \href{http://dx.doi.org/10.1103/RevModPhys.85.1327}{ (\bibinfo{year}{2013})}.

\bibitem{Hohenberg77}
\bibinfo{author}{Hohenberg, P.~C.} \& \bibinfo{author}{Halperin, B.~I.}
\newblock \bibinfo{title}{Theory of dynamic critical phenomena}.
\newblock \href{http://dx.doi.org/10.1103/RevModPhys.49.435}{\emph{Rev. Mod.
  Phys.}}
  \href{http://dx.doi.org/10.1103/RevModPhys.49.435}{\textbf{\bibinfo{volume}{49}},
  \bibinfo{pages}{435--479}}
  \href{http://dx.doi.org/10.1103/RevModPhys.49.435}{ (\bibinfo{year}{1977})}.

\bibitem{Wittkowski14}
\bibinfo{author}{Wittkowski, R.}, \bibinfo{author}{Tiribocchi, A.},
  \bibinfo{author}{Stenhammar, J.}, \bibinfo{author}{Allen, R.~J.},
  \bibinfo{author}{Marenduzzo, D.} \& \bibinfo{author}{Cates, M.~E.}
\newblock \bibinfo{title}{Scalar $\phi^4$ field theory for active-particle
  phase separation}.
\newblock \href{http://dx.doi.org/10.1038/ncomms5351}{\emph{Nat. Commun.}}
  \href{http://dx.doi.org/10.1038/ncomms5351}{\textbf{\bibinfo{volume}{5}},
  \bibinfo{pages}{4351}} \href{http://dx.doi.org/10.1038/ncomms5351}{
  (\bibinfo{year}{2014})}.

\bibitem{Stenhammar13}
\bibinfo{author}{Stenhammar, J.}, \bibinfo{author}{Tiribocchi, A.},
  \bibinfo{author}{Allen, R.~J.}, \bibinfo{author}{Marenduzzo, D.} \&
  \bibinfo{author}{Cates, M.~E.}
\newblock \bibinfo{title}{Continuum theory of phase separation kinetics for
  active {B}rownian particles}.
\newblock \href{http://dx.doi.org/10.1103/PhysRevLett.111.145702}{\emph{Phys.
  Rev. Lett.}}
  \href{http://dx.doi.org/10.1103/PhysRevLett.111.145702}{\textbf{\bibinfo{volume}{111}},
  \bibinfo{pages}{145702}}
  \href{http://dx.doi.org/10.1103/PhysRevLett.111.145702}{
  (\bibinfo{year}{2013})}.

\bibitem{Cates13}
\bibinfo{author}{Cates, M.~E.} \& \bibinfo{author}{Tailleur, J.}
\newblock \bibinfo{title}{When are active {B}rownian particles and
  run-and-tumble particles equivalent? consequences for motility-induced phase
  separation}.
\newblock \href{http://dx.doi.org/10.1209/0295-5075/101/20010}{\emph{EPL}}
  \href{http://dx.doi.org/10.1209/0295-5075/101/20010}{\textbf{\bibinfo{volume}{101}},
  \bibinfo{pages}{20010}} \href{http://dx.doi.org/10.1209/0295-5075/101/20010}{
  (\bibinfo{year}{2013})}.

\bibitem{tailleur08}
\bibinfo{author}{Tailleur, J.} \& \bibinfo{author}{Cates, M.~E.}
\newblock \bibinfo{title}{Statistical mechanics of interacting run-and-tumble
  bacteria}.
\newblock \href{http://dx.doi.org/10.1103/PhysRevLett.100.218103}{\emph{Phys.
  Rev. Lett.}}
  \href{http://dx.doi.org/10.1103/PhysRevLett.100.218103}{\textbf{\bibinfo{volume}{100}},
  \bibinfo{pages}{218103}}
  \href{http://dx.doi.org/10.1103/PhysRevLett.100.218103}{
  (\bibinfo{year}{2008})}.

\bibitem{Purcell77}
\bibinfo{author}{Purcell, E.~M.}
\newblock \bibinfo{title}{Life at low {Reynolds} number}.
\newblock \href{http://dx.doi.org/10.1119/1.10903}{\emph{Am. J. Phys.}}
  \href{http://dx.doi.org/10.1119/1.10903}{\textbf{\bibinfo{volume}{45}},
  \bibinfo{pages}{3--11}} \href{http://dx.doi.org/10.1119/1.10903}{
  (\bibinfo{year}{1977})}.

\bibitem{Spagnolie12}
\bibinfo{author}{Spagnolie, S.~E.} \& \bibinfo{author}{Lauga, E.}
\newblock \bibinfo{title}{Hydrodynamics of self-propulsion near a boundary:
  predictions and accuracy of far-field approximations}.
\newblock \href{http://dx.doi.org/10.1017/jfm.2012.101}{\emph{J. Fluid Mech.}}
  \href{http://dx.doi.org/10.1017/jfm.2012.101}{\textbf{\bibinfo{volume}{700}},
  \bibinfo{pages}{105--147}} \href{http://dx.doi.org/10.1017/jfm.2012.101}{
  (\bibinfo{year}{2012})}.

\bibitem{Winkler18}
\bibinfo{author}{Winkler, R.~G.} \& \bibinfo{author}{Gompper, G.}
\newblock \bibinfo{title}{Hydrodynamics in motile active matter}.
\newblock In \bibinfo{editor}{Andreoni, W.} \& \bibinfo{editor}{Yip, S.} (eds.)
  \emph{\bibinfo{booktitle}{Handbook of Materials Modeling: Methods: Theory and
  Modeling}}, \bibinfo{pages}{1--20} (\bibinfo{publisher}{Springer
  International Publishing}, \bibinfo{address}{Cham}, \bibinfo{year}{2018}).
\newblock \urlprefix\url{https://link.springer.com/referencework/}.

\bibitem{Berke08}
\bibinfo{author}{Berke, A.~P.}, \bibinfo{author}{Turner, L.},
  \bibinfo{author}{Berg, H.~C.} \& \bibinfo{author}{Lauga, E.}
\newblock \bibinfo{title}{Hydrodynamic attraction of swimming microorganisms by
  surfaces}.
\newblock \href{http://dx.doi.org/10.1103/PhysRevLett.101.038102}{\emph{Phys.
  Rev. Lett.}}
  \href{http://dx.doi.org/10.1103/PhysRevLett.101.038102}{\textbf{\bibinfo{volume}{101}},
  \bibinfo{pages}{038102}}
  \href{http://dx.doi.org/10.1103/PhysRevLett.101.038102}{
  (\bibinfo{year}{2008})}.

\bibitem{Lauga09}
\bibinfo{author}{Lauga, E.} \& \bibinfo{author}{Powers, T.~R.}
\newblock \bibinfo{title}{The hydrodynamics of swimming microorganisms}.
\newblock \href{http://dx.doi.org/10.1088/0034-4885/72/9/096601}{\emph{Rep.
  Prog. Phys.}}
  \href{http://dx.doi.org/10.1088/0034-4885/72/9/096601}{\textbf{\bibinfo{volume}{72}},
  \bibinfo{pages}{096601}}
  \href{http://dx.doi.org/10.1088/0034-4885/72/9/096601}{
  (\bibinfo{year}{2009})}.

\bibitem{Elgeti16}
\bibinfo{author}{Elgeti, J.} \& \bibinfo{author}{Gompper, G.}
\newblock \bibinfo{title}{Microswimmers near surfaces}.
\newblock \href{http://dx.doi.org/10.1140/epjst/e2016-60070-6}{\emph{Eur. Phys.
  J. Spec. Top.}}
  \href{http://dx.doi.org/10.1140/epjst/e2016-60070-6}{\textbf{\bibinfo{volume}{225}},
  \bibinfo{pages}{2333--2352}}
  \href{http://dx.doi.org/10.1140/epjst/e2016-60070-6}{
  (\bibinfo{year}{2016})}.

\bibitem{Li09}
\bibinfo{author}{Li, G.} \& \bibinfo{author}{Tang, J.~X.}
\newblock \bibinfo{title}{Accumulation of microswimmers near a surface mediated
  by collision and rotational {B}rownian motion}.
\newblock \href{http://dx.doi.org/10.1103/PhysRevLett.103.078101}{\emph{Phys.
  Rev. Lett.}}
  \href{http://dx.doi.org/10.1103/PhysRevLett.103.078101}{\textbf{\bibinfo{volume}{103}},
  \bibinfo{pages}{078101}}
  \href{http://dx.doi.org/10.1103/PhysRevLett.103.078101}{
  (\bibinfo{year}{2009})}.

\bibitem{Elgeti09}
\bibinfo{author}{Elgeti, J.} \& \bibinfo{author}{Gompper, G.}
\newblock \bibinfo{title}{Self-propelled rods near surfaces}.
\newblock \href{http://dx.doi.org/10.1209/0295-5075/85/38002}{\emph{EPL}}
  \href{http://dx.doi.org/10.1209/0295-5075/85/38002}{\textbf{\bibinfo{volume}{85}},
  \bibinfo{pages}{38002}} \href{http://dx.doi.org/10.1209/0295-5075/85/38002}{
  (\bibinfo{year}{2009})}.

\bibitem{Elgeti15}
\bibinfo{author}{Elgeti, J.} \& \bibinfo{author}{Gompper, G.}
\newblock \bibinfo{title}{Run-and-tumble dynamics of self-propelled particles
  in confinement}.
\newblock \href{http://dx.doi.org/10.1209/0295-5075/109/58003}{\emph{EPL}}
  \href{http://dx.doi.org/10.1209/0295-5075/109/58003}{\textbf{\bibinfo{volume}{109}},
  \bibinfo{pages}{58003}} \href{http://dx.doi.org/10.1209/0295-5075/109/58003}{
  (\bibinfo{year}{2015})}.

\bibitem{Schaar15}
\bibinfo{author}{Schaar, K.}, \bibinfo{author}{Z{\"o}ttl, A.} \&
  \bibinfo{author}{Stark, H.}
\newblock \bibinfo{title}{Detention times of microswimmers close to surfaces:
  Influence of hydrodynamic interactions and noise}.
\newblock \href{http://dx.doi.org/10.1103/PhysRevLett.115.038101}{\emph{Phys.
  Rev. Lett.}}
  \href{http://dx.doi.org/10.1103/PhysRevLett.115.038101}{\textbf{\bibinfo{volume}{115}},
  \bibinfo{pages}{038101}}
  \href{http://dx.doi.org/10.1103/PhysRevLett.115.038101}{
  (\bibinfo{year}{2015})}.

\bibitem{Saintillan07}
\bibinfo{author}{Saintillan, D.} \& \bibinfo{author}{Shelley, M.~J.}
\newblock \bibinfo{title}{Orientational order and instabilities in suspensions
  of self-locomoting rods}.
\newblock \href{http://dx.doi.org/10.1103/PhysRevLett.99.058102}{\emph{Phys.
  Rev. Lett.}}
  \href{http://dx.doi.org/10.1103/PhysRevLett.99.058102}{\textbf{\bibinfo{volume}{99}},
  \bibinfo{pages}{058102}}
  \href{http://dx.doi.org/10.1103/PhysRevLett.99.058102}{
  (\bibinfo{year}{2007})}.

\bibitem{Saintillan08}
\bibinfo{author}{Saintillan, D.} \& \bibinfo{author}{Shelley, M.~J.}
\newblock \bibinfo{title}{Instabilities and pattern formation in active
  particle suspensions: Kinetic theory and continuum simulations}.
\newblock \href{http://dx.doi.org/10.1103/PhysRevLett.100.178103}{\emph{Phys.
  Rev. Lett.}}
  \href{http://dx.doi.org/10.1103/PhysRevLett.100.178103}{\textbf{\bibinfo{volume}{100}},
  \bibinfo{pages}{178103}}
  \href{http://dx.doi.org/10.1103/PhysRevLett.100.178103}{
  (\bibinfo{year}{2008})}.

\bibitem{Sanchez12}
\bibinfo{author}{Sanchez, T.}, \bibinfo{author}{Chen, D. T.~N.},
  \bibinfo{author}{DeCamp, S.~J.}, \bibinfo{author}{Heymann, M.} \&
  \bibinfo{author}{Dogic, Z.}
\newblock \bibinfo{title}{Spontaneous motion in hierarchically assembled active
  matter}.
\newblock \href{http://dx.doi.org/10.1038/nature11591}{\emph{Nature}}
  \href{http://dx.doi.org/10.1038/nature11591}{\textbf{\bibinfo{volume}{491}},
  \bibinfo{pages}{431--434}} \href{http://dx.doi.org/10.1038/nature11591}{
  (\bibinfo{year}{2012})}.

\bibitem{Thampi13}
\bibinfo{author}{Thampi, S.~P.}, \bibinfo{author}{Golestanian, R.} \&
  \bibinfo{author}{Yeomans, J.~M.}
\newblock \bibinfo{title}{Velocity correlations in an active nematic}.
\newblock \href{http://dx.doi.org/10.1103/PhysRevLett.111.118101}{\emph{Phys.
  Rev. Lett.}}
  \href{http://dx.doi.org/10.1103/PhysRevLett.111.118101}{\textbf{\bibinfo{volume}{111}},
  \bibinfo{pages}{118101}}
  \href{http://dx.doi.org/10.1103/PhysRevLett.111.118101}{
  (\bibinfo{year}{2013})}.

\bibitem{Giomi13}
\bibinfo{author}{Giomi, L.}, \bibinfo{author}{Bowick, M.~J.},
  \bibinfo{author}{Ma, X.} \& \bibinfo{author}{Marchetti, M.~C.}
\newblock \bibinfo{title}{Defect annihilation and proliferation in active
  nematics}.
\newblock \href{http://dx.doi.org/10.1103/PhysRevLett.110.228101}{\emph{Phys.
  Rev. Lett.}}
  \href{http://dx.doi.org/10.1103/PhysRevLett.110.228101}{\textbf{\bibinfo{volume}{110}},
  \bibinfo{pages}{228101}}
  \href{http://dx.doi.org/10.1103/PhysRevLett.110.228101}{
  (\bibinfo{year}{2013})}.

\bibitem{Keber14}
\bibinfo{author}{Keber, F.~C.}, \bibinfo{author}{Loiseau, E.},
  \bibinfo{author}{Sanchez, T.}, \bibinfo{author}{DeCamp, S.~J.},
  \bibinfo{author}{Giomi, L.}, \bibinfo{author}{Bowick, M.~J.},
  \bibinfo{author}{Marchetti, M.~C.}, \bibinfo{author}{Dogic, Z.} \&
  \bibinfo{author}{Bausch, A.~R.}
\newblock \bibinfo{title}{Topology and dynamics of active nematic vesicles}.
\newblock \href{http://dx.doi.org/10.1126/science.1254784}{\emph{Science}}
  \href{http://dx.doi.org/10.1126/science.1254784}{\textbf{\bibinfo{volume}{345}},
  \bibinfo{pages}{1135--1139}}
  \href{http://dx.doi.org/10.1126/science.1254784}{ (\bibinfo{year}{2014})}.

\bibitem{Mathijssen19}
\bibinfo{author}{Mathijssen, A. J. T.~M.}, \bibinfo{author}{Culver, J.},
  \bibinfo{author}{Bhamla, M.~S.} \& \bibinfo{author}{Prakash, M.}
\newblock \bibinfo{title}{Collective intercellular communication through
  ultra-fast hydrodynamic trigger waves}.
\newblock \href{http://dx.doi.org/10.1038/s41586-019-1387-9}{\emph{Nature}}
  \href{http://dx.doi.org/10.1038/s41586-019-1387-9}{\textbf{\bibinfo{volume}{571}},
  \bibinfo{pages}{560--564}}
  \href{http://dx.doi.org/10.1038/s41586-019-1387-9}{ (\bibinfo{year}{2019})}.

\bibitem{Qiu14}
\bibinfo{author}{Qiu, T.}, \bibinfo{author}{Lee, T.-C.}, \bibinfo{author}{Mark,
  A.~G.}, \bibinfo{author}{Morozov, K.~I.}, \bibinfo{author}{M{\"u}nster, R.},
  \bibinfo{author}{Mierka, O.}, \bibinfo{author}{Turek, S.},
  \bibinfo{author}{Leshansky, A.~M.} \& \bibinfo{author}{Fischer, P.}
\newblock \bibinfo{title}{Swimming by reciprocal motion at low {Reynolds}
  number}.
\newblock \href{http://dx.doi.org/10.1038/ncomms6119}{\emph{Nat. Commun.}}
  \href{http://dx.doi.org/10.1038/ncomms6119}{\textbf{\bibinfo{volume}{5}},
  \bibinfo{pages}{5119}} \href{http://dx.doi.org/10.1038/ncomms6119}{
  (\bibinfo{year}{2014})}.

\bibitem{Qin15}
\bibinfo{author}{Qin, B.}, \bibinfo{author}{Gopinath, A.},
  \bibinfo{author}{Yang, J.}, \bibinfo{author}{Gollub, J.~P.} \&
  \bibinfo{author}{Arratia, P.~E.}
\newblock \bibinfo{title}{Flagellar kinematics and swimming of algal cells in
  viscoelastic fluids}.
\newblock \href{http://dx.doi.org/10.1038/srep09190}{\emph{Sci. Rep.}}
  \href{http://dx.doi.org/10.1038/srep09190}{\textbf{\bibinfo{volume}{5}},
  \bibinfo{pages}{9190}} \href{http://dx.doi.org/10.1038/srep09190}{
  (\bibinfo{year}{2015})}.

\bibitem{Patteson15}
\bibinfo{author}{Patteson, A.~E.}, \bibinfo{author}{Gopinath, A.},
  \bibinfo{author}{Goulian, M.} \& \bibinfo{author}{Arratia, P.~E.}
\newblock \bibinfo{title}{Running and tumbling with e. coli in polymeric
  solutions}.
\newblock \href{http://dx.doi.org/10.1038/srep15761}{\emph{Sci. Rep.}}
  \href{http://dx.doi.org/10.1038/srep15761}{\textbf{\bibinfo{volume}{5}},
  \bibinfo{pages}{15761}} \href{http://dx.doi.org/10.1038/srep15761}{
  (\bibinfo{year}{2015})}.

\bibitem{Li16}
\bibinfo{author}{Li, G.} \& \bibinfo{author}{Ardekani, A.~M.}
\newblock \bibinfo{title}{Collective motion of microorganisms in a viscoelastic
  fluid}.
\newblock \href{http://dx.doi.org/10.1103/PhysRevLett.117.118001}{\emph{Phys.
  Rev. Lett.}}
  \href{http://dx.doi.org/10.1103/PhysRevLett.117.118001}{\textbf{\bibinfo{volume}{117}},
  \bibinfo{pages}{118001}}
  \href{http://dx.doi.org/10.1103/PhysRevLett.117.118001}{
  (\bibinfo{year}{2016})}.

\bibitem{Lauga07}
\bibinfo{author}{Lauga, E.}
\newblock \bibinfo{title}{Propulsion in a viscoelastic fluid}.
\newblock \href{http://dx.doi.org/10.1063/1.2751388}{\emph{Phys. Fluids}}
  \href{http://dx.doi.org/10.1063/1.2751388}{\textbf{\bibinfo{volume}{19}},
  \bibinfo{pages}{083104}} \href{http://dx.doi.org/10.1063/1.2751388}{
  (\bibinfo{year}{2007})}.

\bibitem{Fu09}
\bibinfo{author}{Fu, H.~C.}, \bibinfo{author}{Wolgemuth, C.~W.} \&
  \bibinfo{author}{Powers, T.~R.}
\newblock \bibinfo{title}{Swimming speeds of filaments in nonlinearly
  viscoelastic fluids}.
\newblock \href{http://dx.doi.org/10.1063/1.3086320}{\emph{Phys. Fluids}}
  \href{http://dx.doi.org/10.1063/1.3086320}{\textbf{\bibinfo{volume}{21}},
  \bibinfo{pages}{033102}} \href{http://dx.doi.org/10.1063/1.3086320}{
  (\bibinfo{year}{2009})}.

\bibitem{Spagnolie13}
\bibinfo{author}{Spagnolie, S.~E.}, \bibinfo{author}{Liu, B.} \&
  \bibinfo{author}{Powers, T.~R.}
\newblock \bibinfo{title}{Locomotion of helical bodies in viscoelastic fluids:
  Enhanced swimming at large helical amplitudes}.
\newblock \href{http://dx.doi.org/10.1103/PhysRevLett.111.068101}{\emph{Phys.
  Rev. Lett.}}
  \href{http://dx.doi.org/10.1103/PhysRevLett.111.068101}{\textbf{\bibinfo{volume}{111}},
  \bibinfo{pages}{068101}}
  \href{http://dx.doi.org/10.1103/PhysRevLett.111.068101}{
  (\bibinfo{year}{2013})}.

\bibitem{Man15}
\bibinfo{author}{Man, Y.} \& \bibinfo{author}{Lauga, E.}
\newblock \bibinfo{title}{Phase-separation models for swimming enhancement in
  complex fluids}.
\newblock \href{http://dx.doi.org/10.1103/PhysRevE.92.023004}{\emph{Phys. Rev.
  E}}
  \href{http://dx.doi.org/10.1103/PhysRevE.92.023004}{\textbf{\bibinfo{volume}{92}},
  \bibinfo{pages}{023004}} \href{http://dx.doi.org/10.1103/PhysRevE.92.023004}{
  (\bibinfo{year}{2015})}.

\bibitem{Liu11}
\bibinfo{author}{Liu, B.}, \bibinfo{author}{Powers, T.~R.} \&
  \bibinfo{author}{Breuer, K.~S.}
\newblock \bibinfo{title}{Force-free swimming of a model helical flagellum in
  viscoelastic fluids}.
\newblock \href{http://dx.doi.org/10.1073/pnas.1113082108}{\emph{Proc. Natl.
  Acad. Sci. USA}}
  \href{http://dx.doi.org/10.1073/pnas.1113082108}{\textbf{\bibinfo{volume}{108}},
  \bibinfo{pages}{19516--19520}}
  \href{http://dx.doi.org/10.1073/pnas.1113082108}{ (\bibinfo{year}{2011})}.

\bibitem{Gagnon14}
\bibinfo{author}{Gagnon, D.~A.}, \bibinfo{author}{Keim, N.~C.} \&
  \bibinfo{author}{Arratia, P.~E.}
\newblock \bibinfo{title}{Undulatory swimming in shear-thinning fluids:
  experiments with caenorhabditis elegans}.
\newblock \href{http://dx.doi.org/10.1017/jfm.2014.539}{\emph{J. Fluid Mech.}}
  \href{http://dx.doi.org/10.1017/jfm.2014.539}{\textbf{\bibinfo{volume}{758}},
  \bibinfo{pages}{R3}} \href{http://dx.doi.org/10.1017/jfm.2014.539}{
  (\bibinfo{year}{2014})}.

\bibitem{Martinez14}
\bibinfo{author}{Martinez, V.~A.}, \bibinfo{author}{Schwarz-Linek, J.},
  \bibinfo{author}{Reufer, M.}, \bibinfo{author}{Wilson, L.~G.},
  \bibinfo{author}{Morozov, A.~N.} \& \bibinfo{author}{Poon, W. C.~K.}
\newblock \bibinfo{title}{Flagellated bacterial motility in polymer solutions}.
\newblock \href{http://dx.doi.org/10.1073/pnas.1415460111}{\emph{Proc. Natl.
  Acad. Sci. USA}}
  \href{http://dx.doi.org/10.1073/pnas.1415460111}{\textbf{\bibinfo{volume}{111}},
  \bibinfo{pages}{17771--17776}}
  \href{http://dx.doi.org/10.1073/pnas.1415460111}{ (\bibinfo{year}{2014})}.

\bibitem{Zottl19}
\bibinfo{author}{Z\"ottl, A.} \& \bibinfo{author}{Yeomans, J.~M.}
\newblock \bibinfo{title}{Enhanced bacterial swimming speeds in macromolecular
  polymer solutions}.
\newblock \href{http://dx.doi.org/10.1038/s41567-019-0454-3}{\emph{Nat. Phys.}}
  \href{http://dx.doi.org/10.1038/s41567-019-0454-3}{\textbf{\bibinfo{volume}{15}},
  \bibinfo{pages}{554--558}}
  \href{http://dx.doi.org/10.1038/s41567-019-0454-3}{ (\bibinfo{year}{2019})}.

\bibitem{McNamara88}
\bibinfo{author}{McNamara, G.~R.} \& \bibinfo{author}{Zanetti, G.}
\newblock \bibinfo{title}{Use of the {B}oltzmann equation to simulate
  lattice-gas automata}.
\newblock \href{http://dx.doi.org/10.1103/PhysRevLett.61.2332}{\emph{Phys. Rev.
  Lett.}}
  \href{http://dx.doi.org/10.1103/PhysRevLett.61.2332}{\textbf{\bibinfo{volume}{61}},
  \bibinfo{pages}{2332--2335}}
  \href{http://dx.doi.org/10.1103/PhysRevLett.61.2332}{
  (\bibinfo{year}{1988})}.

\bibitem{Dunweg09}
\bibinfo{author}{D\"unweg, B.} \& \bibinfo{author}{Ladd, A. J.~C.}
\newblock \bibinfo{title}{Lattice {B}oltzmann simulations of soft matter
  systems}.
\newblock \href{http://dx.doi.org/10.1007/978-3-540-87706-6_2}{\emph{Adv.
  Polym. Sci.}}
  \href{http://dx.doi.org/10.1007/978-3-540-87706-6_2}{\textbf{\bibinfo{volume}{221}},
  \bibinfo{pages}{89--166}}
  \href{http://dx.doi.org/10.1007/978-3-540-87706-6_2}{
  (\bibinfo{year}{2009})}.

\bibitem{Espanol95}
\bibinfo{author}{Espa{\~{n}}ol, P.} \& \bibinfo{author}{Warren, P.}
\newblock \bibinfo{title}{Statistical mechanics of dissipative particle
  dynamics}.
\newblock \href{http://dx.doi.org/10.1209/0295-5075/30/4/001}{\emph{EPL}}
  \href{http://dx.doi.org/10.1209/0295-5075/30/4/001}{\textbf{\bibinfo{volume}{30}},
  \bibinfo{pages}{191--196}}
  \href{http://dx.doi.org/10.1209/0295-5075/30/4/001}{ (\bibinfo{year}{1995})}.

\bibitem{Kapral08}
\bibinfo{author}{Kapral, R.}
\newblock \emph{\bibinfo{title}{Advances in Chemical Physics}}
  (\bibinfo{publisher}{John Wiley}, \bibinfo{address}{UK},
  \bibinfo{year}{2008}).

\bibitem{Gompper09}
\bibinfo{author}{Gompper, G.}, \bibinfo{author}{Ihle, T.},
  \bibinfo{author}{Kroll, D.~M.} \& \bibinfo{author}{Winkler, R.~G.}
\newblock \bibinfo{title}{Multi-particle collision dynamics: A particle-based
  mesoscale simulation approach to the hydrodynamics of complex fluids}.
\newblock \href{http://dx.doi.org/10.1007/978-3-540-87706-6_1}{\emph{Adv.
  Polym. Sci.}}
  \href{http://dx.doi.org/10.1007/978-3-540-87706-6_1}{\textbf{\bibinfo{volume}{221}},
  \bibinfo{pages}{1--87}} \href{http://dx.doi.org/10.1007/978-3-540-87706-6_1}{
  (\bibinfo{year}{2009})}.

\bibitem{Ishikawa08}
\bibinfo{author}{Ishikawa, T.} \& \bibinfo{author}{Pedley, T.~J.}
\newblock \bibinfo{title}{Coherent structures in monolayers of swimming
  particles}.
\newblock \href{http://dx.doi.org/10.1103/PhysRevLett.100.088103}{\emph{Phys.
  Rev. Lett.}}
  \href{http://dx.doi.org/10.1103/PhysRevLett.100.088103}{\textbf{\bibinfo{volume}{100}},
  \bibinfo{pages}{088103}}
  \href{http://dx.doi.org/10.1103/PhysRevLett.100.088103}{
  (\bibinfo{year}{2008})}.

\bibitem{Mathijssen16}
\bibinfo{author}{Mathijssen, A. J. T.~M.}, \bibinfo{author}{Doostmohammadi,
  A.}, \bibinfo{author}{Yeomans, J.~M.} \& \bibinfo{author}{Shendruk, T.~N.}
\newblock \bibinfo{title}{Hydrodynamics of micro-swimmers in films}.
\newblock \href{http://dx.doi.org/10.1017/jfm.2016.479}{\emph{J. Fluid Mech.}}
  \href{http://dx.doi.org/10.1017/jfm.2016.479}{\textbf{\bibinfo{volume}{806}},
  \bibinfo{pages}{35--70}} \href{http://dx.doi.org/10.1017/jfm.2016.479}{
  (\bibinfo{year}{2016})}.

\bibitem{Singh15}
\bibinfo{author}{Singh, R.}, \bibinfo{author}{Ghose, S.} \&
  \bibinfo{author}{Adhikari, R.}
\newblock \bibinfo{title}{Many-body microhydrodynamics of colloidal particles
  with active boundary layers}.
\newblock \href{http://dx.doi.org/10.1088/1742-5468/2015/06/p06017}{\emph{J.
  Stat. Mech. Theor. Exp.}}
  \href{http://dx.doi.org/10.1088/1742-5468/2015/06/p06017}{\textbf{\bibinfo{volume}{2015}},
  \bibinfo{pages}{P06017}}
  \href{http://dx.doi.org/10.1088/1742-5468/2015/06/p06017}{
  (\bibinfo{year}{2015})}.

\bibitem{Elgeti10}
\bibinfo{author}{Elgeti, J.}, \bibinfo{author}{Kaupp, U.~B.} \&
  \bibinfo{author}{Gompper, G.}
\newblock \bibinfo{title}{Hydrodynamics of sperm cells near surfaces}.
\newblock \href{http://dx.doi.org/10.1016/j.bpj.2010.05.015}{\emph{Biophys.
  J.}}
  \href{http://dx.doi.org/10.1016/j.bpj.2010.05.015}{\textbf{\bibinfo{volume}{99}},
  \bibinfo{pages}{1018--1026}}
  \href{http://dx.doi.org/10.1016/j.bpj.2010.05.015}{ (\bibinfo{year}{2010})}.

\bibitem{Watari10}
\bibinfo{author}{Watari, N.} \& \bibinfo{author}{Larson, R.~G.}
\newblock \bibinfo{title}{The hydrodynamics of a run-and-tumble bacterium
  propelled by polymorphic helical flagella}.
\newblock \href{http://dx.doi.org/10.1016/j.bpj.2009.09.044}{\emph{Biophys.
  J.}}
  \href{http://dx.doi.org/10.1016/j.bpj.2009.09.044}{\textbf{\bibinfo{volume}{98}},
  \bibinfo{pages}{12--17}} \href{http://dx.doi.org/10.1016/j.bpj.2009.09.044}{
  (\bibinfo{year}{2010})}.

\bibitem{Shum10}
\bibinfo{author}{Shum, H.}, \bibinfo{author}{Gaffney, E.~A.} \&
  \bibinfo{author}{Smith, D.~J.}
\newblock \bibinfo{title}{Modelling bacterial behaviour close to a no-slip
  plane boundary: the influence of bacterial geometry}.
\newblock \href{http://dx.doi.org/10.1098/rspa.2009.0520}{\emph{Proc. R. Soc.
  A}}
  \href{http://dx.doi.org/10.1098/rspa.2009.0520}{\textbf{\bibinfo{volume}{466}},
  \bibinfo{pages}{1725--1748}} \href{http://dx.doi.org/10.1098/rspa.2009.0520}{
  (\bibinfo{year}{2010})}.

\bibitem{Pimponi16}
\bibinfo{author}{Pimponi, D.}, \bibinfo{author}{Chinappi, M.},
  \bibinfo{author}{Gualtieri, P.} \& \bibinfo{author}{Casciola, C.~M.}
\newblock \bibinfo{title}{Hydrodynamics of flagellated microswimmers near
  free-slip interfaces}.
\newblock \href{http://dx.doi.org/10.1017/jfm.2015.738}{\emph{J. Fluid Mech.}}
  \href{http://dx.doi.org/10.1017/jfm.2015.738}{\textbf{\bibinfo{volume}{789}},
  \bibinfo{pages}{514--533}} \href{http://dx.doi.org/10.1017/jfm.2015.738}{
  (\bibinfo{year}{2016})}.

\bibitem{Lighthill52}
\bibinfo{author}{Lighthill, M.~J.}
\newblock \bibinfo{title}{On the squirming motion of nearly spherical
  deformable bodies through liquids at very small {Reynolds} numbers}.
\newblock \href{http://dx.doi.org/10.1002/cpa.3160050201}{\emph{Comm. Pure
  Appl. Math.}}
  \href{http://dx.doi.org/10.1002/cpa.3160050201}{\textbf{\bibinfo{volume}{5}},
  \bibinfo{pages}{109--118}} \href{http://dx.doi.org/10.1002/cpa.3160050201}{
  (\bibinfo{year}{1952})}.

\bibitem{Blake71}
\bibinfo{author}{Blake, J.~R.}
\newblock \bibinfo{title}{A spherical envelope approach to ciliary propulsion}.
\newblock \href{http://dx.doi.org/10.1017/S002211207100048X}{\emph{J. Fluid
  Mech.}}
  \href{http://dx.doi.org/10.1017/S002211207100048X}{\textbf{\bibinfo{volume}{46}},
  \bibinfo{pages}{199--208}}
  \href{http://dx.doi.org/10.1017/S002211207100048X}{ (\bibinfo{year}{1971})}.

\bibitem{Ishikawa06}
\bibinfo{author}{Ishikawa, T.}, \bibinfo{author}{Simmonds, M.~P.} \&
  \bibinfo{author}{Pedley, T.~J.}
\newblock \bibinfo{title}{Hydrodynamic interaction of two swimming model
  micro-organisms}.
\newblock \href{http://dx.doi.org/10.1017/S0022112006002631}{\emph{J. Fluid
  Mech.}}
  \href{http://dx.doi.org/10.1017/S0022112006002631}{\textbf{\bibinfo{volume}{568}},
  \bibinfo{pages}{119--160}}
  \href{http://dx.doi.org/10.1017/S0022112006002631}{ (\bibinfo{year}{2006})}.

\bibitem{Pedley16}
\bibinfo{author}{Pedley, T.~J.}
\newblock \bibinfo{title}{Spherical squirmers: models for swimming
  micro-organisms}.
\newblock \href{http://dx.doi.org/10.1093/imamat/hxw030}{\emph{IMA J. Applied
  Mathematics}}
  \href{http://dx.doi.org/10.1093/imamat/hxw030}{\textbf{\bibinfo{volume}{81}},
  \bibinfo{pages}{488--521}} \href{http://dx.doi.org/10.1093/imamat/hxw030}{
  (\bibinfo{year}{2016})}.

\bibitem{Llopis10}
\bibinfo{author}{Llopis, I.} \& \bibinfo{author}{Pagonabarraga, I.}
\newblock \bibinfo{title}{Hydrodynamic interactions in squirmer motion:
  Swimming with a neighbour and close to a wall}.
\newblock \href{http://dx.doi.org/10.1016/j.jnnfm.2010.01.023}{\emph{J.
  Non-Newtonian Fluid Mech.}}
  \href{http://dx.doi.org/10.1016/j.jnnfm.2010.01.023}{\textbf{\bibinfo{volume}{165}},
  \bibinfo{pages}{946--952}}
  \href{http://dx.doi.org/10.1016/j.jnnfm.2010.01.023}{
  (\bibinfo{year}{2010})}.

\bibitem{Gotze10}
\bibinfo{author}{G\"otze, I.~O.} \& \bibinfo{author}{Gompper, G.}
\newblock \bibinfo{title}{Mesoscale simulations of hydrodynamic squirmer
  interactions}.
\newblock \href{http://dx.doi.org/10.1103/PhysRevE.82.041921}{\emph{Phys. Rev.
  E}}
  \href{http://dx.doi.org/10.1103/PhysRevE.82.041921}{\textbf{\bibinfo{volume}{82}},
  \bibinfo{pages}{041921}} \href{http://dx.doi.org/10.1103/PhysRevE.82.041921}{
  (\bibinfo{year}{2010})}.

\bibitem{Evans11}
\bibinfo{author}{Evans, A.~A.}, \bibinfo{author}{Ishikawa, T.},
  \bibinfo{author}{Yamaguchi, T.} \& \bibinfo{author}{Lauga, E.}
\newblock \bibinfo{title}{Orientational order in concentrated suspensions of
  spherical microswimmers}.
\newblock \href{http://dx.doi.org/10.1063/1.3660268}{\emph{Phys. Fluids}}
  \href{http://dx.doi.org/10.1063/1.3660268}{\textbf{\bibinfo{volume}{23}},
  \bibinfo{pages}{111702}} \href{http://dx.doi.org/10.1063/1.3660268}{
  (\bibinfo{year}{2011})}.

\bibitem{Alarcon13}
\bibinfo{author}{Alarcon, F.} \& \bibinfo{author}{Pagonabarraga, I.}
\newblock \bibinfo{title}{Spontaneous aggregation and global polar ordering in
  squirmer suspensions}.
\newblock \href{http://dx.doi.org/10.1016/j.molliq.2012.12.009}{\emph{J. Mol.
  Liq.}}
  \href{http://dx.doi.org/10.1016/j.molliq.2012.12.009}{\textbf{\bibinfo{volume}{185}},
  \bibinfo{pages}{56--61}}
  \href{http://dx.doi.org/10.1016/j.molliq.2012.12.009}{
  (\bibinfo{year}{2013})}.

\bibitem{Molina13}
\bibinfo{author}{Molina, J.~J.}, \bibinfo{author}{Nakayama, Y.} \&
  \bibinfo{author}{Yamamoto, R.}
\newblock \bibinfo{title}{Hydrodynamic interactions of self-propelled
  swimmers}.
\newblock \href{http://dx.doi.org/10.1039/C3SM00140G}{\emph{Soft Matter}}
  \href{http://dx.doi.org/10.1039/C3SM00140G}{\textbf{\bibinfo{volume}{9}},
  \bibinfo{pages}{4923--4936}} \href{http://dx.doi.org/10.1039/C3SM00140G}{
  (\bibinfo{year}{2013})}.

\bibitem{Yoshinaga17}
\bibinfo{author}{Yoshinaga, N.} \& \bibinfo{author}{Liverpool, T.~B.}
\newblock \bibinfo{title}{Hydrodynamic interactions in dense active
  suspensions: From polar order to dynamical clusters}.
\newblock \href{http://dx.doi.org/10.1103/PhysRevE.96.020603}{\emph{Phys. Rev.
  E}}
  \href{http://dx.doi.org/10.1103/PhysRevE.96.020603}{\textbf{\bibinfo{volume}{96}},
  \bibinfo{pages}{020603}} \href{http://dx.doi.org/10.1103/PhysRevE.96.020603}{
  (\bibinfo{year}{2017})}.

\bibitem{Ishimoto13}
\bibinfo{author}{Ishimoto, K.} \& \bibinfo{author}{Gaffney, E.~A.}
\newblock \bibinfo{title}{Squirmer dynamics near a boundary}.
\newblock \href{http://dx.doi.org/10.1103/PhysRevE.88.062702}{\emph{Phys. Rev.
  E}}
  \href{http://dx.doi.org/10.1103/PhysRevE.88.062702}{\textbf{\bibinfo{volume}{88}},
  \bibinfo{pages}{062702}} \href{http://dx.doi.org/10.1103/PhysRevE.88.062702}{
  (\bibinfo{year}{2013})}.

\bibitem{Lintuvuori16}
\bibinfo{author}{Lintuvuori, J.~S.}, \bibinfo{author}{Brown, A.~T.},
  \bibinfo{author}{Stratford, K.} \& \bibinfo{author}{Marenduzzo, D.}
\newblock \bibinfo{title}{Hydrodynamic oscillations and variable swimming speed
  in squirmers close to repulsive walls}.
\newblock \href{http://dx.doi.org/10.1039/C6SM01353H}{\emph{Soft Matter}}
  \href{http://dx.doi.org/10.1039/C6SM01353H}{\textbf{\bibinfo{volume}{12}},
  \bibinfo{pages}{7959--7968}} \href{http://dx.doi.org/10.1039/C6SM01353H}{
  (\bibinfo{year}{2016})}.

\bibitem{Theers16}
\bibinfo{author}{Theers, M.}, \bibinfo{author}{Westphal, E.},
  \bibinfo{author}{Gompper, G.} \& \bibinfo{author}{Winkler, R.~G.}
\newblock \bibinfo{title}{Modeling a spheroidal microswimmer and cooperative
  swimming in a narrow slit}.
\newblock \href{http://dx.doi.org/10.1039/C6SM01424K}{\emph{Soft Matter}}
  \href{http://dx.doi.org/10.1039/C6SM01424K}{\textbf{\bibinfo{volume}{12}},
  \bibinfo{pages}{7372--7385}} \href{http://dx.doi.org/10.1039/C6SM01424K}{
  (\bibinfo{year}{2016})}.

\bibitem{Theers18}
\bibinfo{author}{Theers, M.}, \bibinfo{author}{Westphal, E.},
  \bibinfo{author}{Qi, K.}, \bibinfo{author}{Winkler, R.~G.} \&
  \bibinfo{author}{Gompper, G.}
\newblock \bibinfo{title}{Clustering of microswimmers: interplay of shape and
  hydrodynamics}.
\newblock \href{http://dx.doi.org/10.1039/C8SM01390J}{\emph{Soft Matter}}
  \href{http://dx.doi.org/10.1039/C8SM01390J}{\textbf{\bibinfo{volume}{14}},
  \bibinfo{pages}{8590--8603}} \href{http://dx.doi.org/10.1039/C8SM01390J}{
  (\bibinfo{year}{2018})}.

\bibitem{Keller77}
\bibinfo{author}{Keller, S.~R.} \& \bibinfo{author}{Wu, T.~Y.}
\newblock \bibinfo{title}{A porous prolate-spheroidal model for ciliated
  micro-organisms}.
\newblock \href{http://dx.doi.org/10.1017/S0022112077001669}{\emph{J. Fluid
  Mech.}}
  \href{http://dx.doi.org/10.1017/S0022112077001669}{\textbf{\bibinfo{volume}{80}},
  \bibinfo{pages}{259--278}}
  \href{http://dx.doi.org/10.1017/S0022112077001669}{ (\bibinfo{year}{1977})}.

\bibitem{Theers16b}
\bibinfo{author}{Theers, M.}, \bibinfo{author}{Westphal, E.},
  \bibinfo{author}{Gompper, G.} \& \bibinfo{author}{Winkler, R.~G.}
\newblock \bibinfo{title}{From local to hydrodynamic friction in {B}rownian
  motion: A multiparticle collision dynamics simulation study}.
\newblock \href{http://dx.doi.org/10.1103/PhysRevE.93.032604}{\emph{Phys. Rev.
  E}}
  \href{http://dx.doi.org/10.1103/PhysRevE.93.032604}{\textbf{\bibinfo{volume}{93}},
  \bibinfo{pages}{032604}} \href{http://dx.doi.org/10.1103/PhysRevE.93.032604}{
  (\bibinfo{year}{2016})}.

\bibitem{Nash10}
\bibinfo{author}{Nash, R.~W.}, \bibinfo{author}{Adhikari, R.},
  \bibinfo{author}{Tailleur, J.} \& \bibinfo{author}{Cates, M.~E.}
\newblock \bibinfo{title}{Run-and-tumble particles with hydrodynamics:
  Sedimentation, trapping, and upstream swimming}.
\newblock \href{http://dx.doi.org/10.1103/PhysRevLett.104.258101}{\emph{Phys.
  Rev. Lett.}}
  \href{http://dx.doi.org/10.1103/PhysRevLett.104.258101}{\textbf{\bibinfo{volume}{104}},
  \bibinfo{pages}{258101}}
  \href{http://dx.doi.org/10.1103/PhysRevLett.104.258101}{
  (\bibinfo{year}{2010})}.

\bibitem{Hernandez-Ortiz05}
\bibinfo{author}{Hernandez-Ortiz, J.~P.}, \bibinfo{author}{Stoltz, C.~G.} \&
  \bibinfo{author}{Graham, M.~D.}
\newblock \bibinfo{title}{Transport and collective dynamics in suspensions of
  confined swimming particles}.
\newblock \href{http://dx.doi.org/10.1103/PhysRevLett.95.204501}{\emph{Phys.
  Rev. Lett.}}
  \href{http://dx.doi.org/10.1103/PhysRevLett.95.204501}{\textbf{\bibinfo{volume}{95}},
  \bibinfo{pages}{204501}}
  \href{http://dx.doi.org/10.1103/PhysRevLett.95.204501}{
  (\bibinfo{year}{2005})}.

\bibitem{deGraaf16}
\bibinfo{author}{de~Graaf, J.}, \bibinfo{author}{Menke, H.},
  \bibinfo{author}{Mathijssen, A. J. T.~M.}, \bibinfo{author}{Fabritius, M.},
  \bibinfo{author}{Holm, C.} \& \bibinfo{author}{Shendruk, T.~N.}
\newblock \bibinfo{title}{Lattice {B}oltzmann hydrodynamics of anisotropic
  active matter}.
\newblock \href{http://dx.doi.org/10.1063/1.4944962}{\emph{J. Chem. Phys.}}
  \href{http://dx.doi.org/10.1063/1.4944962}{\textbf{\bibinfo{volume}{144}},
  \bibinfo{pages}{134106}} \href{http://dx.doi.org/10.1063/1.4944962}{
  (\bibinfo{year}{2016})}.

\bibitem{Menzel16}
\bibinfo{author}{Menzel, A.~M.}, \bibinfo{author}{Saha, A.},
  \bibinfo{author}{Hoell, C.} \& \bibinfo{author}{L\"owen, H.}
\newblock \bibinfo{title}{Dynamical density functional theory for
  microswimmers}.
\newblock \href{http://dx.doi.org/10.1063/1.4939630}{\emph{J. Chem. Phys.}}
  \href{http://dx.doi.org/10.1063/1.4939630}{\textbf{\bibinfo{volume}{144}},
  \bibinfo{pages}{024115}} \href{http://dx.doi.org/10.1063/1.4939630}{
  (\bibinfo{year}{2016})}.

\bibitem{Lighthill76}
\bibinfo{author}{Lighthill, J.}
\newblock \bibinfo{title}{Flagellar hydrodynamics}.
\newblock \href{http://dx.doi.org/10.1137/1018040}{\emph{SIAM Rev.}}
  \href{http://dx.doi.org/10.1137/1018040}{\textbf{\bibinfo{volume}{18}},
  \bibinfo{pages}{161--230}} \href{http://dx.doi.org/10.1137/1018040}{
  (\bibinfo{year}{1976})}.

\bibitem{Saggiorato17}
\bibinfo{author}{Saggiorato, G.}, \bibinfo{author}{Alvarez, L.},
  \bibinfo{author}{Jikeli, J.~F.}, \bibinfo{author}{Kaupp, U.~B.},
  \bibinfo{author}{Gompper, G.} \& \bibinfo{author}{Elgeti, J.}
\newblock \bibinfo{title}{Human sperm steer with second harmonics of the
  flagellar beat}.
\newblock \href{http://dx.doi.org/10.1038/s41467-017-01462-y}{\emph{Nat.
  Commun.}}
  \href{http://dx.doi.org/10.1038/s41467-017-01462-y}{\textbf{\bibinfo{volume}{8}},
  \bibinfo{pages}{1415}} \href{http://dx.doi.org/10.1038/s41467-017-01462-y}{
  (\bibinfo{year}{2017})}.

\bibitem{Shum15}
\bibinfo{author}{Shum, H.} \& \bibinfo{author}{Gaffney, E.~A.}
\newblock \bibinfo{title}{Hydrodynamic analysis of flagellated bacteria
  swimming near one and between two no-slip plane boundaries}.
\newblock \href{http://dx.doi.org/10.1103/PhysRevE.91.033012}{\emph{Phys. Rev.
  E}}
  \href{http://dx.doi.org/10.1103/PhysRevE.91.033012}{\textbf{\bibinfo{volume}{91}},
  \bibinfo{pages}{033012}} \href{http://dx.doi.org/10.1103/PhysRevE.91.033012}{
  (\bibinfo{year}{2015})}.

\bibitem{Lauga16}
\bibinfo{author}{Lauga, E.}
\newblock \bibinfo{title}{Bacterial hydrodynamics}.
\newblock
  \href{http://dx.doi.org/10.1146/annurev-fluid-122414-034606}{\emph{Annu. Rev.
  Fluid Mech.}}
  \href{http://dx.doi.org/10.1146/annurev-fluid-122414-034606}{\textbf{\bibinfo{volume}{48}},
  \bibinfo{pages}{105--130}}
  \href{http://dx.doi.org/10.1146/annurev-fluid-122414-034606}{
  (\bibinfo{year}{2016})}.

\bibitem{Rode19}
\bibinfo{author}{Rode, S.}, \bibinfo{author}{Elgeti, J.} \&
  \bibinfo{author}{Gompper, G.}
\newblock \bibinfo{title}{Sperm motility in modulated microchannels}.
\newblock \href{http://dx.doi.org/10.1088/1367-2630/aaf544}{\emph{New J.
  Phys.}}
  \href{http://dx.doi.org/10.1088/1367-2630/aaf544}{\textbf{\bibinfo{volume}{21}},
  \bibinfo{pages}{013016}} \href{http://dx.doi.org/10.1088/1367-2630/aaf544}{
  (\bibinfo{year}{2019})}.

\bibitem{Reigh12}
\bibinfo{author}{Reigh, S.~Y.}, \bibinfo{author}{Winkler, R.~G.} \&
  \bibinfo{author}{Gompper, G.}
\newblock \bibinfo{title}{Synchronization and bundling of anchored bacterial
  flagella}.
\newblock \href{http://dx.doi.org/10.1039/C2SM07378A}{\emph{Soft Matter}}
  \href{http://dx.doi.org/10.1039/C2SM07378A}{\textbf{\bibinfo{volume}{8}},
  \bibinfo{pages}{4363--4372}} \href{http://dx.doi.org/10.1039/C2SM07378A}{
  (\bibinfo{year}{2012})}.

\bibitem{Reichert05}
\bibinfo{author}{Reichert, M.} \& \bibinfo{author}{Stark, H.}
\newblock \bibinfo{title}{Synchronization of rotating helices by hydrodynamic
  interactions}.
\newblock \href{http://dx.doi.org/10.1140/epje/i2004-10152-7}{\emph{Eur. Phys.
  J. E}}
  \href{http://dx.doi.org/10.1140/epje/i2004-10152-7}{\textbf{\bibinfo{volume}{17}},
  \bibinfo{pages}{493--500}}
  \href{http://dx.doi.org/10.1140/epje/i2004-10152-7}{ (\bibinfo{year}{2005})}.

\bibitem{Vogel12}
\bibinfo{author}{Vogel, R.} \& \bibinfo{author}{Stark, H.}
\newblock \bibinfo{title}{Motor-driven bacterial flagella and buckling
  instabilities}.
\newblock \href{http://dx.doi.org/10.1140/epje/i2012-12015-0}{\emph{Eur. Phys.
  J. E}}
  \href{http://dx.doi.org/10.1140/epje/i2012-12015-0}{\textbf{\bibinfo{volume}{35}},
  \bibinfo{pages}{15}} \href{http://dx.doi.org/10.1140/epje/i2012-12015-0}{
  (\bibinfo{year}{2012})}.

\bibitem{Janssen11}
\bibinfo{author}{Janssen, P. J.~A.} \& \bibinfo{author}{Graham, M.~D.}
\newblock \bibinfo{title}{Coexistence of tight and loose bundled states in a
  model of bacterial flagellar dynamics}.
\newblock \href{http://dx.doi.org/10.1103/PhysRevE.84.011910}{\emph{Phys. Rev.
  E}}
  \href{http://dx.doi.org/10.1103/PhysRevE.84.011910}{\textbf{\bibinfo{volume}{84}},
  \bibinfo{pages}{011910}} \href{http://dx.doi.org/10.1103/PhysRevE.84.011910}{
  (\bibinfo{year}{2011})}.

\bibitem{Hu15b}
\bibinfo{author}{Hu, J.}, \bibinfo{author}{Wysocki, A.},
  \bibinfo{author}{Winkler, R.~G.} \& \bibinfo{author}{Gompper, G.}
\newblock \bibinfo{title}{Physical sensing of surface properties by
  microswimmers - directing bacterial motion via wall slip}.
\newblock \href{http://dx.doi.org/10.1038/srep09586}{\emph{Sci. Rep.}}
  \href{http://dx.doi.org/10.1038/srep09586}{\textbf{\bibinfo{volume}{5}},
  \bibinfo{pages}{9586}} \href{http://dx.doi.org/10.1038/srep09586}{
  (\bibinfo{year}{2015})}.

\bibitem{Lemelle13}
\bibinfo{author}{Lemelle, L.}, \bibinfo{author}{Palierne, J.-F.},
  \bibinfo{author}{Chatre, E.}, \bibinfo{author}{Vaillant, C.} \&
  \bibinfo{author}{Place, C.}
\newblock \bibinfo{title}{Curvature reversal of the circular motion of swimming
  bacteria probes for slip at solid/liquid interfaces}.
\newblock \href{http://dx.doi.org/10.1039/C3SM51426A}{\emph{Soft Matter}}
  \href{http://dx.doi.org/10.1039/C3SM51426A}{\textbf{\bibinfo{volume}{9}},
  \bibinfo{pages}{9759--9762}} \href{http://dx.doi.org/10.1039/C3SM51426A}{
  (\bibinfo{year}{2013})}.

\bibitem{Lauga06}
\bibinfo{author}{Lauga, E.}, \bibinfo{author}{DiLuzio, W.~R.},
  \bibinfo{author}{Whitesides, G.~M.} \& \bibinfo{author}{Stone, H.~A.}
\newblock \bibinfo{title}{Swimming in circles: Motion of bacteria near solid
  boundaries}.
\newblock \href{http://dx.doi.org/10.1529/biophysj.105.069401}{\emph{Biophys.
  J.}}
  \href{http://dx.doi.org/10.1529/biophysj.105.069401}{\textbf{\bibinfo{volume}{90}},
  \bibinfo{pages}{400--412}}
  \href{http://dx.doi.org/10.1529/biophysj.105.069401}{
  (\bibinfo{year}{2006})}.

\bibitem{DiLeonardo11}
\bibinfo{author}{Di~Leonardo, R.}, \bibinfo{author}{Dell~Arciprete, D.},
  \bibinfo{author}{Angelani, L.} \& \bibinfo{author}{Iebba, V.}
\newblock \bibinfo{title}{Swimming with an image}.
\newblock \href{http://dx.doi.org/10.1103/PhysRevLett.106.038101}{\emph{Phys.
  Rev. Lett.}}
  \href{http://dx.doi.org/10.1103/PhysRevLett.106.038101}{\textbf{\bibinfo{volume}{106}},
  \bibinfo{pages}{038101}}
  \href{http://dx.doi.org/10.1103/PhysRevLett.106.038101}{
  (\bibinfo{year}{2011})}.

\bibitem{Dombrowski04}
\bibinfo{author}{Dombrowski, C.}, \bibinfo{author}{Cisneros, L.},
  \bibinfo{author}{Chatkaew, S.}, \bibinfo{author}{Goldstein, R.~E.} \&
  \bibinfo{author}{Kessler, J.~O.}
\newblock \bibinfo{title}{Self-concentration and large-scale coherence in
  bacterial dynamics}.
\newblock \href{http://dx.doi.org/10.1103/PhysRevLett.93.098103}{\emph{Phys.
  Rev. Lett.}}
  \href{http://dx.doi.org/10.1103/PhysRevLett.93.098103}{\textbf{\bibinfo{volume}{93}},
  \bibinfo{pages}{098103}}
  \href{http://dx.doi.org/10.1103/PhysRevLett.93.098103}{
  (\bibinfo{year}{2004})}.

\bibitem{Bialke12}
\bibinfo{author}{Bialk\'e, J.}, \bibinfo{author}{Speck, T.} \&
  \bibinfo{author}{L\"owen, H.}
\newblock \bibinfo{title}{Crystallization in a dense suspension of
  self-propelled particles}.
\newblock \href{http://dx.doi.org/10.1103/PhysRevLett.108.168301}{\emph{Phys.
  Rev. Lett.}}
  \href{http://dx.doi.org/10.1103/PhysRevLett.108.168301}{\textbf{\bibinfo{volume}{108}},
  \bibinfo{pages}{168301}}
  \href{http://dx.doi.org/10.1103/PhysRevLett.108.168301}{
  (\bibinfo{year}{2012})}.

\bibitem{Matas-Navarro14}
\bibinfo{author}{Matas-Navarro, R.}, \bibinfo{author}{Golestanian, R.},
  \bibinfo{author}{Liverpool, T.~B.} \& \bibinfo{author}{Fielding, S.~M.}
\newblock \bibinfo{title}{Hydrodynamic suppression of phase separation in
  active suspensions}.
\newblock \href{http://dx.doi.org/10.1103/PhysRevE.90.032304}{\emph{Phys. Rev.
  E}}
  \href{http://dx.doi.org/10.1103/PhysRevE.90.032304}{\textbf{\bibinfo{volume}{90}},
  \bibinfo{pages}{032304}} \href{http://dx.doi.org/10.1103/PhysRevE.90.032304}{
  (\bibinfo{year}{2014})}.

\bibitem{Gaspard19}
\bibinfo{author}{Gaspard, P.} \& \bibinfo{author}{Kapral, R.}
\newblock \bibinfo{title}{Thermodynamics and statistical mechanics of
  chemically powered synthetic nanomotors}.
\newblock \href{http://dx.doi.org/10.1080/23746149.2019.1602480}{\emph{Advances
  in Physics: X}}
  \href{http://dx.doi.org/10.1080/23746149.2019.1602480}{\textbf{\bibinfo{volume}{4}},
  \bibinfo{pages}{1602480}}
  \href{http://dx.doi.org/10.1080/23746149.2019.1602480}{
  (\bibinfo{year}{2019})}.

\bibitem{Bayati19}
\bibinfo{author}{Bayati, P.}, \bibinfo{author}{Popescu, M.~N.},
  \bibinfo{author}{Uspal, W.~E.}, \bibinfo{author}{Dietrich, S.} \&
  \bibinfo{author}{Najafi, A.}
\newblock \bibinfo{title}{Dynamics near planar walls for various model
  self-phoretic particles}.
\newblock \href{http://dx.doi.org/10.1039/C9SM00488B}{\emph{Soft Matter}}
  \href{http://dx.doi.org/10.1039/C9SM00488B}{\textbf{\bibinfo{volume}{15}},
  \bibinfo{pages}{5644--5672}} \href{http://dx.doi.org/10.1039/C9SM00488B}{
  (\bibinfo{year}{2019})}.

\bibitem{Ruckner07}
\bibinfo{author}{R\"uckner, G.} \& \bibinfo{author}{Kapral, R.}
\newblock \bibinfo{title}{Chemically powered nanodimers}.
\newblock \href{http://dx.doi.org/10.1103/PhysRevLett.98.150603}{\emph{Phys.
  Rev. Lett.}}
  \href{http://dx.doi.org/10.1103/PhysRevLett.98.150603}{\textbf{\bibinfo{volume}{98}},
  \bibinfo{pages}{150603}}
  \href{http://dx.doi.org/10.1103/PhysRevLett.98.150603}{
  (\bibinfo{year}{2007})}.

\bibitem{Yang11}
\bibinfo{author}{Yang, M.} \& \bibinfo{author}{Ripoll, M.}
\newblock \bibinfo{title}{Simulations of thermophoretic nanoswimmers}.
\newblock \href{http://dx.doi.org/10.1103/PhysRevE.84.061401}{\emph{Phys. Rev.
  E}}
  \href{http://dx.doi.org/10.1103/PhysRevE.84.061401}{\textbf{\bibinfo{volume}{84}},
  \bibinfo{pages}{061401}} \href{http://dx.doi.org/10.1103/PhysRevE.84.061401}{
  (\bibinfo{year}{2011})}.

\bibitem{Saha14}
\bibinfo{author}{Saha, S.}, \bibinfo{author}{Golestanian, R.} \&
  \bibinfo{author}{Ramaswamy, S.}
\newblock \bibinfo{title}{Clusters, asters, and collective oscillations in
  chemotactic colloids}.
\newblock \href{http://dx.doi.org/10.1103/PhysRevE.89.062316}{\emph{Phys. Rev.
  E}}
  \href{http://dx.doi.org/10.1103/PhysRevE.89.062316}{\textbf{\bibinfo{volume}{89}},
  \bibinfo{pages}{062316}} \href{http://dx.doi.org/10.1103/PhysRevE.89.062316}{
  (\bibinfo{year}{2014})}.

\bibitem{Michelin14}
\bibinfo{author}{Michelin, S.} \& \bibinfo{author}{Lauga, E.}
\newblock \bibinfo{title}{Phoretic self-propulsion at finite {P{\'e}clet}
  numbers}.
\newblock \href{http://dx.doi.org/10.1017/jfm.2014.158}{\emph{J. Fluid Mech.}}
  \href{http://dx.doi.org/10.1017/jfm.2014.158}{\textbf{\bibinfo{volume}{747}},
  \bibinfo{pages}{572}} \href{http://dx.doi.org/10.1017/jfm.2014.158}{
  (\bibinfo{year}{2014})}.

\bibitem{Liebchen15}
\bibinfo{author}{Liebchen, B.}, \bibinfo{author}{Marenduzzo, D.},
  \bibinfo{author}{Pagonabarraga, I.} \& \bibinfo{author}{Cates, M.~E.}
\newblock \bibinfo{title}{Clustering and pattern formation in chemorepulsive
  active colloids}.
\newblock \href{http://dx.doi.org/10.1103/PhysRevLett.115.258301}{\emph{Phys.
  Rev. Lett.}}
  \href{http://dx.doi.org/10.1103/PhysRevLett.115.258301}{\textbf{\bibinfo{volume}{115}},
  \bibinfo{pages}{258301}}
  \href{http://dx.doi.org/10.1103/PhysRevLett.115.258301}{
  (\bibinfo{year}{2015})}.

\bibitem{Stark18}
\bibinfo{author}{Stark, H.}
\newblock \bibinfo{title}{Artificial chemotaxis of self-phoretic active
  colloids: Collective behavior}.
\newblock \href{http://dx.doi.org/10.1021/acs.accounts.8b00259}{\emph{Acc.
  Chem. Res.}}
  \href{http://dx.doi.org/10.1021/acs.accounts.8b00259}{\textbf{\bibinfo{volume}{51}},
  \bibinfo{pages}{2681--2688}}
  \href{http://dx.doi.org/10.1021/acs.accounts.8b00259}{
  (\bibinfo{year}{2018})}.

\bibitem{Moran17}
\bibinfo{author}{Moran, J.~L.} \& \bibinfo{author}{Posner, J.~D.}
\newblock \bibinfo{title}{Phoretic self-propulsion}.
\newblock
  \href{http://dx.doi.org/10.1146/annurev-fluid-122414-034456}{\emph{Annu. Rev.
  Fluid Mech.}}
  \href{http://dx.doi.org/10.1146/annurev-fluid-122414-034456}{\textbf{\bibinfo{volume}{49}},
  \bibinfo{pages}{511--540}}
  \href{http://dx.doi.org/10.1146/annurev-fluid-122414-034456}{
  (\bibinfo{year}{2017})}.

\bibitem{Howse07}
\bibinfo{author}{Howse, J.~R.}, \bibinfo{author}{Jones, R. A.~L.},
  \bibinfo{author}{Ryan, A.~J.}, \bibinfo{author}{Gough, T.},
  \bibinfo{author}{Vafabakhsh, R.} \& \bibinfo{author}{Golestanian, R.}
\newblock \bibinfo{title}{Self-motile colloidal particles: From directed
  propulsion to random walk}.
\newblock \href{http://dx.doi.org/10.1103/PhysRevLett.99.048102}{\emph{Phys.
  Rev. Lett.}}
  \href{http://dx.doi.org/10.1103/PhysRevLett.99.048102}{\textbf{\bibinfo{volume}{99}},
  \bibinfo{pages}{048102}}
  \href{http://dx.doi.org/10.1103/PhysRevLett.99.048102}{
  (\bibinfo{year}{2007})}.

\bibitem{Uspal15}
\bibinfo{author}{Uspal, W.~E.}, \bibinfo{author}{Popescu, M.~N.},
  \bibinfo{author}{Dietrich, S.} \& \bibinfo{author}{Tasinkevych, M.}
\newblock \bibinfo{title}{Self-propulsion of a catalytically active particle
  near a planar wall: from reflection to sliding and hovering}.
\newblock \href{http://dx.doi.org/10.1039/C4SM02317J}{\emph{Soft Matter}}
  \href{http://dx.doi.org/10.1039/C4SM02317J}{\textbf{\bibinfo{volume}{11}},
  \bibinfo{pages}{434}} \href{http://dx.doi.org/10.1039/C4SM02317J}{
  (\bibinfo{year}{2015})}.

\bibitem{Ishimoto15}
\bibinfo{author}{Ishimoto, K.} \& \bibinfo{author}{Gaffney, E.~A.}
\newblock \bibinfo{title}{Fluid flow and sperm guidance: a simulation study of
  hydrodynamic sperm rheotaxis}.
\newblock \href{http://dx.doi.org/10.1098/rsif.2015.0172}{\emph{J. Roy. Soc.
  Interface}}
  \href{http://dx.doi.org/10.1098/rsif.2015.0172}{\textbf{\bibinfo{volume}{12}},
  \bibinfo{pages}{20150172}} \href{http://dx.doi.org/10.1098/rsif.2015.0172}{
  (\bibinfo{year}{2015})}.

\bibitem{Koh16}
\bibinfo{author}{Koh, J. B.~Y.}, \bibinfo{author}{Shen, X.} \&
  \bibinfo{author}{Marcos}.
\newblock \bibinfo{title}{Theoretical modeling in microscale locomotion}.
\newblock
  \href{http://dx.doi.org/10.1007/s10404-016-1761-y}{\emph{Microfluidics and
  Nanofluidics}}
  \href{http://dx.doi.org/10.1007/s10404-016-1761-y}{\textbf{\bibinfo{volume}{20}},
  \bibinfo{pages}{98}} \href{http://dx.doi.org/10.1007/s10404-016-1761-y}{
  (\bibinfo{year}{2016})}.

\bibitem{Uspal15b}
\bibinfo{author}{Uspal, W.~E.}, \bibinfo{author}{Popescu, M.~N.},
  \bibinfo{author}{Dietrich, S.} \& \bibinfo{author}{Tasinkevych, M.}
\newblock \bibinfo{title}{Rheotaxis of spherical active particles near a planar
  wall}.
\newblock \href{http://dx.doi.org/10.1039/C5SM01088H}{\emph{Soft Matter}}
  \href{http://dx.doi.org/10.1039/C5SM01088H}{\textbf{\bibinfo{volume}{11}},
  \bibinfo{pages}{6613--6632}} \href{http://dx.doi.org/10.1039/C5SM01088H}{
  (\bibinfo{year}{2015})}.

\bibitem{Mathijssen19b}
\bibinfo{author}{Mathijssen, A.}, \bibinfo{author}{Figueroa-Morales, N.},
  \bibinfo{author}{Junot, G.}, \bibinfo{author}{Clement, E.},
  \bibinfo{author}{Lindner, A.} \& \bibinfo{author}{Z{\"o}ttl, A.}
\newblock \bibinfo{title}{Oscillatory surface rheotaxis of swimming {E. coli}
  bacteria}.
\newblock \href{http://dx.doi.org/10.1038/s41467-019-11360-0}{\emph{Nat.
  Commun.}}
  \href{http://dx.doi.org/10.1038/s41467-019-11360-0}{\textbf{\bibinfo{volume}{20}},
  \bibinfo{pages}{3434}} \href{http://dx.doi.org/10.1038/s41467-019-11360-0}{
  (\bibinfo{year}{2019})}.

\bibitem{Friedrich07}
\bibinfo{author}{Friedrich, B.~M.} \& \bibinfo{author}{J{\"u}licher, F.}
\newblock \bibinfo{title}{Chemotaxis of sperm cells}.
\newblock \href{http://dx.doi.org/0.1073/pnas.0703530104}{\emph{Proc. Natl.
  Acad. Sci. USA}}
  \href{http://dx.doi.org/0.1073/pnas.0703530104}{\textbf{\bibinfo{volume}{104}},
  \bibinfo{pages}{13256}} \href{http://dx.doi.org/0.1073/pnas.0703530104}{
  (\bibinfo{year}{2007})}.

\bibitem{Tu13}
\bibinfo{author}{Tu, Y.}
\newblock \bibinfo{title}{Quantitative modeling of bacterial chemotaxis: Signal
  amplification and accurate adaptation}.
\newblock
  \href{http://dx.doi.org/10.1146/annurev-biophys-083012-130358}{\emph{Annu.
  Rev. Biophys.}}
  \href{http://dx.doi.org/10.1146/annurev-biophys-083012-130358}{\textbf{\bibinfo{volume}{42}},
  \bibinfo{pages}{337--359}}
  \href{http://dx.doi.org/10.1146/annurev-biophys-083012-130358}{
  (\bibinfo{year}{2013})}.

\bibitem{Camley16}
\bibinfo{author}{Camley, B.~A.}, \bibinfo{author}{Zimmermann, J.},
  \bibinfo{author}{Levine, H.} \& \bibinfo{author}{Rappel, W.-J.}
\newblock \bibinfo{title}{Emergent collective chemotaxis without single-cell
  gradient sensing}.
\newblock \href{http://dx.doi.org/10.1103/PhysRevLett.116.098101}{\emph{Phys.
  Rev. Lett.}}
  \href{http://dx.doi.org/10.1103/PhysRevLett.116.098101}{\textbf{\bibinfo{volume}{116}},
  \bibinfo{pages}{098101}}
  \href{http://dx.doi.org/10.1103/PhysRevLett.116.098101}{
  (\bibinfo{year}{2016})}.

\bibitem{tenHagen14}
\bibinfo{author}{ten Hagen, B.}, \bibinfo{author}{K\"ummel, F.},
  \bibinfo{author}{Wittkowski, R.}, \bibinfo{author}{Takagi, D.},
  \bibinfo{author}{L\"owen, H.} \& \bibinfo{author}{Bechinger, C.}
\newblock \bibinfo{title}{Gravitaxis of asymmetric self-propelled colloidal
  particles}.
\newblock \href{http://dx.doi.org/10.1038/ncomms5829}{\emph{Nat. Commun.}}
  \href{http://dx.doi.org/10.1038/ncomms5829}{\textbf{\bibinfo{volume}{5}},
  \bibinfo{pages}{4829}} \href{http://dx.doi.org/10.1038/ncomms5829}{
  (\bibinfo{year}{2014})}.

\bibitem{Kuhr17}
\bibinfo{author}{Kuhr, J.-T.}, \bibinfo{author}{Blaschke, J.},
  \bibinfo{author}{R{\"u}hle, F.} \& \bibinfo{author}{Stark, H.}
\newblock \bibinfo{title}{Collective sedimentation of squirmers under gravity}.
\newblock \href{http://dx.doi.org/10.1039/c7sm01180f}{\emph{Soft Matter}}
  \href{http://dx.doi.org/10.1039/c7sm01180f}{\textbf{\bibinfo{volume}{13}},
  \bibinfo{pages}{7548--7555}} \href{http://dx.doi.org/10.1039/c7sm01180f}{
  (\bibinfo{year}{2017})}.

\bibitem{Cohen14}
\bibinfo{author}{Cohen, J.~A.} \& \bibinfo{author}{Golestanian, R.}
\newblock \bibinfo{title}{Emergent cometlike swarming of optically driven
  thermally active colloids}.
\newblock \href{http://dx.doi.org/10.1103/PhysRevLett.112.068302}{\emph{Phys.
  Rev. Lett.}}
  \href{http://dx.doi.org/10.1103/PhysRevLett.112.068302}{\textbf{\bibinfo{volume}{112}},
  \bibinfo{pages}{068302}}
  \href{http://dx.doi.org/10.1103/PhysRevLett.112.068302}{
  (\bibinfo{year}{2014})}.

\bibitem{Martin72}
\bibinfo{author}{Martin, P.~C.}, \bibinfo{author}{Parodi, O.} \&
  \bibinfo{author}{Pershan, P.~S.}
\newblock \bibinfo{title}{Unified hydrodynamic theory for crystals, liquid
  crystals, and normal fluids}.
\newblock \href{http://dx.doi.org/10.1103/PhysRevA.6.2401}{\emph{Phys. Rev. A}}
  \href{http://dx.doi.org/10.1103/PhysRevA.6.2401}{\textbf{\bibinfo{volume}{6}},
  \bibinfo{pages}{2401--2420}}
  \href{http://dx.doi.org/10.1103/PhysRevA.6.2401}{ (\bibinfo{year}{1972})}.

\bibitem{Prost15}
\bibinfo{author}{Prost, J.}, \bibinfo{author}{J\"ulicher, F.} \&
  \bibinfo{author}{Joanny, J.-F.}
\newblock \bibinfo{title}{Active gel physics}.
\newblock \href{http://dx.doi.org/10.1038/nphys3224}{\emph{Nat. Phys.}}
  \href{http://dx.doi.org/10.1038/nphys3224}{\textbf{\bibinfo{volume}{11}},
  \bibinfo{pages}{111--117}} \href{http://dx.doi.org/10.1038/nphys3224}{
  (\bibinfo{year}{2015})}.

\bibitem{Julicher18}
\bibinfo{author}{J\"ulicher, F.}, \bibinfo{author}{Grill, S.~W.} \&
  \bibinfo{author}{Salbreux, G.}
\newblock \bibinfo{title}{Hydrodynamic theory of active matter}.
\newblock \href{http://dx.doi.org/10.1088/1361-6633/aab6bb}{\emph{Rep. Prog.
  Phys.}}
  \href{http://dx.doi.org/10.1088/1361-6633/aab6bb}{\textbf{\bibinfo{volume}{81}},
  \bibinfo{pages}{076601}} \href{http://dx.doi.org/10.1088/1361-6633/aab6bb}{
  (\bibinfo{year}{2018})}.

\bibitem{Carenza19}
\bibinfo{author}{Carenza, L.~N.}, \bibinfo{author}{Gonnella, G.},
  \bibinfo{author}{Lamura, A.}, \bibinfo{author}{Negro, G.} \&
  \bibinfo{author}{Tiribocchi, A.}
\newblock \bibinfo{title}{Lattice boltzmann methods and active fluids}.
\newblock \href{http://dx.doi.org/10.1140/epje/i2019-11843-6}{\emph{The
  European Physical Journal E}}
  \href{http://dx.doi.org/10.1140/epje/i2019-11843-6}{\textbf{\bibinfo{volume}{42}},
  \bibinfo{pages}{81}} \href{http://dx.doi.org/10.1140/epje/i2019-11843-6}{
  (\bibinfo{year}{2019})}.

\bibitem{Simha02}
\bibinfo{author}{Aditi~Simha, R.} \& \bibinfo{author}{Ramaswamy, S.}
\newblock \bibinfo{title}{Hydrodynamic fluctuations and instabilities in
  ordered suspensions of self-propelled particles}.
\newblock \href{http://dx.doi.org/10.1103/PhysRevLett.89.058101}{\emph{Phys.
  Rev. Lett.}}
  \href{http://dx.doi.org/10.1103/PhysRevLett.89.058101}{\textbf{\bibinfo{volume}{89}},
  \bibinfo{pages}{058101}}
  \href{http://dx.doi.org/10.1103/PhysRevLett.89.058101}{
  (\bibinfo{year}{2002})}.

\bibitem{Hatwalne04}
\bibinfo{author}{Hatwalne, Y.}, \bibinfo{author}{Ramaswamy, S.},
  \bibinfo{author}{Rao, M.} \& \bibinfo{author}{Simha, R.~A.}
\newblock \bibinfo{title}{Rheology of active-particle suspensions}.
\newblock \href{http://dx.doi.org/10.1103/PhysRevLett.92.118101}{\emph{Phys.
  Rev. Lett.}}
  \href{http://dx.doi.org/10.1103/PhysRevLett.92.118101}{\textbf{\bibinfo{volume}{92}},
  \bibinfo{pages}{118101}}
  \href{http://dx.doi.org/10.1103/PhysRevLett.92.118101}{
  (\bibinfo{year}{2004})}.

\bibitem{Kruse04}
\bibinfo{author}{Kruse, K.}, \bibinfo{author}{Joanny, J.~F.},
  \bibinfo{author}{J\"ulicher, F.}, \bibinfo{author}{Prost, J.} \&
  \bibinfo{author}{Sekimoto, K.}
\newblock \bibinfo{title}{Asters, vortices, and rotating spirals in active gels
  of polar filaments}.
\newblock \href{http://dx.doi.org/10.1103/PhysRevLett.92.078101}{\emph{Phys.
  Rev. Lett.}}
  \href{http://dx.doi.org/10.1103/PhysRevLett.92.078101}{\textbf{\bibinfo{volume}{92}},
  \bibinfo{pages}{078101}}
  \href{http://dx.doi.org/10.1103/PhysRevLett.92.078101}{
  (\bibinfo{year}{2004})}.

\bibitem{Giomi08}
\bibinfo{author}{Giomi, L.}, \bibinfo{author}{Marchetti, M.~C.} \&
  \bibinfo{author}{Liverpool, T.~B.}
\newblock \bibinfo{title}{Complex spontaneous flows and concentration banding
  in active polar films}.
\newblock \href{http://dx.doi.org/10.1103/PhysRevLett.101.198101}{\emph{Phys.
  Rev. Lett.}}
  \href{http://dx.doi.org/10.1103/PhysRevLett.101.198101}{\textbf{\bibinfo{volume}{101}},
  \bibinfo{pages}{198101}}
  \href{http://dx.doi.org/10.1103/PhysRevLett.101.198101}{
  (\bibinfo{year}{2008})}.

\bibitem{Baskaran09}
\bibinfo{author}{Baskaran, A.} \& \bibinfo{author}{Marchetti, M.~C.}
\newblock \bibinfo{title}{Statistical mechanics and hydrodynamics of bacterial
  suspensions}.
\newblock \href{http://dx.doi.org/10.1073/pnas.0906586106}{\emph{Proc. Natl.
  Acad. Sci. USA}}
  \href{http://dx.doi.org/10.1073/pnas.0906586106}{\textbf{\bibinfo{volume}{106}},
  \bibinfo{pages}{15567--15572}}
  \href{http://dx.doi.org/10.1073/pnas.0906586106}{ (\bibinfo{year}{2009})}.

\bibitem{Linkmann19}
\bibinfo{author}{Linkmann, M.}, \bibinfo{author}{Marchetti, M.~C.},
  \bibinfo{author}{Boffetta, G.} \& \bibinfo{author}{Eckhardt, B.}
\newblock \bibinfo{title}{{Condensate formation and multiscale dynamics in
  two-dimensional active suspensions}}.
\newblock \emph{arXiv e-prints} \bibinfo{pages}{arXiv:1905.06267}
  (\bibinfo{year}{2019}).

\bibitem{Marenduzzo07}
\bibinfo{author}{Marenduzzo, D.}, \bibinfo{author}{Orlandini, E.},
  \bibinfo{author}{Cates, M.~E.} \& \bibinfo{author}{Yeomans, J.~M.}
\newblock \bibinfo{title}{Steady-state hydrodynamic instabilities of active
  liquid crystals: Hybrid lattice {B}oltzmann simulations}.
\newblock \href{http://dx.doi.org/10.1103/PhysRevE.76.031921}{\emph{Phys. Rev.
  E}}
  \href{http://dx.doi.org/10.1103/PhysRevE.76.031921}{\textbf{\bibinfo{volume}{76}},
  \bibinfo{pages}{031921}} \href{http://dx.doi.org/10.1103/PhysRevE.76.031921}{
  (\bibinfo{year}{2007})}.

\bibitem{Marenduzzo07b}
\bibinfo{author}{Marenduzzo, D.}, \bibinfo{author}{Orlandini, E.} \&
  \bibinfo{author}{Yeomans, J.~M.}
\newblock \bibinfo{title}{Hydrodynamics and rheology of active liquid crystals:
  A numerical investigation}.
\newblock \href{http://dx.doi.org/10.1103/PhysRevLett.98.118102}{\emph{Phys.
  Rev. Lett.}}
  \href{http://dx.doi.org/10.1103/PhysRevLett.98.118102}{\textbf{\bibinfo{volume}{98}},
  \bibinfo{pages}{118102}}
  \href{http://dx.doi.org/10.1103/PhysRevLett.98.118102}{
  (\bibinfo{year}{2007})}.

\bibitem{Giomi11}
\bibinfo{author}{Giomi, L.}, \bibinfo{author}{Mahadevan, L.},
  \bibinfo{author}{Chakraborty, B.} \& \bibinfo{author}{Hagan, M.~F.}
\newblock \bibinfo{title}{Excitable patterns in active nematics}.
\newblock \href{http://dx.doi.org/10.1103/PhysRevLett.106.218101}{\emph{Phys.
  Rev. Lett.}}
  \href{http://dx.doi.org/10.1103/PhysRevLett.106.218101}{\textbf{\bibinfo{volume}{106}},
  \bibinfo{pages}{218101}}
  \href{http://dx.doi.org/10.1103/PhysRevLett.106.218101}{
  (\bibinfo{year}{2011})}.

\bibitem{Doostmohammadi18}
\bibinfo{author}{Doostmohammadi, A.}, \bibinfo{author}{Ignes-Mullol, J.},
  \bibinfo{author}{Yeomans, J.~M.} \& \bibinfo{author}{Sagues, F.}
\newblock \bibinfo{title}{Active nematics}.
\newblock \href{http://dx.doi.org/10.1038/s41467-018-05666-8}{\emph{Nat.
  Commun.}}
  \href{http://dx.doi.org/10.1038/s41467-018-05666-8}{\textbf{\bibinfo{volume}{9}},
  \bibinfo{pages}{3246}} \href{http://dx.doi.org/10.1038/s41467-018-05666-8}{
  (\bibinfo{year}{2018})}.

\bibitem{Dunkel13}
\bibinfo{author}{Dunkel, J.}, \bibinfo{author}{Heidenreich, S.},
  \bibinfo{author}{B{\"a}r, M.} \& \bibinfo{author}{Goldstein, R.~E.}
\newblock \bibinfo{title}{Minimal continuum theories of structure formation in
  dense active fluids}.
\newblock \href{http://dx.doi.org/10.1088/1367-2630/15/4/045016}{\emph{New J.
  Phys.}}
  \href{http://dx.doi.org/10.1088/1367-2630/15/4/045016}{\textbf{\bibinfo{volume}{15}},
  \bibinfo{pages}{045016}}
  \href{http://dx.doi.org/10.1088/1367-2630/15/4/045016}{
  (\bibinfo{year}{2013})}.

\bibitem{Slomka15}
\bibinfo{author}{Slomka, J.} \& \bibinfo{author}{Dunkel, J.}
\newblock \bibinfo{title}{Generalized navier-stokes equations for active
  suspensions}.
\newblock \href{http://dx.doi.org/10.1140/epjst/e2015-02463-2}{\emph{Eur. Phys.
  J. Spec. Top.}}
  \href{http://dx.doi.org/10.1140/epjst/e2015-02463-2}{\textbf{\bibinfo{volume}{224}},
  \bibinfo{pages}{1349--1358}}
  \href{http://dx.doi.org/10.1140/epjst/e2015-02463-2}{
  (\bibinfo{year}{2015})}.

\bibitem{Dunkel13b}
\bibinfo{author}{Dunkel, J.}, \bibinfo{author}{Heidenreich, S.},
  \bibinfo{author}{Drescher, K.}, \bibinfo{author}{Wensink, H.~H.},
  \bibinfo{author}{B\"ar, M.} \& \bibinfo{author}{Goldstein, R.~E.}
\newblock \bibinfo{title}{Fluid dynamics of bacterial turbulence}.
\newblock \href{http://dx.doi.org/10.1103/PhysRevLett.110.228102}{\emph{Phys.
  Rev. Lett.}}
  \href{http://dx.doi.org/10.1103/PhysRevLett.110.228102}{\textbf{\bibinfo{volume}{110}},
  \bibinfo{pages}{228102}}
  \href{http://dx.doi.org/10.1103/PhysRevLett.110.228102}{
  (\bibinfo{year}{2013})}.

\bibitem{Wensink12}
\bibinfo{author}{Wensink, H.~H.}, \bibinfo{author}{Dunkel, J.},
  \bibinfo{author}{Heidenreich, S.}, \bibinfo{author}{Drescher, K.},
  \bibinfo{author}{Goldstein, R.~E.}, \bibinfo{author}{L{\"o}wen, H.} \&
  \bibinfo{author}{Yeomans, J.~M.}
\newblock \bibinfo{title}{Meso-scale turbulence in living fluids}.
\newblock \href{http://dx.doi.org/10.1073/pnas.1202032109}{\emph{Proc. Natl.
  Acad. Sci. USA}}
  \href{http://dx.doi.org/10.1073/pnas.1202032109}{\textbf{\bibinfo{volume}{109}},
  \bibinfo{pages}{14308--14313}}
  \href{http://dx.doi.org/10.1073/pnas.1202032109}{ (\bibinfo{year}{2012})}.

\bibitem{Slomka17}
\bibinfo{author}{Slomka, J.} \& \bibinfo{author}{Dunkel, J.}
\newblock \bibinfo{title}{Geometry-dependent viscosity reduction in sheared
  active fluids}.
\newblock \href{http://dx.doi.org/10.1103/PhysRevFluids.2.043102}{\emph{Phys.
  Rev. Fluids}}
  \href{http://dx.doi.org/10.1103/PhysRevFluids.2.043102}{\textbf{\bibinfo{volume}{2}},
  \bibinfo{pages}{043102}}
  \href{http://dx.doi.org/10.1103/PhysRevFluids.2.043102}{
  (\bibinfo{year}{2017})}.

\bibitem{Slomka17b}
\bibinfo{author}{Slomka, J.} \& \bibinfo{author}{Dunkel, J.}
\newblock \bibinfo{title}{Spontaneous mirror-symmetry breaking induces inverse
  energy cascade in 3d active fluids}.
\newblock \href{http://dx.doi.org/10.1073/pnas.1614721114}{\emph{Proc. Natl.
  Acad. Sci. USA}}
  \href{http://dx.doi.org/10.1073/pnas.1614721114}{\textbf{\bibinfo{volume}{114}},
  \bibinfo{pages}{2119--2124}}
  \href{http://dx.doi.org/10.1073/pnas.1614721114}{ (\bibinfo{year}{2017})}.

\bibitem{Tiribocchi15}
\bibinfo{author}{Tiribocchi, A.}, \bibinfo{author}{Wittkowski, R.},
  \bibinfo{author}{Marenduzzo, D.} \& \bibinfo{author}{Cates, M.~E.}
\newblock \bibinfo{title}{Active model h: Scalar active matter in a
  momentum-conserving fluid}.
\newblock \href{http://dx.doi.org/10.1103/PhysRevLett.115.188302}{\emph{Phys.
  Rev. Lett.}}
  \href{http://dx.doi.org/10.1103/PhysRevLett.115.188302}{\textbf{\bibinfo{volume}{115}},
  \bibinfo{pages}{188302}}
  \href{http://dx.doi.org/10.1103/PhysRevLett.115.188302}{
  (\bibinfo{year}{2015})}.

\bibitem{Cates19}
\bibinfo{author}{{Cates}, M.~E.}
\newblock \bibinfo{title}{{Active Field Theories}}.
\newblock \emph{arXiv e-prints} \bibinfo{pages}{arXiv:1904.01330}
  (\bibinfo{year}{2019}).

\bibitem{Mogilner96}
\bibinfo{author}{Mogilner, A.} \& \bibinfo{author}{Oster, G.}
\newblock \bibinfo{title}{Cell motility driven by actin polymerization}.
\newblock \href{http://dx.doi.org/10.1016/S0006-3495(96)79496-1}{\emph{Biophys.
  J.}}
  \href{http://dx.doi.org/10.1016/S0006-3495(96)79496-1}{\textbf{\bibinfo{volume}{71}},
  \bibinfo{pages}{3030--3045}}
  \href{http://dx.doi.org/10.1016/S0006-3495(96)79496-1}{
  (\bibinfo{year}{1996})}.

\bibitem{Dogterom05}
\bibinfo{author}{Dogterom, M.}, \bibinfo{author}{Kerssemakers, J.~W.},
  \bibinfo{author}{Romet-Lemonne, G.} \& \bibinfo{author}{Janson, M.~E.}
\newblock \bibinfo{title}{Force generation by dynamic microtubules}.
\newblock \href{http://dx.doi.org/10.1016/j.ceb.2004.12.011}{\emph{Curr. Opin.
  Cell Biol.}}
  \href{http://dx.doi.org/10.1016/j.ceb.2004.12.011}{\textbf{\bibinfo{volume}{17}},
  \bibinfo{pages}{67--74}} \href{http://dx.doi.org/10.1016/j.ceb.2004.12.011}{
  (\bibinfo{year}{2005})}.

\bibitem{Mogilner08}
\bibinfo{author}{Mogilner, A.}
\newblock \bibinfo{title}{Mathematics of cell motility: have we got its
  number?}
\newblock \href{http://dx.doi.org/10.1007/s00285-008-0182-2}{\emph{J. Math.
  Biol.}}
  \href{http://dx.doi.org/10.1007/s00285-008-0182-2}{\textbf{\bibinfo{volume}{58}},
  \bibinfo{pages}{105}} \href{http://dx.doi.org/10.1007/s00285-008-0182-2}{
  (\bibinfo{year}{2008})}.

\bibitem{Erlenkamper09}
\bibinfo{author}{Erlenk\"amper, C.} \& \bibinfo{author}{Kruse, K.}
\newblock \bibinfo{title}{Uncorrelated changes of subunit stability can
  generate length-dependent disassembly of treadmilling filaments}.
\newblock \href{http://dx.doi.org/10.1088/1478-3975/6/4/046016}{\emph{Phys.
  Biol.}}
  \href{http://dx.doi.org/10.1088/1478-3975/6/4/046016}{\textbf{\bibinfo{volume}{6}},
  \bibinfo{pages}{046016}}
  \href{http://dx.doi.org/10.1088/1478-3975/6/4/046016}{
  (\bibinfo{year}{2009})}.

\bibitem{Howard01}
\bibinfo{author}{Howard, J.}
\newblock \emph{\bibinfo{title}{Mechanics of motor proteins and the
  cytoskeleton}} (\bibinfo{publisher}{Sinauer Associates},
  \bibinfo{address}{Sunderland}, \bibinfo{year}{2001}).

\bibitem{Nielsen10}
\bibinfo{author}{Nielsen, S.~O.}, \bibinfo{author}{Bulo, R.~E.},
  \bibinfo{author}{Moore, P.~B.} \& \bibinfo{author}{Ensing, B.}
\newblock \bibinfo{title}{Recent progress in adaptive multiscale molecular
  dynamics simulations of soft matter}.
\newblock \href{http://dx.doi.org/10.1039/C004111D}{\emph{Phys. Chem. Chem.
  Phys.}}
  \href{http://dx.doi.org/10.1039/C004111D}{\textbf{\bibinfo{volume}{12}},
  \bibinfo{pages}{12401--12414}} \href{http://dx.doi.org/10.1039/C004111D}{
  (\bibinfo{year}{2010})}.

\bibitem{Ekimoto18}
\bibinfo{author}{Ekimoto, T.} \& \bibinfo{author}{Ikeguchi, M.}
\newblock \bibinfo{title}{Multiscale molecular dynamics simulations of rotary
  motor proteins}.
\newblock \href{http://dx.doi.org/10.1007/s12551-017-0373-4}{\emph{Biophys.
  Rev.}}
  \href{http://dx.doi.org/10.1007/s12551-017-0373-4}{\textbf{\bibinfo{volume}{10}},
  \bibinfo{pages}{605--615}}
  \href{http://dx.doi.org/10.1007/s12551-017-0373-4}{ (\bibinfo{year}{2018})}.

\bibitem{Kolomeisky07}
\bibinfo{author}{Kolomeisky, A.~B.} \& \bibinfo{author}{Fisher, M.~E.}
\newblock \bibinfo{title}{Molecular motors: A theorist's perspective}.
\newblock
  \href{http://dx.doi.org/10.1146/annurev.physchem.58.032806.104532}{\emph{Annu.
  Rev. Phys. Chem.}}
  \href{http://dx.doi.org/10.1146/annurev.physchem.58.032806.104532}{\textbf{\bibinfo{volume}{58}},
  \bibinfo{pages}{675--695}}
  \href{http://dx.doi.org/10.1146/annurev.physchem.58.032806.104532}{
  (\bibinfo{year}{2007})}.

\bibitem{Chowdhury13}
\bibinfo{author}{Chowdhury, D.}
\newblock \bibinfo{title}{Stochastic mechano-chemical kinetics of molecular
  motors: A multidisciplinary enterprise from a physicist's perspective}.
\newblock \href{http://dx.doi.org/10.1016/j.physrep.2013.03.005}{\emph{Phys.
  Rep.}}
  \href{http://dx.doi.org/10.1016/j.physrep.2013.03.005}{\textbf{\bibinfo{volume}{529}},
  \bibinfo{pages}{1--197}}
  \href{http://dx.doi.org/10.1016/j.physrep.2013.03.005}{
  (\bibinfo{year}{2013})}.

\bibitem{Klumpp05}
\bibinfo{author}{Klumpp, S.} \& \bibinfo{author}{Lipowsky, R.}
\newblock \bibinfo{title}{Cooperative cargo transport by several molecular
  motors}.
\newblock \href{http://dx.doi.org/10.1073/pnas.0507363102}{\emph{Proc. Natl.
  Acad. Sci. USA}}
  \href{http://dx.doi.org/10.1073/pnas.0507363102}{\textbf{\bibinfo{volume}{102}},
  \bibinfo{pages}{17284--17289}}
  \href{http://dx.doi.org/10.1073/pnas.0507363102}{ (\bibinfo{year}{2005})}.

\bibitem{Appert-Rolland15}
\bibinfo{author}{Appert-Rolland, C.}, \bibinfo{author}{Ebbinghaus, M.} \&
  \bibinfo{author}{Santen, L.}
\newblock \bibinfo{title}{Intracellular transport driven by cytoskeletal
  motors: General mechanisms and defects}.
\newblock \href{http://dx.doi.org/10.1016/j.physrep.2015.07.001}{\emph{Phys.
  Rep.}}
  \href{http://dx.doi.org/10.1016/j.physrep.2015.07.001}{\textbf{\bibinfo{volume}{593}},
  \bibinfo{pages}{1--59}}
  \href{http://dx.doi.org/10.1016/j.physrep.2015.07.001}{
  (\bibinfo{year}{2015})}.

\bibitem{Bausch06}
\bibinfo{author}{Bausch, A.~R.} \& \bibinfo{author}{Kroy, K.}
\newblock \bibinfo{title}{A bottom-up approach to cell mechanics}.
\newblock \href{http://dx.doi.org/10.1038/nphys260}{\emph{Nat. Phys.}}
  \href{http://dx.doi.org/10.1038/nphys260}{\textbf{\bibinfo{volume}{2}},
  \bibinfo{pages}{231--238}} \href{http://dx.doi.org/10.1038/nphys260}{
  (\bibinfo{year}{2006})}.

\bibitem{Huber13}
\bibinfo{author}{Huber, F.}, \bibinfo{author}{Schnau{\ss}, J.},
  \bibinfo{author}{R\"onicke, S.}, \bibinfo{author}{Rauch, P.},
  \bibinfo{author}{M\"uller, K.}, \bibinfo{author}{F\"utterer, C.} \&
  \bibinfo{author}{K\"as, J.}
\newblock \bibinfo{title}{Emergent complexity of the cytoskeleton: from single
  filaments to tissue}.
\newblock \href{http://dx.doi.org/10.1080/00018732.2013.771509}{\emph{Adv.
  Phys.}}
  \href{http://dx.doi.org/10.1080/00018732.2013.771509}{\textbf{\bibinfo{volume}{62}},
  \bibinfo{pages}{1--112}}
  \href{http://dx.doi.org/10.1080/00018732.2013.771509}{
  (\bibinfo{year}{2013})}.

\bibitem{Broedersz14}
\bibinfo{author}{Broedersz, C.~P.} \& \bibinfo{author}{MacKintosh, F.~C.}
\newblock \bibinfo{title}{Modeling semiflexible polymer networks}.
\newblock \href{http://dx.doi.org/10.1103/RevModPhys.86.995}{\emph{Rev. Mod.
  Phys.}}
  \href{http://dx.doi.org/10.1103/RevModPhys.86.995}{\textbf{\bibinfo{volume}{86}},
  \bibinfo{pages}{995--1036}}
  \href{http://dx.doi.org/10.1103/RevModPhys.86.995}{ (\bibinfo{year}{2014})}.

\bibitem{Nedelec07}
\bibinfo{author}{Nedelec, F.} \& \bibinfo{author}{Foethke, D.}
\newblock \bibinfo{title}{Collective langevin dynamics of flexible cytoskeletal
  fibers}.
\newblock \href{http://dx.doi.org/10.1088/1367-2630/9/11/427}{\emph{New J.
  Phys.}}
  \href{http://dx.doi.org/10.1088/1367-2630/9/11/427}{\textbf{\bibinfo{volume}{9}},
  \bibinfo{pages}{427--427}}
  \href{http://dx.doi.org/10.1088/1367-2630/9/11/427}{ (\bibinfo{year}{2007})}.

\bibitem{Mohapatra16}
\bibinfo{author}{Mohapatra, L.}, \bibinfo{author}{Goode, B.~L.},
  \bibinfo{author}{Jelenkovic, P.}, \bibinfo{author}{Phillips, R.} \&
  \bibinfo{author}{Kondev, J.}
\newblock \bibinfo{title}{Design principles of length control of cytoskeletal
  structures}.
\newblock
  \href{http://dx.doi.org/10.1146/annurev-biophys-070915-094206}{\emph{Annu.
  Rev. Biophys.}}
  \href{http://dx.doi.org/10.1146/annurev-biophys-070915-094206}{\textbf{\bibinfo{volume}{45}},
  \bibinfo{pages}{85--116}}
  \href{http://dx.doi.org/10.1146/annurev-biophys-070915-094206}{
  (\bibinfo{year}{2016})}.

\bibitem{Mogilner10}
\bibinfo{author}{Mogilner, A.} \& \bibinfo{author}{Craig, E.}
\newblock \bibinfo{title}{Towards a quantitative understanding of mitotic
  spindle assembly and mechanics}.
\newblock \href{http://dx.doi.org/10.1242/jcs.062208}{\emph{J. Cell Sci.}}
  \href{http://dx.doi.org/10.1242/jcs.062208}{\textbf{\bibinfo{volume}{123}},
  \bibinfo{pages}{3435--3445}} \href{http://dx.doi.org/10.1242/jcs.062208}{
  (\bibinfo{year}{2010})}.

\bibitem{Pavin16}
\bibinfo{author}{Pavin, N.} \& \bibinfo{author}{Toli\'c, I.~M.}
\newblock \bibinfo{title}{Self-organization and forces in the mitotic spindle}.
\newblock
  \href{http://dx.doi.org/10.1146/annurev-biophys-062215-010934}{\emph{Annu.
  Rev. Biophys.}}
  \href{http://dx.doi.org/10.1146/annurev-biophys-062215-010934}{\textbf{\bibinfo{volume}{45}},
  \bibinfo{pages}{279--298}}
  \href{http://dx.doi.org/10.1146/annurev-biophys-062215-010934}{
  (\bibinfo{year}{2016})}.

\bibitem{Broedersz11}
\bibinfo{author}{Broedersz, C.~P.} \& \bibinfo{author}{MacKintosh, F.~C.}
\newblock \bibinfo{title}{Molecular motors stiffen non-affine semiflexible
  polymer networks}.
\newblock \href{http://dx.doi.org/10.1039/C0SM01004A}{\emph{Soft Matter}}
  \href{http://dx.doi.org/10.1039/C0SM01004A}{\textbf{\bibinfo{volume}{7}},
  \bibinfo{pages}{3186--3191}} \href{http://dx.doi.org/10.1039/C0SM01004A}{
  (\bibinfo{year}{2011})}.

\bibitem{Ronceray16}
\bibinfo{author}{Ronceray, P.}, \bibinfo{author}{Broedersz, C.~P.} \&
  \bibinfo{author}{Lenz, M.}
\newblock \bibinfo{title}{Fiber networks amplify active stress}.
\newblock \href{http://dx.doi.org/10.1073/pnas.1514208113}{\emph{Proc. Natl.
  Acad. Sci. USA}}
  \href{http://dx.doi.org/10.1073/pnas.1514208113}{\textbf{\bibinfo{volume}{113}},
  \bibinfo{pages}{2827--2832}}
  \href{http://dx.doi.org/10.1073/pnas.1514208113}{ (\bibinfo{year}{2016})}.

\bibitem{Julicher07}
\bibinfo{author}{J{\"u}licher, F.}, \bibinfo{author}{Kruse, K.},
  \bibinfo{author}{Prost, J.} \& \bibinfo{author}{Joanny, J.-F.}
\newblock \bibinfo{title}{Active behavior of the cytoskeleton}.
\newblock \href{http://dx.doi.org/10.1016/j.physrep.2007.02.018}{\emph{Phys.
  Rep.}}
  \href{http://dx.doi.org/10.1016/j.physrep.2007.02.018}{\textbf{\bibinfo{volume}{449}},
  \bibinfo{pages}{3--28}}
  \href{http://dx.doi.org/10.1016/j.physrep.2007.02.018}{
  (\bibinfo{year}{2007})}.

\bibitem{Joanny09}
\bibinfo{author}{Joanny, J.~F.} \& \bibinfo{author}{Prost, J.}
\newblock \bibinfo{title}{Active gels as a description of the actin-myosin
  cytoskeleton}.
\newblock \href{http://dx.doi.org/10.2976/1.3054712}{\emph{HFSP J.}}
  \href{http://dx.doi.org/10.2976/1.3054712}{\textbf{\bibinfo{volume}{3}},
  \bibinfo{pages}{94--104}} \href{http://dx.doi.org/10.2976/1.3054712}{
  (\bibinfo{year}{2009})}.

\bibitem{Jilkine11}
\bibinfo{author}{Jilkine, A.} \& \bibinfo{author}{Edelstein-Keshet, L.}
\newblock \bibinfo{title}{A comparison of mathematical models for polarization
  of single eukaryotic cells in response to guided cues}.
\newblock \href{http://dx.doi.org/10.1371/journal.pcbi.1001121}{\emph{PLoS
  Comput. Biol.}}
  \href{http://dx.doi.org/10.1371/journal.pcbi.1001121}{\textbf{\bibinfo{volume}{7}},
  \bibinfo{pages}{e1001121}}
  \href{http://dx.doi.org/10.1371/journal.pcbi.1001121}{
  (\bibinfo{year}{2011})}.

\bibitem{Holmes12}
\bibinfo{author}{Holmes, W.~R.} \& \bibinfo{author}{Edelstein-Keshet, L.}
\newblock \bibinfo{title}{A comparison of computational models for eukaryotic
  cell shape and motility}.
\newblock \href{http://dx.doi.org/10.1371/journal.pcbi.1002793}{\emph{PLoS
  Comput. Biol.}}
  \href{http://dx.doi.org/10.1371/journal.pcbi.1002793}{\textbf{\bibinfo{volume}{8}},
  \bibinfo{pages}{e1002793}}
  \href{http://dx.doi.org/10.1371/journal.pcbi.1002793}{
  (\bibinfo{year}{2012})}.

\bibitem{Danuser13}
\bibinfo{author}{Danuser, G.}, \bibinfo{author}{Allard, J.} \&
  \bibinfo{author}{Mogilner, A.}
\newblock \bibinfo{title}{Mathematical modeling of eukaryotic cell migration:
  Insights beyond experiments}.
\newblock
  \href{http://dx.doi.org/10.1146/annurev-cellbio-101512-122308}{\emph{Annu.
  Rev. Cell Dev. Biol.}}
  \href{http://dx.doi.org/10.1146/annurev-cellbio-101512-122308}{\textbf{\bibinfo{volume}{29}},
  \bibinfo{pages}{501--528}}
  \href{http://dx.doi.org/10.1146/annurev-cellbio-101512-122308}{
  (\bibinfo{year}{2013})}.

\bibitem{teBoekhorst16}
\bibinfo{author}{te~Boekhorst, V.}, \bibinfo{author}{Preziosi, L.} \&
  \bibinfo{author}{Friedl, P.}
\newblock \bibinfo{title}{Plasticity of cell migration in vivo and in silico}.
\newblock
  \href{http://dx.doi.org/10.1146/annurev-cellbio-111315-125201}{\emph{Annu.
  Rev. Cell Dev. Biol.}}
  \href{http://dx.doi.org/10.1146/annurev-cellbio-111315-125201}{\textbf{\bibinfo{volume}{32}},
  \bibinfo{pages}{491--526}}
  \href{http://dx.doi.org/10.1146/annurev-cellbio-111315-125201}{
  (\bibinfo{year}{2016})}.

\bibitem{Doubrovinski11}
\bibinfo{author}{Doubrovinski, K.} \& \bibinfo{author}{Kruse, K.}
\newblock \bibinfo{title}{Cell motility resulting from spontaneous
  polymerization waves}.
\newblock \href{http://dx.doi.org/10.1103/PhysRevLett.107.258103}{\emph{Phys.
  Rev. Lett.}}
  \href{http://dx.doi.org/10.1103/PhysRevLett.107.258103}{\textbf{\bibinfo{volume}{107}},
  \bibinfo{pages}{258103}}
  \href{http://dx.doi.org/10.1103/PhysRevLett.107.258103}{
  (\bibinfo{year}{2011})}.

\bibitem{Wolgemuth11}
\bibinfo{author}{Wolgemuth, C.~W.}, \bibinfo{author}{Stajic, J.} \&
  \bibinfo{author}{Mogilner, A.}
\newblock \bibinfo{title}{Redundant mechanisms for stable cell locomotion
  revealed by minimal models}.
\newblock \href{http://dx.doi.org/10.1016/j.bpj.2011.06.032}{\emph{Biophys.
  J.}}
  \href{http://dx.doi.org/10.1016/j.bpj.2011.06.032}{\textbf{\bibinfo{volume}{101}},
  \bibinfo{pages}{545--553}}
  \href{http://dx.doi.org/10.1016/j.bpj.2011.06.032}{ (\bibinfo{year}{2011})}.

\bibitem{Ziebert12}
\bibinfo{author}{Ziebert, F.}, \bibinfo{author}{Swaminathan, S.} \&
  \bibinfo{author}{Aranson, I.~S.}
\newblock \bibinfo{title}{Model for self-polarization and motility of
  keratocyte fragments}.
\newblock
  \href{http://dx.doi.org/http://doi.org/10.1098/rsif.2011.043}{\emph{J. Royal
  Soc. Interface}}
  \href{http://dx.doi.org/http://doi.org/10.1098/rsif.2011.043}{\textbf{\bibinfo{volume}{9}},
  \bibinfo{pages}{1084--1092}}
  \href{http://dx.doi.org/http://doi.org/10.1098/rsif.2011.043}{
  (\bibinfo{year}{2012})}.

\bibitem{Ziebert16}
\bibinfo{author}{Ziebert, F.} \& \bibinfo{author}{Aranson, I.~S.}
\newblock \bibinfo{title}{Computational approaches to substrate-based cell
  motility}.
\newblock \href{http://dx.doi.org/10.1038/npjcompumats.2016.19}{\emph{npj
  Comput. Mater.}}
  \href{http://dx.doi.org/10.1038/npjcompumats.2016.19}{\textbf{\bibinfo{volume}{2}},
  \bibinfo{pages}{16019}}
  \href{http://dx.doi.org/10.1038/npjcompumats.2016.19}{
  (\bibinfo{year}{2016})}.

\bibitem{Linsmeier16}
\bibinfo{author}{Linsmeier, I.}, \bibinfo{author}{Banerjee, S.},
  \bibinfo{author}{Oakes, P.~W.}, \bibinfo{author}{Jung, W.},
  \bibinfo{author}{Kim, T.} \& \bibinfo{author}{Murrell, M.~P.}
\newblock \bibinfo{title}{Disordered actomyosin networks are sufficient to
  produce cooperative and telescopic contractility}.
\newblock \href{http://dx.doi.org/10.1038/ncomms12615 (2016)}{\emph{Nat.
  Commun.}} \href{http://dx.doi.org/10.1038/ncomms12615
  (2016)}{\textbf{\bibinfo{volume}{7}}, \bibinfo{pages}{12615}}
  \href{http://dx.doi.org/10.1038/ncomms12615 (2016)}{ (\bibinfo{year}{2016})}.

\bibitem{Singer-Loginova08}
\bibinfo{author}{Singer-Loginova, I.} \& \bibinfo{author}{Singer, H.~M.}
\newblock \bibinfo{title}{The phase field technique for modeling multiphase
  materials}.
\newblock \href{http://dx.doi.org/10.1088/0034-4885/71/10/106501}{\emph{Rep.
  Prog. Phys.}}
  \href{http://dx.doi.org/10.1088/0034-4885/71/10/106501}{\textbf{\bibinfo{volume}{71}},
  \bibinfo{pages}{106501}}
  \href{http://dx.doi.org/10.1088/0034-4885/71/10/106501}{
  (\bibinfo{year}{2008})}.

\bibitem{Nonomura12}
\bibinfo{author}{Nonomura, M.}
\newblock \bibinfo{title}{Study on multicellular systems using a phase field
  model}.
\newblock \href{http://dx.doi.org/10.1371/journal.pone.0033501}{\emph{PLoS
  ONE}}
  \href{http://dx.doi.org/10.1371/journal.pone.0033501}{\textbf{\bibinfo{volume}{7}},
  \bibinfo{pages}{1--9}} \href{http://dx.doi.org/10.1371/journal.pone.0033501}{
  (\bibinfo{year}{2012})}.

\bibitem{Camley14}
\bibinfo{author}{Camley, B.~A.}, \bibinfo{author}{Zhang, Y.},
  \bibinfo{author}{Zhao, Y.}, \bibinfo{author}{Li, B.},
  \bibinfo{author}{Ben-Jacob, E.}, \bibinfo{author}{Levine, H.} \&
  \bibinfo{author}{Rappel, W.-J.}
\newblock \bibinfo{title}{Polarity mechanisms such as contact inhibition of
  locomotion regulate persistent rotational motion of mammalian cells on
  micropatterns}.
\newblock \href{http://dx.doi.org/10.1073/pnas.1414498111}{\emph{Proc. Natl.
  Acad. Sci. USA}}
  \href{http://dx.doi.org/10.1073/pnas.1414498111}{\textbf{\bibinfo{volume}{111}},
  \bibinfo{pages}{14770--14775}}
  \href{http://dx.doi.org/10.1073/pnas.1414498111}{ (\bibinfo{year}{2014})}.

\bibitem{Najem16}
\bibinfo{author}{Najem, S.} \& \bibinfo{author}{Grant, M.}
\newblock \bibinfo{title}{Phase-field model for collective cell migration}.
\newblock \href{http://dx.doi.org/10.1103/PhysRevE.93.052405}{\emph{Phys. Rev.
  E}}
  \href{http://dx.doi.org/10.1103/PhysRevE.93.052405}{\textbf{\bibinfo{volume}{93}},
  \bibinfo{pages}{052405}} \href{http://dx.doi.org/10.1103/PhysRevE.93.052405}{
  (\bibinfo{year}{2016})}.

\bibitem{Camley17}
\bibinfo{author}{Camley, B.~A.} \& \bibinfo{author}{Rappel, W.-J.}
\newblock \bibinfo{title}{Physical models of collective cell motility: from
  cell to tissue}.
\newblock \href{http://dx.doi.org/10.1088/1361-6463/aa56fe}{\emph{J. Phys. D}}
  \href{http://dx.doi.org/10.1088/1361-6463/aa56fe}{\textbf{\bibinfo{volume}{50}},
  \bibinfo{pages}{113002}} \href{http://dx.doi.org/10.1088/1361-6463/aa56fe}{
  (\bibinfo{year}{2017})}.

\bibitem{Mueller19}
\bibinfo{author}{Mueller, R.}, \bibinfo{author}{Yeomans, J.~M.} \&
  \bibinfo{author}{Doostmohammadi, A.}
\newblock \bibinfo{title}{Emergence of active nematic behavior in monolayers of
  isotropic cells}.
\newblock \href{http://dx.doi.org/10.1103/PhysRevLett.122.048004}{\emph{Phys.
  Rev. Lett.}}
  \href{http://dx.doi.org/10.1103/PhysRevLett.122.048004}{\textbf{\bibinfo{volume}{122}},
  \bibinfo{pages}{048004}}
  \href{http://dx.doi.org/10.1103/PhysRevLett.122.048004}{
  (\bibinfo{year}{2019})}.

\bibitem{Wenzel19}
\bibinfo{author}{Wenzel, D.}, \bibinfo{author}{Praetorius, S.} \&
  \bibinfo{author}{Voigt, A.}
\newblock \bibinfo{title}{Topological and geometrical quantities in active
  cellular structures}.
\newblock \href{http://dx.doi.org/10.1063/1.5085766}{\emph{J. Chem. Phys.}}
  \href{http://dx.doi.org/10.1063/1.5085766}{\textbf{\bibinfo{volume}{150}},
  \bibinfo{pages}{164108}} \href{http://dx.doi.org/10.1063/1.5085766}{
  (\bibinfo{year}{2019})}.

\bibitem{Abaurrea17}
\bibinfo{author}{Abaurrea-Velasco, C.}, \bibinfo{author}{Ghahnaviyeh, S.~D.},
  \bibinfo{author}{Pishkenari, H.~N.}, \bibinfo{author}{Auth, T.} \&
  \bibinfo{author}{Gompper, G.}
\newblock \bibinfo{title}{Complex self-propelled rings: a minimal model for
  cell motility}.
\newblock \href{http://dx.doi.org/10.1039/C7SM00439G}{\emph{Soft Matter}}
  \href{http://dx.doi.org/10.1039/C7SM00439G}{\textbf{\bibinfo{volume}{13}},
  \bibinfo{pages}{5865--5876}} \href{http://dx.doi.org/10.1039/C7SM00439G}{
  (\bibinfo{year}{2017})}.

\bibitem{Abaurrea19}
\bibinfo{author}{Abaurrea-Velasco, C.}, \bibinfo{author}{Auth, T.} \&
  \bibinfo{author}{Gompper, G.}
\newblock \bibinfo{title}{{Self-organized motility of vesicles with internal
  active filaments}}.
\newblock \emph{arXiv e-prints} \bibinfo{pages}{arXiv:1812.09932}
  (\bibinfo{year}{2019}).

\bibitem{Preziosi10}
\bibinfo{author}{Preziosi, L.}, \bibinfo{author}{Ambrosi, D.} \&
  \bibinfo{author}{Verdier, C.}
\newblock \bibinfo{title}{An elasto-visco-plastic model of cell aggregates}.
\newblock \href{http://dx.doi.org/10.1016/j.jtbi.2009.08.023}{\emph{J. Theor.
  Biol.}}
  \href{http://dx.doi.org/10.1016/j.jtbi.2009.08.023}{\textbf{\bibinfo{volume}{262}},
  \bibinfo{pages}{35--47}} \href{http://dx.doi.org/10.1016/j.jtbi.2009.08.023}{
  (\bibinfo{year}{2010})}.

\bibitem{Rodriguez94}
\bibinfo{author}{Rodriguez, E.~K.}, \bibinfo{author}{Hoger, A.} \&
  \bibinfo{author}{McCulloch, A.~D.}
\newblock \bibinfo{title}{Stress-dependent finite growth in soft elastic
  tissues}.
\newblock \href{http://dx.doi.org/10.1016/0021-9290(94)90021-3}{\emph{J.
  Biomech.}}
  \href{http://dx.doi.org/10.1016/0021-9290(94)90021-3}{\textbf{\bibinfo{volume}{27}},
  \bibinfo{pages}{455--467}}
  \href{http://dx.doi.org/10.1016/0021-9290(94)90021-3}{
  (\bibinfo{year}{1994})}.

\bibitem{Drasdo05}
\bibinfo{author}{Drasdo, D.} \& \bibinfo{author}{H{\"o}hme, S.}
\newblock \bibinfo{title}{A single-cell-based model of tumor growthin vitro:
  monolayers and spheroids}.
\newblock \href{http://dx.doi.org/10.1088/1478-3975/2/3/001}{\emph{Phys.
  Biol.}}
  \href{http://dx.doi.org/10.1088/1478-3975/2/3/001}{\textbf{\bibinfo{volume}{2}},
  \bibinfo{pages}{133--147}}
  \href{http://dx.doi.org/10.1088/1478-3975/2/3/001}{ (\bibinfo{year}{2005})}.

\bibitem{Basan11}
\bibinfo{author}{Basan, M.}, \bibinfo{author}{Prost, J.},
  \bibinfo{author}{Joanny, J.-F.} \& \bibinfo{author}{Elgeti, J.}
\newblock \bibinfo{title}{Dissipative particle dynamics simulations for
  biological tissues: rheology and competition}.
\newblock \href{http://dx.doi.org/10.1088/1478-3975/8/2/026014}{\emph{Phys.
  Biol.}}
  \href{http://dx.doi.org/10.1088/1478-3975/8/2/026014}{\textbf{\bibinfo{volume}{8}},
  \bibinfo{pages}{026014}}
  \href{http://dx.doi.org/10.1088/1478-3975/8/2/026014}{
  (\bibinfo{year}{2011})}.

\bibitem{MalmiKakkada18}
\bibinfo{author}{Malmi-Kakkada, A.~N.}, \bibinfo{author}{Li, X.},
  \bibinfo{author}{Samanta, H.~S.}, \bibinfo{author}{Sinha, S.} \&
  \bibinfo{author}{Thirumalai, D.}
\newblock \bibinfo{title}{Cell growth rate dictates the onset of glass to
  fluidlike transition and long time superdiffusion in an evolving cell
  colony}.
\newblock \href{http://dx.doi.org/10.1103/PhysRevX.8.021025}{\emph{Phys. Rev.
  X}}
  \href{http://dx.doi.org/10.1103/PhysRevX.8.021025}{\textbf{\bibinfo{volume}{8}},
  \bibinfo{pages}{021025}} \href{http://dx.doi.org/10.1103/PhysRevX.8.021025}{
  (\bibinfo{year}{2018})}.

\bibitem{MatozFernandez17}
\bibinfo{author}{Matoz-Fernandez, D.~A.}, \bibinfo{author}{Martens, K.},
  \bibinfo{author}{Sknepnek, R.}, \bibinfo{author}{Barrat, J.~L.} \&
  \bibinfo{author}{Henkes, S.}
\newblock \bibinfo{title}{Cell division and death inhibit glassy behaviour of
  confluent tissues}.
\newblock \href{http://dx.doi.org/10.1039/C6SM02580C}{\emph{Soft Matter}}
  \href{http://dx.doi.org/10.1039/C6SM02580C}{\textbf{\bibinfo{volume}{13}},
  \bibinfo{pages}{3205--3212}} \href{http://dx.doi.org/10.1039/C6SM02580C}{
  (\bibinfo{year}{2017})}.

\bibitem{Graner92}
\bibinfo{author}{Graner, F.} \& \bibinfo{author}{Glazier, J.~A.}
\newblock \bibinfo{title}{Simulation of biological cell sorting using a
  two-dimensional extended potts model}.
\newblock \href{http://dx.doi.org/10.1103/PhysRevLett.69.2013}{\emph{Phys. Rev.
  Lett.}}
  \href{http://dx.doi.org/10.1103/PhysRevLett.69.2013}{\textbf{\bibinfo{volume}{69}},
  \bibinfo{pages}{2013--2016}}
  \href{http://dx.doi.org/10.1103/PhysRevLett.69.2013}{
  (\bibinfo{year}{1992})}.

\bibitem{Glazier93}
\bibinfo{author}{Glazier, J.~A.} \& \bibinfo{author}{Graner, F.}
\newblock \bibinfo{title}{Simulation of the differential adhesion driven
  rearrangement of biological cells}.
\newblock \href{http://dx.doi.org/10.1103/PhysRevE.47.2128}{\emph{Phys. Rev.
  E}}
  \href{http://dx.doi.org/10.1103/PhysRevE.47.2128}{\textbf{\bibinfo{volume}{47}},
  \bibinfo{pages}{2128--2154}}
  \href{http://dx.doi.org/10.1103/PhysRevE.47.2128}{ (\bibinfo{year}{1993})}.

\bibitem{Chen07}
\bibinfo{author}{Chen, N.}, \bibinfo{author}{Glazier, J.~A.},
  \bibinfo{author}{Izaguirre, J.~A.} \& \bibinfo{author}{Alber, M.~S.}
\newblock \bibinfo{title}{A parallel implementation of the cellular potts model
  for simulation of cell-based morphogenesis}.
\newblock \href{http://dx.doi.org/10.1016/j.cpc.2007.03.007}{\emph{Comput.
  Phys. Commun.}}
  \href{http://dx.doi.org/10.1016/j.cpc.2007.03.007}{\textbf{\bibinfo{volume}{176}},
  \bibinfo{pages}{670--681}}
  \href{http://dx.doi.org/10.1016/j.cpc.2007.03.007}{ (\bibinfo{year}{2007})}.

\bibitem{Mare01}
\bibinfo{author}{Maree, A. F.~M.} \& \bibinfo{author}{Hogeweg, P.}
\newblock \bibinfo{title}{How amoeboids self-organize into a fruiting body:
  Multicellular coordination in dictyostelium discoideum}.
\newblock \href{http://dx.doi.org/10.1073/pnas.061535198}{\emph{Proc. Natl.
  Acad. Sci. USA}}
  \href{http://dx.doi.org/10.1073/pnas.061535198}{\textbf{\bibinfo{volume}{98}},
  \bibinfo{pages}{3879--3883}} \href{http://dx.doi.org/10.1073/pnas.061535198}{
  (\bibinfo{year}{2001})}.

\bibitem{Merks08}
\bibinfo{author}{Merks, R. M.~H.}, \bibinfo{author}{Perryn, E.~D.},
  \bibinfo{author}{Shirinifard, A.} \& \bibinfo{author}{Glazier, J.~A.}
\newblock \bibinfo{title}{Contact-inhibited chemotaxis in de novo and sprouting
  blood-vessel growth}.
\newblock \href{http://dx.doi.org/10.1371/journal.pcbi.1000163}{\emph{PLoS
  Comp. Biol.}}
  \href{http://dx.doi.org/10.1371/journal.pcbi.1000163}{\textbf{\bibinfo{volume}{4}},
  \bibinfo{pages}{e1000163}}
  \href{http://dx.doi.org/10.1371/journal.pcbi.1000163}{
  (\bibinfo{year}{2008})}.

\bibitem{Maree06}
\bibinfo{author}{Maree, A. F.~M.}, \bibinfo{author}{Jilkine, A.},
  \bibinfo{author}{Dawes, A.}, \bibinfo{author}{Grieneisen, V.~A.} \&
  \bibinfo{author}{Edelstein-Keshet, L.}
\newblock \bibinfo{title}{Polarization and movement of keratocytes: A
  multiscale modelling approach}.
\newblock \href{http://dx.doi.org/10.1007/s11538-006-9131-7}{\emph{Bull. Math.
  Biol.}}
  \href{http://dx.doi.org/10.1007/s11538-006-9131-7}{\textbf{\bibinfo{volume}{68}},
  \bibinfo{pages}{1169--1211}}
  \href{http://dx.doi.org/10.1007/s11538-006-9131-7}{ (\bibinfo{year}{2006})}.

\bibitem{Albert14}
\bibinfo{author}{Albert, P.~J.} \& \bibinfo{author}{Schwarz, U.~S.}
\newblock \bibinfo{title}{Dynamics of cell shape and forces on micropatterned
  substrates predicted by a cellular potts model}.
\newblock \href{http://dx.doi.org/10.1016/j.bpj.2014.04.036}{\emph{Biophys.
  J.}}
  \href{http://dx.doi.org/10.1016/j.bpj.2014.04.036}{\textbf{\bibinfo{volume}{106}},
  \bibinfo{pages}{2340--2352}}
  \href{http://dx.doi.org/10.1016/j.bpj.2014.04.036}{ (\bibinfo{year}{2014})}.

\bibitem{Segerer15}
\bibinfo{author}{Segerer, F.~J.}, \bibinfo{author}{Th\"uroff, F.},
  \bibinfo{author}{Piera~Alberola, A.}, \bibinfo{author}{Frey, E.} \&
  \bibinfo{author}{R\"adler, J.~O.}
\newblock \bibinfo{title}{Emergence and persistence of collective cell
  migration on small circular micropatterns}.
\newblock \href{http://dx.doi.org/10.1103/PhysRevLett.114.228102}{\emph{Phys.
  Rev. Lett.}}
  \href{http://dx.doi.org/10.1103/PhysRevLett.114.228102}{\textbf{\bibinfo{volume}{114}},
  \bibinfo{pages}{228102}}
  \href{http://dx.doi.org/10.1103/PhysRevLett.114.228102}{
  (\bibinfo{year}{2015})}.

\bibitem{Hufnagel07}
\bibinfo{author}{Hufnagel, L.}, \bibinfo{author}{Teleman, A.~A.},
  \bibinfo{author}{Rouault, H.}, \bibinfo{author}{Cohen, S.~M.} \&
  \bibinfo{author}{Shraiman, B.~I.}
\newblock \bibinfo{title}{On the mechanism of wing size determination in fly
  development}.
\newblock \href{http://dx.doi.org/10.1073/pnas.0607134104}{\emph{Proc. Natl.
  Acad. Sci. USA}}
  \href{http://dx.doi.org/10.1073/pnas.0607134104}{\textbf{\bibinfo{volume}{104}},
  \bibinfo{pages}{3835--3840}}
  \href{http://dx.doi.org/10.1073/pnas.0607134104}{ (\bibinfo{year}{2007})}.

\bibitem{Fletcher14}
\bibinfo{author}{Fletcher, A.~G.}, \bibinfo{author}{Osterfield, M.},
  \bibinfo{author}{Baker, R.~E.} \& \bibinfo{author}{Shvartsman, S.~Y.}
\newblock \bibinfo{title}{Vertex models of epithelial morphogenesis}.
\newblock \href{http://dx.doi.org/10.1016/j.bpj.2013.11.4498}{\emph{Biophys.
  J.}}
  \href{http://dx.doi.org/10.1016/j.bpj.2013.11.4498}{\textbf{\bibinfo{volume}{106}},
  \bibinfo{pages}{2291--2304}}
  \href{http://dx.doi.org/10.1016/j.bpj.2013.11.4498}{ (\bibinfo{year}{2014})}.

\bibitem{Sussman18}
\bibinfo{author}{Sussman, D.~M.}, \bibinfo{author}{Schwarz, J.~M.},
  \bibinfo{author}{Marchetti, M.~C.} \& \bibinfo{author}{Manning, M.~L.}
\newblock \bibinfo{title}{Soft yet sharp interfaces in a vertex model of
  confluent tissue}.
\newblock \href{http://dx.doi.org/10.1103/PhysRevLett.120.058001}{\emph{Phys.
  Rev. Lett.}}
  \href{http://dx.doi.org/10.1103/PhysRevLett.120.058001}{\textbf{\bibinfo{volume}{120}},
  \bibinfo{pages}{058001}}
  \href{http://dx.doi.org/10.1103/PhysRevLett.120.058001}{
  (\bibinfo{year}{2018})}.

\bibitem{Barton17}
\bibinfo{author}{Barton, D.~L.}, \bibinfo{author}{Henkes, S.},
  \bibinfo{author}{Weijer, C.~J.} \& \bibinfo{author}{Sknepnek, R.}
\newblock \bibinfo{title}{Active vertex model for cell-resolution description
  of epithelial tissue mechanics}.
\newblock \href{http://dx.doi.org/10.1371/journal.pcbi.1005569}{\emph{PLOS
  Comput. Biol.}}
  \href{http://dx.doi.org/10.1371/journal.pcbi.1005569}{\textbf{\bibinfo{volume}{13}},
  \bibinfo{pages}{e1005569}}
  \href{http://dx.doi.org/10.1371/journal.pcbi.1005569}{
  (\bibinfo{year}{2017})}.

\bibitem{Oswald17}
\bibinfo{author}{Oswald, L.}, \bibinfo{author}{Grosser, S.},
  \bibinfo{author}{Smith, D.~M.} \& \bibinfo{author}{K\"as, J.~A.}
\newblock \bibinfo{title}{Jamming transitions in cancer}.
\newblock \href{http://dx.doi.org/10.1088/1361-6463/aa8e83}{\emph{J. Phys. D}}
  \href{http://dx.doi.org/10.1088/1361-6463/aa8e83}{\textbf{\bibinfo{volume}{50}},
  \bibinfo{pages}{483001}} \href{http://dx.doi.org/10.1088/1361-6463/aa8e83}{
  (\bibinfo{year}{2017})}.

\bibitem{Chiang16}
\bibinfo{author}{Chiang, M.} \& \bibinfo{author}{Marenduzzo, D.}
\newblock \bibinfo{title}{Glass transitions in the cellular potts model}.
\newblock \href{http://dx.doi.org/10.1209/0295-5075/116/28009}{\emph{Europhys.
  Lett.}}
  \href{http://dx.doi.org/10.1209/0295-5075/116/28009}{\textbf{\bibinfo{volume}{116}},
  \bibinfo{pages}{28009}} \href{http://dx.doi.org/10.1209/0295-5075/116/28009}{
  (\bibinfo{year}{2016})}.

\bibitem{Bi16}
\bibinfo{author}{Bi, D.}, \bibinfo{author}{Yang, X.},
  \bibinfo{author}{Marchetti, M.~C.} \& \bibinfo{author}{Manning, M.~L.}
\newblock \bibinfo{title}{Motility-driven glass and jamming transitions in
  biological tissues}.
\newblock \href{http://dx.doi.org/10.1103/PhysRevX.6.021011}{\emph{Phys. Rev.
  X}}
  \href{http://dx.doi.org/10.1103/PhysRevX.6.021011}{\textbf{\bibinfo{volume}{6}},
  \bibinfo{pages}{021011}} \href{http://dx.doi.org/10.1103/PhysRevX.6.021011}{
  (\bibinfo{year}{2016})}.

\bibitem{Fung10}
\bibinfo{author}{Fung, Y.-C.}
\newblock \emph{\bibinfo{title}{Biomechanics: Mechanical Properties of Living
  Tissues, 2nd ed.}} (\bibinfo{publisher}{Springer New York},
  \bibinfo{year}{2010}).

\bibitem{Dunlop10}
\bibinfo{author}{Dunlop, J. W.~C.}, \bibinfo{author}{Fischer, F.~D.},
  \bibinfo{author}{Gamsj\"ager, E.} \& \bibinfo{author}{Fratzl, P.}
\newblock \bibinfo{title}{A theoretical model for tissue growth in confined
  geometries}.
\newblock \href{http://dx.doi.org/10.1016/j.jmps.2010.04.008}{\emph{J. Mech.
  Phys. Solids}}
  \href{http://dx.doi.org/10.1016/j.jmps.2010.04.008}{\textbf{\bibinfo{volume}{58}},
  \bibinfo{pages}{1073--1087}}
  \href{http://dx.doi.org/10.1016/j.jmps.2010.04.008}{ (\bibinfo{year}{2010})}.

\bibitem{Ambrosi12}
\bibinfo{author}{Ambrosi, D.}, \bibinfo{author}{Preziosi, L.} \&
  \bibinfo{author}{Vitale, G.}
\newblock \bibinfo{title}{The interplay between stress and growth in solid
  tumors}.
\newblock \href{http://dx.doi.org/10.1016/j.mechrescom.2012.01.002}{\emph{Mech.
  Res. Commun.}}
  \href{http://dx.doi.org/10.1016/j.mechrescom.2012.01.002}{\textbf{\bibinfo{volume}{42}},
  \bibinfo{pages}{87--91}}
  \href{http://dx.doi.org/10.1016/j.mechrescom.2012.01.002}{
  (\bibinfo{year}{2012})}.

\bibitem{Shraiman05}
\bibinfo{author}{Shraiman, B.~I.}
\newblock \bibinfo{title}{Mechanical feedback as a possible regulator of tissue
  growth}.
\newblock \href{http://dx.doi.org/10.1073/pnas.0404782102}{\emph{Proc. Natl.
  Acad. Sci. USA}}
  \href{http://dx.doi.org/10.1073/pnas.0404782102}{\textbf{\bibinfo{volume}{102}},
  \bibinfo{pages}{3318--3323}}
  \href{http://dx.doi.org/10.1073/pnas.0404782102}{ (\bibinfo{year}{2005})}.

\bibitem{Byrne03}
\bibinfo{author}{Byrne, H.} \& \bibinfo{author}{Preziosi, L.}
\newblock \bibinfo{title}{Modelling solid tumour growth using the theory of
  mixtures}.
\newblock \href{http://dx.doi.org/10.1093/imammb/20.4.341}{\emph{Math. Med.
  Biol.}}
  \href{http://dx.doi.org/10.1093/imammb/20.4.341}{\textbf{\bibinfo{volume}{20}},
  \bibinfo{pages}{341--366}} \href{http://dx.doi.org/10.1093/imammb/20.4.341}{
  (\bibinfo{year}{2003})}.

\bibitem{Tracqui09}
\bibinfo{author}{Tracqui, P.}
\newblock \bibinfo{title}{Biophysical models of tumour growth}.
\newblock \href{http://dx.doi.org/10.1088/0034-4885/72/5/056701}{\emph{Rep.
  Prog. Phys.}}
  \href{http://dx.doi.org/10.1088/0034-4885/72/5/056701}{\textbf{\bibinfo{volume}{72}},
  \bibinfo{pages}{056701}}
  \href{http://dx.doi.org/10.1088/0034-4885/72/5/056701}{
  (\bibinfo{year}{2009})}.

\bibitem{Rieger15}
\bibinfo{author}{Rieger, H.} \& \bibinfo{author}{Welter, M.}
\newblock \bibinfo{title}{Integrative models of vascular remodeling during
  tumor growth}.
\newblock \href{http://dx.doi.org/10.1002/wsbm.1295}{\emph{WIREs Syst. Biol.
  Med.}}
  \href{http://dx.doi.org/10.1002/wsbm.1295}{\textbf{\bibinfo{volume}{7}},
  \bibinfo{pages}{113--129}} \href{http://dx.doi.org/10.1002/wsbm.1295}{
  (\bibinfo{year}{2015})}.

\bibitem{Fredrich19}
\bibinfo{author}{Fredrich, T.}, \bibinfo{author}{Rieger, H.},
  \bibinfo{author}{Chignola, R.} \& \bibinfo{author}{Milotti, E.}
\newblock \bibinfo{title}{Fine-grained simulations of the microenvironment of
  vascularized tumours}.
\newblock \href{http://dx.doi.org/10.1038/s41598-019-48252-8}{\emph{Sci. Rep.}}
  \href{http://dx.doi.org/10.1038/s41598-019-48252-8}{\textbf{\bibinfo{volume}{9}},
  \bibinfo{pages}{11698}} \href{http://dx.doi.org/10.1038/s41598-019-48252-8}{
  (\bibinfo{year}{2019})}.

\bibitem{Romanczuk09}
\bibinfo{author}{Romanczuk, P.}, \bibinfo{author}{Couzin, I.~D.} \&
  \bibinfo{author}{Schimansky-Geier, L.}
\newblock \bibinfo{title}{Collective motion due to individual escape and
  pursuit response}.
\newblock \href{http://dx.doi.org/10.1103/PhysRevLett.102.010602}{\emph{Phys.
  Rev. Lett.}}
  \href{http://dx.doi.org/10.1103/PhysRevLett.102.010602}{\textbf{\bibinfo{volume}{102}},
  \bibinfo{pages}{010602}}
  \href{http://dx.doi.org/10.1103/PhysRevLett.102.010602}{
  (\bibinfo{year}{2009})}.

\bibitem{Simpson06}
\bibinfo{author}{Simpson, S.~J.}, \bibinfo{author}{Sword, G.~A.},
  \bibinfo{author}{Lorch, P.~D.} \& \bibinfo{author}{Couzin, I.~D.}
\newblock \bibinfo{title}{Cannibal crickets on a forced march for protein and
  salt}.
\newblock \href{http://dx.doi.org/10.1073/pnas.0508915103}{\emph{Proc. Natl.
  Acad. Sci. USA}}
  \href{http://dx.doi.org/10.1073/pnas.0508915103}{\textbf{\bibinfo{volume}{103}},
  \bibinfo{pages}{4152--4156}}
  \href{http://dx.doi.org/10.1073/pnas.0508915103}{ (\bibinfo{year}{2006})}.

\bibitem{Agudo19}
\bibinfo{author}{Agudo-Canalejo, J.} \& \bibinfo{author}{Golestanian, R.}
\newblock \bibinfo{title}{Active phase separation in mixtures of chemically
  interacting particles}.
\newblock \href{http://dx.doi.org/10.1103/PhysRevLett.123.018101}{\emph{Phys.
  Rev. Lett.}}
  \href{http://dx.doi.org/10.1103/PhysRevLett.123.018101}{\textbf{\bibinfo{volume}{123}},
  \bibinfo{pages}{018101}}
  \href{http://dx.doi.org/10.1103/PhysRevLett.123.018101}{
  (\bibinfo{year}{2019})}.

\bibitem{Pearce14}
\bibinfo{author}{Pearce, D. J.~G.}, \bibinfo{author}{Miller, A.~M.},
  \bibinfo{author}{Rowlands, G.} \& \bibinfo{author}{Turner, M.~S.}
\newblock \bibinfo{title}{Role of projection in the control of bird flocks}.
\newblock \href{http://dx.doi.org/10.1073/pnas.1402202111}{\emph{Proc. Natl.
  Acad. Sci. USA}}
  \href{http://dx.doi.org/10.1073/pnas.1402202111}{\textbf{\bibinfo{volume}{111}},
  \bibinfo{pages}{10422--10426}}
  \href{http://dx.doi.org/10.1073/pnas.1402202111}{ (\bibinfo{year}{2014})}.

\bibitem{Lavergne19}
\bibinfo{author}{Lavergne, F.~A.}, \bibinfo{author}{Wendehenne, H.},
  \bibinfo{author}{B\"auerle, T.} \& \bibinfo{author}{Bechinger, C.}
\newblock \bibinfo{title}{Group formation and cohesion of active particles with
  visual perception-dependent motility}.
\newblock \href{http://dx.doi.org/10.1126/science.aau5347}{\emph{Science}}
  \href{http://dx.doi.org/10.1126/science.aau5347}{\textbf{\bibinfo{volume}{364}},
  \bibinfo{pages}{70--74}} \href{http://dx.doi.org/10.1126/science.aau5347}{
  (\bibinfo{year}{2019})}.

\bibitem{Ballerini08}
\bibinfo{author}{Ballerini, M.}, \bibinfo{author}{Cabibbo, N.},
  \bibinfo{author}{Candelier, R.}, \bibinfo{author}{Cavagna, A.},
  \bibinfo{author}{Cisbani, E.}, \bibinfo{author}{Giardina, I.},
  \bibinfo{author}{Lecomte, V.}, \bibinfo{author}{Orlandi, A.},
  \bibinfo{author}{Parisi, G.}, \bibinfo{author}{Procaccini, A.},
  \bibinfo{author}{Viale, M.} \& \bibinfo{author}{Zdravkovic, V.}
\newblock \bibinfo{title}{Interaction ruling animal collective behavior depends
  on topological rather than metric distance: Evidence from a field study}.
\newblock \href{http://dx.doi.org/10.1073/pnas.0711437105}{\emph{Proc. Natl.
  Acad. Sci. USA}}
  \href{http://dx.doi.org/10.1073/pnas.0711437105}{\textbf{\bibinfo{volume}{105}},
  \bibinfo{pages}{1232--1237}}
  \href{http://dx.doi.org/10.1073/pnas.0711437105}{ (\bibinfo{year}{2008})}.

\bibitem{Bajec09}
\bibinfo{author}{Bajec, I.~L.} \& \bibinfo{author}{Heppner, F.~H.}
\newblock \bibinfo{title}{Organized flight in birds}.
\newblock \href{http://dx.doi.org/10.1016/j.anbehav.2009.07.007}{\emph{Animal
  Behav.}}
  \href{http://dx.doi.org/10.1016/j.anbehav.2009.07.007}{\textbf{\bibinfo{volume}{78}},
  \bibinfo{pages}{777--789}}
  \href{http://dx.doi.org/10.1016/j.anbehav.2009.07.007}{
  (\bibinfo{year}{2009})}.

\bibitem{Mijalkov16}
\bibinfo{author}{Mijalkov, M.}, \bibinfo{author}{McDaniel, A.},
  \bibinfo{author}{Wehr, J.} \& \bibinfo{author}{Volpe, G.}
\newblock \bibinfo{title}{Engineering sensorial delay to control phototaxis and
  emergent collective behaviors}.
\newblock \href{http://dx.doi.org/10.1103/PhysRevX.6.011008}{\emph{Phys. Rev.
  X}}
  \href{http://dx.doi.org/10.1103/PhysRevX.6.011008}{\textbf{\bibinfo{volume}{6}},
  \bibinfo{pages}{011008}} \href{http://dx.doi.org/10.1103/PhysRevX.6.011008}{
  (\bibinfo{year}{2016})}.

\bibitem{Charlesworth19}
\bibinfo{author}{Charlesworth, H.~J.} \& \bibinfo{author}{Turner, M.~S.}
\newblock \bibinfo{title}{Intrinsically motivated collective motion}.
\newblock \href{http://dx.doi.org/10.1073/pnas.1822069116}{\emph{Proc. Natl.
  Acad. Sci. USA}}
  \href{http://dx.doi.org/10.1073/pnas.1822069116}{\textbf{\bibinfo{volume}{116}},
  \bibinfo{pages}{15362--15367}}
  \href{http://dx.doi.org/10.1073/pnas.1822069116}{ (\bibinfo{year}{2019})}.

\bibitem{Khadka18}
\bibinfo{author}{Khadka, U.}, \bibinfo{author}{Holubec, V.},
  \bibinfo{author}{Yang, H.} \& \bibinfo{author}{Cichos, F.}
\newblock \bibinfo{title}{Active particles bound by information flows}.
\newblock \href{http://dx.doi.org/10.1038/s41467-018-06445-1}{\emph{Nat.
  Commun.}}
  \href{http://dx.doi.org/10.1038/s41467-018-06445-1}{\textbf{\bibinfo{volume}{9}},
  \bibinfo{pages}{3864}} \href{http://dx.doi.org/10.1038/s41467-018-06445-1}{
  (\bibinfo{year}{2018})}.

\bibitem{Mann15}
\bibinfo{author}{Mann, R.~P.} \& \bibinfo{author}{Garnett, R.}
\newblock \bibinfo{title}{The entropic basis of collective behaviour}.
\newblock \href{http://dx.doi.org/10.1098/rsif.2015.0037}{\emph{J. R. Soc.
  Interface}}
  \href{http://dx.doi.org/10.1098/rsif.2015.0037}{\textbf{\bibinfo{volume}{12}},
  \bibinfo{pages}{20150037}} \href{http://dx.doi.org/10.1098/rsif.2015.0037}{
  (\bibinfo{year}{2015})}.

\bibitem{Ward08}
\bibinfo{author}{Ward, A. J.~W.}, \bibinfo{author}{Sumpter, D. J.~T.},
  \bibinfo{author}{Couzin, I.~D.}, \bibinfo{author}{Hart, P. J.~B.} \&
  \bibinfo{author}{Krause, J.}
\newblock \bibinfo{title}{Quorum decision-making facilitates information
  transfer in fish shoals}.
\newblock \href{http://dx.doi.org/10.1073/pnas.0710344105}{\emph{Proc. Natl.
  Acad. Sci. USA}}
  \href{http://dx.doi.org/10.1073/pnas.0710344105}{\textbf{\bibinfo{volume}{105}},
  \bibinfo{pages}{6948--6953}}
  \href{http://dx.doi.org/10.1073/pnas.0710344105}{ (\bibinfo{year}{2008})}.

\bibitem{Abaurrea18}
\bibinfo{author}{Abaurrea~Velasco, C.}, \bibinfo{author}{Abkenar, M.},
  \bibinfo{author}{Gompper, G.} \& \bibinfo{author}{Auth, T.}
\newblock \bibinfo{title}{Collective behavior of self-propelled rods with
  quorum sensing}.
\newblock \href{http://dx.doi.org/10.1103/PhysRevE.98.022605}{\emph{Phys. Rev.
  E}}
  \href{http://dx.doi.org/10.1103/PhysRevE.98.022605}{\textbf{\bibinfo{volume}{98}},
  \bibinfo{pages}{022605}} \href{http://dx.doi.org/10.1103/PhysRevE.98.022605}{
  (\bibinfo{year}{2018})}.

\bibitem{Castellano09}
\bibinfo{author}{Castellano, C.}, \bibinfo{author}{Fortunato, S.} \&
  \bibinfo{author}{Loreto, V.}
\newblock \bibinfo{title}{Statistical physics of social dynamics}.
\newblock \href{http://dx.doi.org/10.1103/RevModPhys.81.591}{\emph{Rev. Mod.
  Phys.}}
  \href{http://dx.doi.org/10.1103/RevModPhys.81.591}{\textbf{\bibinfo{volume}{81}},
  \bibinfo{pages}{591--646}}
  \href{http://dx.doi.org/10.1103/RevModPhys.81.591}{ (\bibinfo{year}{2009})}.

\bibitem{King08}
\bibinfo{author}{King, A.~J.}, \bibinfo{author}{Douglas, C.~M.},
  \bibinfo{author}{Huchard, E.}, \bibinfo{author}{Isaac, N.~J.} \&
  \bibinfo{author}{Cowlishaw, G.}
\newblock \bibinfo{title}{Dominance and affiliation mediate despotism in a
  social primate}.
\newblock \href{http://dx.doi.org/10.1016/j.cub.2008.10.048}{\emph{Curr.
  Biol.}}
  \href{http://dx.doi.org/10.1016/j.cub.2008.10.048}{\textbf{\bibinfo{volume}{18}},
  \bibinfo{pages}{1833--1838}}
  \href{http://dx.doi.org/10.1016/j.cub.2008.10.048}{ (\bibinfo{year}{2008})}.

\bibitem{Couzin05}
\bibinfo{author}{Couzin, I.~D.}, \bibinfo{author}{Krause, J.},
  \bibinfo{author}{Franks, N.~R.} \& \bibinfo{author}{Levin, S.~A.}
\newblock \bibinfo{title}{Effective leadership and decision-making in animal
  groups on the move}.
\newblock \href{http://dx.doi.org/10.1038/nature03236}{\emph{Nature}}
  \href{http://dx.doi.org/10.1038/nature03236}{\textbf{\bibinfo{volume}{433}},
  \bibinfo{pages}{513--516}} \href{http://dx.doi.org/10.1038/nature03236}{
  (\bibinfo{year}{2005})}.

\bibitem{Freeman09}
\bibinfo{author}{Freeman, R.} \& \bibinfo{author}{Biro, D.}
\newblock \bibinfo{title}{Modelling group navigation: Dominance and democracy
  in homing pigeons}.
\newblock \href{http://dx.doi.org/10.1017/S0373463308005080}{\emph{J.
  Navigation}}
  \href{http://dx.doi.org/10.1017/S0373463308005080}{\textbf{\bibinfo{volume}{62}},
  \bibinfo{pages}{33--40}} \href{http://dx.doi.org/10.1017/S0373463308005080}{
  (\bibinfo{year}{2009})}.

\bibitem{Baeuerle18}
\bibinfo{author}{B{\"a}uerle, T.}, \bibinfo{author}{Fischer, A.},
  \bibinfo{author}{Speck, T.} \& \bibinfo{author}{Bechinger, C.}
\newblock \bibinfo{title}{Self-organization of active particles by quorum
  sensing rules}.
\newblock \href{http://dx.doi.org/10.1038/s41467-018-05675-7}{\emph{Nat.
  Commun.}}
  \href{http://dx.doi.org/10.1038/s41467-018-05675-7}{\textbf{\bibinfo{volume}{9}},
  \bibinfo{pages}{3232}} \href{http://dx.doi.org/10.1038/s41467-018-05675-7}{
  (\bibinfo{year}{2018})}.

\bibitem{Moussaid11}
\bibinfo{author}{Moussa{\"\i}d, M.}, \bibinfo{author}{Helbing, D.} \&
  \bibinfo{author}{Theraulaz, G.}
\newblock \bibinfo{title}{How simple rules determine pedestrian behavior and
  crowd disasters}.
\newblock \href{http://dx.doi.org/10.1073/pnas.1016507108}{\emph{Proc. Natl.
  Acad. Sci. USA}}
  \href{http://dx.doi.org/10.1073/pnas.1016507108}{\textbf{\bibinfo{volume}{108}},
  \bibinfo{pages}{6884--6888}}
  \href{http://dx.doi.org/10.1073/pnas.1016507108}{ (\bibinfo{year}{2011})}.

\bibitem{Faria10}
\bibinfo{author}{Faria, J.~J.}, \bibinfo{author}{Dyer, J.~R.},
  \bibinfo{author}{Tosh, C.~R.} \& \bibinfo{author}{Krause, J.}
\newblock \bibinfo{title}{Leadership and social information use in human
  crowds}.
\newblock \href{http://dx.doi.org/10.1016/j.anbehav.2009.12.039}{\emph{Animal
  Behav.}}
  \href{http://dx.doi.org/10.1016/j.anbehav.2009.12.039}{\textbf{\bibinfo{volume}{79}},
  \bibinfo{pages}{895--901}}
  \href{http://dx.doi.org/10.1016/j.anbehav.2009.12.039}{
  (\bibinfo{year}{2010})}.

\bibitem{Bain19}
\bibinfo{author}{Bain, N.} \& \bibinfo{author}{Bartolo, D.}
\newblock \bibinfo{title}{Dynamic response and hydrodynamics of polarized
  crowds}.
\newblock \href{http://dx.doi.org/10.1126/science.aat9891}{\emph{Science}}
  \href{http://dx.doi.org/10.1126/science.aat9891}{\textbf{\bibinfo{volume}{363}},
  \bibinfo{pages}{46--49}} \href{http://dx.doi.org/10.1126/science.aat9891}{
  (\bibinfo{year}{2019})}.

\bibitem{Kim18}
\bibinfo{author}{Kim, M.-C.}, \bibinfo{author}{Silberberg, Y.~R.},
  \bibinfo{author}{Abeyaratne, R.}, \bibinfo{author}{Kamm, R.~D.} \&
  \bibinfo{author}{Asada, H.~H.}
\newblock \bibinfo{title}{Computational modeling of three-dimensional
  ecm-rigidity sensing to guide directed cell migration}.
\newblock \href{http://dx.doi.org/10.1073/pnas.1717230115}{\emph{Proc. Natl.
  Acad. Sci. USA}}
  \href{http://dx.doi.org/10.1073/pnas.1717230115}{\textbf{\bibinfo{volume}{115}},
  \bibinfo{pages}{E390--E399}}
  \href{http://dx.doi.org/10.1073/pnas.1717230115}{ (\bibinfo{year}{2018})}.

\bibitem{Paluch13}
\bibinfo{author}{Paluch, E.~K.} \& \bibinfo{author}{Raz, E.}
\newblock \bibinfo{title}{The role and regulation of blebs in cell migration}.
\newblock \href{http://dx.doi.org/10.1016/j.ceb.2013.05.005}{\emph{Curr. Opin.
  Cell Biol.}}
  \href{http://dx.doi.org/10.1016/j.ceb.2013.05.005}{\textbf{\bibinfo{volume}{25}},
  \bibinfo{pages}{582--590}}
  \href{http://dx.doi.org/10.1016/j.ceb.2013.05.005}{ (\bibinfo{year}{2013})}.

\bibitem{Tozluoglu13}
\bibinfo{author}{Tozluoglu, M.}, \bibinfo{author}{Tournier, A.~L.},
  \bibinfo{author}{Jenkins, R.~P.}, \bibinfo{author}{Hooper, S.},
  \bibinfo{author}{Bates, P.~A.} \& \bibinfo{author}{Sahai, E.}
\newblock \bibinfo{title}{Matrix geometry determines optimal cancer cell
  migration strategy and modulates response to interventions}.
\newblock \href{http://dx.doi.org/10.1038/ncb2775}{\emph{Nat. Cell Biol.}}
  \href{http://dx.doi.org/10.1038/ncb2775}{\textbf{\bibinfo{volume}{15}},
  \bibinfo{pages}{751}} \href{http://dx.doi.org/10.1038/ncb2775}{
  (\bibinfo{year}{2013})}.

\bibitem{Besser07}
\bibinfo{author}{Besser, A.} \& \bibinfo{author}{Schwarz, U.~S.}
\newblock \bibinfo{title}{Coupling biochemistry and mechanics in cell adhesion:
  a model for inhomogeneous stress fiber contraction}.
\newblock \href{http://dx.doi.org/10.1088/1367-2630/9/11/425}{\emph{New J.
  Phys.}}
  \href{http://dx.doi.org/10.1088/1367-2630/9/11/425}{\textbf{\bibinfo{volume}{9}},
  \bibinfo{pages}{425--425}}
  \href{http://dx.doi.org/10.1088/1367-2630/9/11/425}{ (\bibinfo{year}{2007})}.

\bibitem{Nishikawa17}
\bibinfo{author}{Nishikawa, M.}, \bibinfo{author}{Naganathan, S.~R.},
  \bibinfo{author}{J\"ulicher, F.} \& \bibinfo{author}{Grill, S.~W.}
\newblock \bibinfo{title}{Controlling contractile instabilities in the
  actomyosin cortex}.
\newblock \href{http://dx.doi.org/10.7554/eLife.19595}{\emph{eLife}}
  \href{http://dx.doi.org/10.7554/eLife.19595}{\textbf{\bibinfo{volume}{6}},
  \bibinfo{pages}{e19595}} \href{http://dx.doi.org/10.7554/eLife.19595}{
  (\bibinfo{year}{2017})}.

\bibitem{Gross19}
\bibinfo{author}{Gross, P.}, \bibinfo{author}{Kumar, K.~V.},
  \bibinfo{author}{Goehring, N.~W.}, \bibinfo{author}{Bois, J.~S.},
  \bibinfo{author}{Hoege, C.}, \bibinfo{author}{J{\"u}licher, F.} \&
  \bibinfo{author}{Grill, S.~W.}
\newblock \bibinfo{title}{Guiding self-organized pattern formation in cell
  polarity establishment}.
\newblock \href{http://dx.doi.org/10.1038/s41567-018-0358-7}{\emph{Nat. Phys.}}
  \href{http://dx.doi.org/10.1038/s41567-018-0358-7}{\textbf{\bibinfo{volume}{15}},
  \bibinfo{pages}{293--300}}
  \href{http://dx.doi.org/10.1038/s41567-018-0358-7}{ (\bibinfo{year}{2019})}.

\end{thebibliography}

\centering
\includegraphics[width=0.88\textwidth]{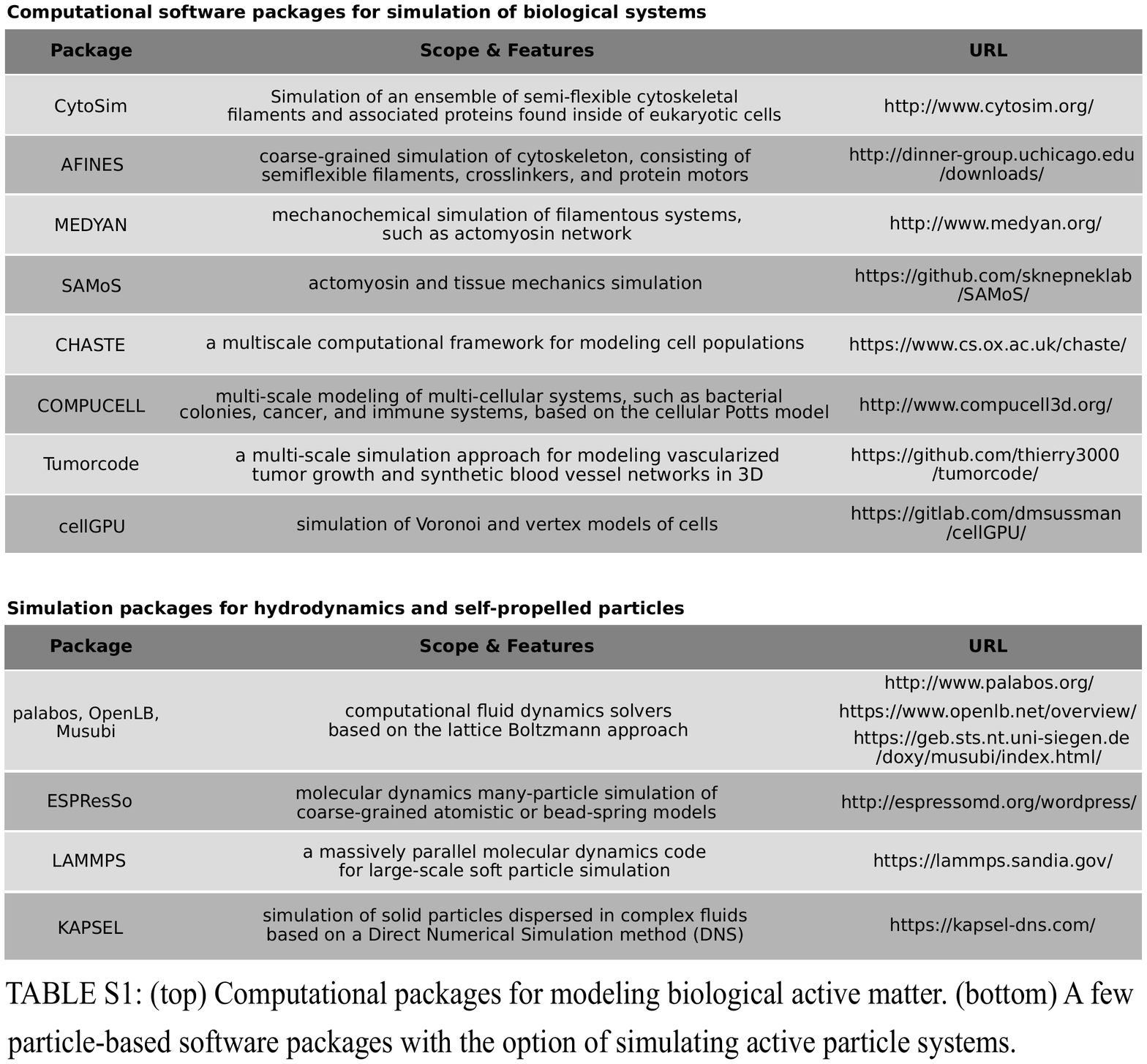}

\end{document}